\documentclass[useAMS,usenatbib,psfig]{mn2e}
\pdfoutput=1
\usepackage{graphicx} 
\usepackage{grffile}
\usepackage{longtable}
\usepackage{rotating}
\usepackage{tablefootnote}
\title[The VMC Survey. XIII. Type II Cepheids in the LMC]{The
  VMC Survey. XIII. Type II Cepheids in the Large Magellanic Cloud\thanks{Based on
    observations made with VISTA at ESO under programme ID 179.B-2003.}}
\author[V. Ripepi et al.]{V. Ripepi$^{1}$\thanks{E-mail:
ripepi@oacn.inaf.it},
M. I. Moretti$^{1,2,3}$,  
M. Marconi$^{1}$, 
G. Clementini$^{2}$,
M-R. L. Cioni$^{4,5}$, 
\and  
R. de Grijs$^{6,7}$, 
J. P. Emerson$^{8}$,
M.A.T. Groenewegen$^{9}$,
V. D. Ivanov$^{10}$,
\and
T. Muraveva$^{2}$,
 A. E. Piatti$^{11,12}$,
S. Subramanian$^{13}$ 
\\
$^{1}$ INAF-Osservatorio Astronomico di Capodimonte, Via Moiariello
 16, 80131, Naples, Italy \\
$^{2}$ INAF-Osservatorio Astronomico di Bologna, via Ranzani 1, 40127, 
Bologna, Italy \\ 
$^{3}$ Scuola Normale Superiore, Piazza dei Cavalieri, 7, 56126, Pisa,
Italy\\ 
$^{4}$ University of Hertfordshire, Physics Astronomy and
 Mathematics, College Lane, Hatfield AL10 9AB, UK \\
$^{5}$ Leibnitz-Institut f\"{u}r Astrophysik Potsdam, An der
Sternwarte 16, 14482 Potsdam, Germany \\
$^{6}$ Kavli Institute for Astronomy and Astrophysics, Peking 
University, Yi He Yuan Lu 5, Hai Dian District, Beijing 100871, China \\
$^{7}$ Department of Astronomy, Peking University, Yi He Yuan Lu 5, Hai Dian District, Beijing 100871, China\\
$^{8}$ Astronomy Unit, School of Physics \& Astronomy, Queen Mary 
University of London, Mile End Road, London E1 4NS, UK\\
$^{9}$ Koninklijke Sterrenwacht van Belgi\"e, Ringlaan 3, 1180, Brussel, Belgium\\
$^{10}$ European Southern Observatory, Ave. Alonso de Cordova 3107, Casilla 19, Chile \\
$^{11}$ Observatorio Astron\'omico, Universidad Nacional de C\'odoba, Laprida 854, 5000,
C\'ordoba, Argentina. \\
$^{12}$ Consejo Nacional de Investigaciones Cient\'{\i}ficas y T\'ecnicas, Av. Rivadavia 1917, C1033AAJ,
Buenos Aires, Argentina \\
$^{13}$ Indian Institute of Astrophysics, Koramangala, Bangalore, 560 034, India
}
\begin{document}

\date{}

\pagerange{\pageref{firstpage}--\pageref{lastpage}} \pubyear{2002}

\maketitle

\label{firstpage}

\begin{abstract}
{
The VISTA survey of the Magellanic
  Clouds System (VMC) is collecting deep
  $K_\mathrm{s}$--band time--series photometry of the pulsating
  variable stars hosted in the system formed by the two Magellanic
  Clouds and the Bridge connecting them.
  In this paper we have analysed a sample of 130 Large Magellanic
  Cloud (LMC) Type II Cepheids (T2CEPs) found in tiles with complete
  or near complete VMC observations for which identification and 
optical magnitudes were obtained from the OGLE~III survey. 
 We present $J$ and $K_\mathrm{s}$ light curves for all 130
  pulsators, including 41 BL Her, 62 W Vir (12 pW Vir) and 27 RV Tau
  variables. 
We complement our near-infrared photometry with the $V$ magnitudes from the OGLE~III survey, allowing us to build 
a variety of Period-Luminosity ($PL$), Period-Luminosity-Colour ($PLC$) and
Period-Wesenheit ($PW$) relationships, including any combination of the $V, J, K_\mathrm{s}$ filters and valid for BL Her and W Vir classes. These relationships were calibrated  in terms of the LMC distance modulus, while an
independent absolute calibration of the $PL(K_\mathrm{s})$ and the $PW(K_\mathrm{s},V)$ was derived on the basis
of distances obtained from  $Hubble~Space~Telescope$ parallaxes and Baade-Wesselink technique. 
When applied to the LMC and to the Galactic Globular Clusters hosting T2CEPs, these relations seem to show that: 1) the two population II standard candles RR Lyrae and T2CEPs give results in excellent agreement with each other; 2) there
is a discrepancy of $\sim$0.1 mag between
population II standard candles and Classical Cepheids when the distances are 
gauged in a similar way for all the quoted pulsators. However,
given the uncertainties, this discrepancy is within the formal 1$\sigma$ uncertainties.
}
\end{abstract}


\section{Introduction}

The Magellanic Clouds (MCs) are fundamental benchmarks in the
framework of stellar populations and galactic evolution investigations
\citep[see e.g.][]{HZ04,HZ09,Ripepi2014b}. The ongoing interaction with the Milky
Way also allows us to study in detail the complex mechanisms that rule
the interaction among galaxies \citep[see
e.g.][]{putman98,muller04,stanim04,bekki07,Venzmer2012,For2013}. 
Additionally, the MCs are more metal poor than our Galaxy and host a
large population of young populous clusters, thus they are useful to test the physical and
 numerical assumptions at the basis of stellar evolution 
codes \citep[see e.g.][]{Matteucci2002,bro04,nl12}.

The Large Magellanic Cloud (LMC) is also fundamental in the context of
the extragalactic distance scale. Indeed, it represents the first critical 
step on which the calibration of Classical Cepheid (CC) 
Period-Luminosity ($PL$) relations and in turn of secondary
distance indicators relies \citep[see e.g.][and references therein]{f01,Rie11,w12}.
At the same time, the LMC hosts several  thousand of  RR Lyrae variables,
which represent  the most important Population II standard
candles  through the well known $M_V(\rm RR)$--[Fe/H]  and 
near-infrared (NIR) metal dependent $PL$ relations. Moreover, the LMC
contains tens of thousands of intermediate-age Red Clump stars, which can 
profitably used as accurate distance indicators \citep[see e.g.][]{Laney2012,Subramanian2013}. 
Hence, the LMC is the ideal place 
to compare the distance scales derived from Population I and
II indicators \citep[see e.g.][and references therein]{Clementini2003,w12,degrijs2014}.
In particular, NIR observations of pulsating stars \citep[see
e.g.][and references therein]{Ripepi12a,Moretti14,Ripepi2014a} provide
stringent constraints to the calibration of their distance scale
thanks to the existence of well defined $PL$, Period-Luminosity-Color
($PLC$) and Period-Wesenheit ($PW$) relations at these wavelengths 
\citep[see][for the definition of Wesenheit functions]{m82,mf91}. 
 
The {\it VISTA\footnote{Visible and Infrared Survey Telescope for Astronomy} near-infrared $YJK_\mathrm{s}$ survey of the Magellanic
  Clouds system} \citep[VMC; ][]{Cioni11} aims at observing a wide
area across the Magellanic system, including the relatively unexplored
Bridge connecting the two Clouds. This ESO public survey relies on the
VIRCAM camera \citep{Dalton_etal06} of the ESO VISTA telescope
\citep{Emerson_etal06} to obtain deep NIR photometric data in the $Y$,
$J$ and $K_\mathrm{s}$ filters.  The main aims are: i) 
to reconstruct the spatially-resolved star-formation history (SFH) and
ii) to infer an accurate 3D map of the whole Magellanic system.
The properties of pulsating stars observed by VMC and adopted as
tracers of three different stellar populations, namely CCs (younger
than few hundreds  Myr), RR Lyrae stars (older than 9-10 Gyr) and
Anomalous Cepheids (traditionally associated to an intermediate age
population with few Gyr), have been discussed in recent papers by our
team \citep{Ripepi12a,Ripepi12b,Moretti14,Ripepi2014a}. In these
papers, relevant results on the calibration of the distance scales for
all these important standard candles have been provided.  

An additional class of Population II pulsating stars is represented by
the so-called Type II Cepheids \citep[T2CEPs, see e.g.][]{c98,Sandage2006}.
These objects show periods from $\sim$1 to $\sim$ 20 days and are observed in
Galactic Globular Clusters (GGCs) with few RR Lyrae stars and blue
horizontal branch (HB) morphology.  They are brighter but less massive
than RR Lyrae stars for similar metal content \citep[see
e.g.][]{Caputo2004}.  
T2CEPs are often separated into BL Herculis stars (BL Her; periods
between 1 and 4 days) and W Virginis 
stars (W Vir; periods between 4 and 20 days) and, as discussed by several
authors
\citep[e.g.][]{Wallerstein1984,Gingold1985,Harris1985,Bono1997,Wallerstein2002},
originate from hot, low-mass stellar structures, starting their
central He burning on the blue side of the RR Lyrae gap.
Moreover, according to several authors \citep[see e.g.][and references
therein]{Feast2008,Feast2010} RV Tauri stars, with periods from about
20 to 150 days and often irregular light curves, are considered as an
additional subgroup of the Type II Cepheid class. Their evolutionary
phase corresponds to the post Asymptotic Giant Branch phase path towards 
planetary nebula status.  This feature corresponds to the latest
evolution of intermediate mass stellar structures and for this reason
the claimed link with the low mass W Vir stars should be considered
with caution.

In addition to the three quoted groups, \citet{sos08} suggested the
existence of a new sub-class of T2CEPs, the so-called peculiar W Vir
(pWVir) stars. These objects show peculiar light curves and, at
constant period, are usually brighter than normal T2CEPs. It is likely
that pWVir belong to binary systems, however the true nature of these
variables remains uncertain.

\citet{Nemec1994} derived metal-dependent period-luminosity ($PL$)
relations in various optical photometric bands both in the fundamental and in
the first overtone modes but subsequently \citet{Kubiak2003} found that
all the observed T2CEPs in the OGLE II \citep[Optical
Gravitational Lensing Experiment; ][]{Udalski1992} sample, with periods in the
range $\sim$0.7 to about 10 days, satisfy the same $PL$ relation.  This
result was then confirmed by \citet{Pritzl2003} and \citet{Matsunaga2006} for
GGCs, by \citet{Groenewegen2008} for the Galactic Bulge and again by \citet{sos08}
on the basis of OGLE~III data.  

From the theoretical point of view, 
\citet{Dicriscienzo2007} and \citet{Marconi2007} have investigated the properties
of BL Her stars,  by adopting an updated evolutionary and
pulsational scenario for metallicities in the range of Z = 0.0001 to Z =
0.004. The predicted $PL$ and $PW$ relations derived on the basis of
these models were found to be in good agreement with the slopes
determined by the variables observed in GGCs.  Moreover, the distances
obtained from the theoretical relations for T2CEPs agree within the
errors with the RR Lyrae-based values.

In the NIR bands, a tight $PL$ for 46 T2CEPs hosted in GGCs was found
by \citet{Matsunaga2006}. Such relations were calibrated by
\citet{Feast2008} by means of pulsation parallaxes of nearby T2CEPs 
and used to estimate the distances of the LMC and the Galactic Centre.
Subsequent investigations \citep{Matsunaga2009,Matsunaga2011}
confirmed the existence of such tight $PL$ relations in the
$J,H,K_\mathrm{s}$ bands for the T2CEPs belonging to the LMC and Small
Magellanic Cloud (SMC) found by the OGLE~III collaboration
\citep{sos08}. 
However, the NIR observations at the base of these studies consist of only two
epochs for each variable light curve obtained with the Infrared Survey Facility (IRSF)
1.4m telescope in South Africa. The average magnitudes of the T2CEPs
analysed in that paper 
were derived by comparison with the OGLE~III $I$-band photometry.   

In the context of the VMC survey, we
present here the NIR results for a significant sample of T2CEPs
in the LMC, based on high precision and well-sampled $K_\mathrm{s}$-
band light curves.

The VMC data for the T2CEPs are presented in Section 2.  The $PL$, $PLC$ and
$PW$ relations involving $J$ and  $K_\mathrm{s}$ bands are calculated in Section 3.
Section 4 includes the absolute calibration of such relations and a
comparison with the literature. In Sections 5 we discuss the
results; a concise summary (Sect. 6) concludes the paper.

\begin{figure}
\includegraphics[width=8.8cm]{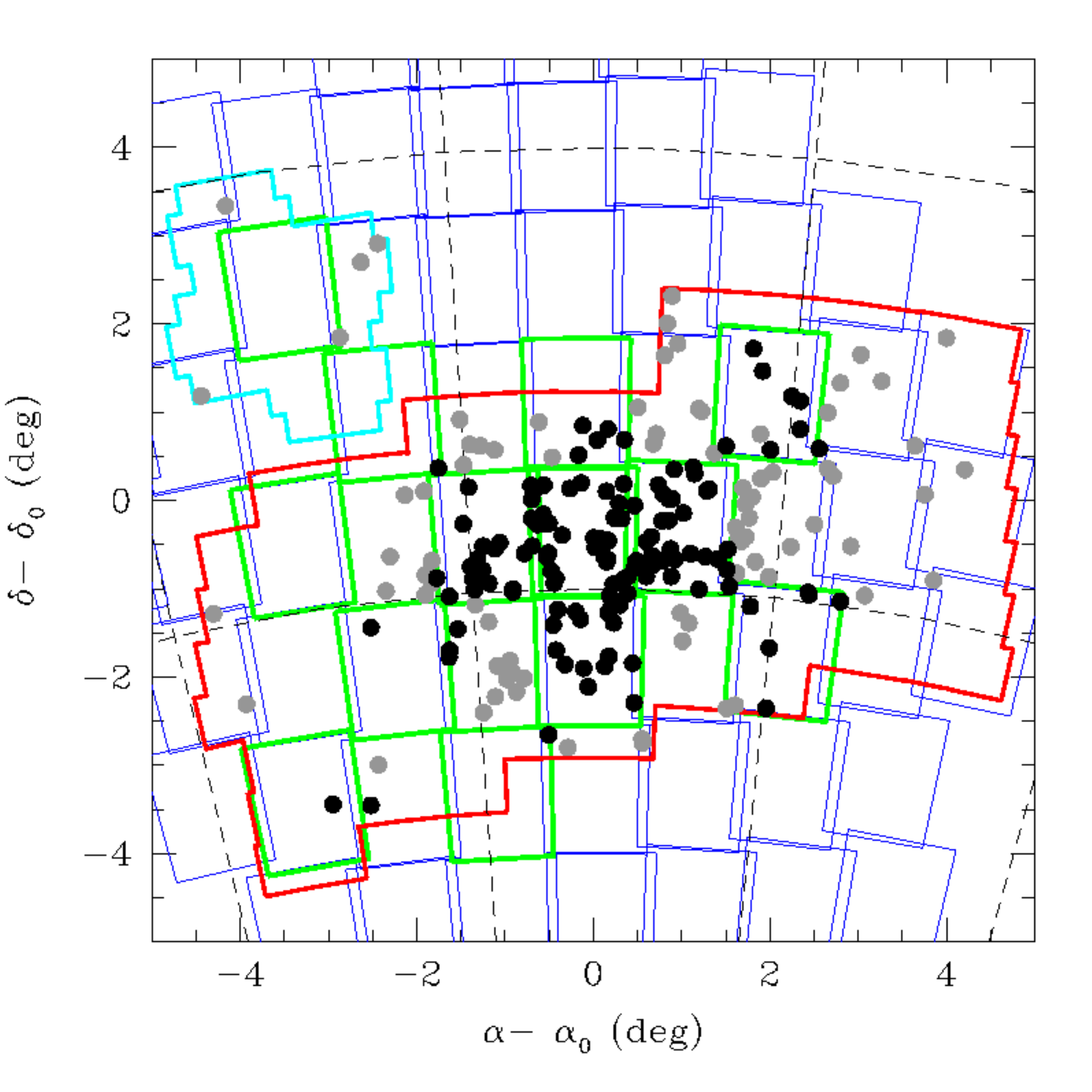}
\caption{Distribution of the known T2CEPs over the LMC (projected on the
  sky adopting $\alpha_0 = 81.0$ deg and $\delta_0 = -69.0$ deg). 
Grey symbols show all the T2CEPs detected by the OGLE
collaboration, whereas  black filled circles present the T2CEPs falling in
  the VMC tiles and studied in this paper.  Thin blue and thick green
  squares (distorted by the projection into the sky) show part of the
  VMC tiles in the LMC and the 13 tiles treated in this paper,
  respectively. The thick red and light blue lines show the areas covered by OGLE
  III and IV (released to date), respectively.}
\label{figMap}
\end{figure}

\begin{table}
\scriptsize 
\caption{Number of T2CEPs in the 13 VMC tiles  analysed in this paper, 
according to OGLE~III/IV.}
\label{tabLog}
\begin{center}
\begin{tabular}{cccc}
\hline 
\noalign{\smallskip} 
Tile & RA (centre)  & DEC (centre) & n$_{\rm T2CEP}$ \\
LMC       & J(2000)   &   J(2000) &  \\
\noalign{\smallskip}
\hline 
\noalign{\smallskip}  
LMC\,4\_6   & 05:38:00.41  & $-$72:17:20.0 & 1   \\  
LMC\,4\_8   & 06:06:32.95  & $-$72:08:31.2 & 2   \\  
LMC\,5\_3   & 04:58:11.66 &  $-$70:35:28.0 & 6   \\ 
LMC\,5\_5   & 05:24:30.34 &  $-$70:48:34.2 & 17   \\  
LMC\,5\_7   & 05:51:04.87 &  $-$70:47:31.2 & 4   \\  
LMC\,6\_4   & 05:12:55.80 &  $-$69:16:39.4 & 33   \\ 
LMC\,6\_5   & 05:25:16.27 &  $-$69:21:08.3 & 31 \\ 
LMC\,6\_6   & 05:37:40.01 &  $-$69:22:18.1 & 20   \\ 
LMC\,6\_8   & 06:02:22.00 &  $-$69:14:42.4 & 0 \\
LMC\,7\_3   & 05:02:55.20 &  $-$67:42:14.8 & 9    \\ 
LMC\,7\_5   & 05:25:58.44 &  $-$67:53:42.0 & 6    \\ 
LMC\,7\_7  &  05:49:12.19 &  $-$67:52:45.5 & 1   \\
LMC\,8\_8  & 05:59:23.14 & $-$66:20:28.7 & 0 \\
\noalign{\smallskip}
\hline 
\noalign{\smallskip}
\end{tabular}
\end{center}
\end{table}

\section{Type II Cepheids in the VMC survey}

T2CEPs in the LMC were identified and studied in the
$V,I$ optical bands by \citet[][]{sos08} in the context of
the OGLE~III project\footnote{data available at http://ogle.astrouw.edu.pl}. 
We have also considered the recent  early release of the 
OGLE IV survey \citep{sos12}, including the South Ecliptic Pole (SEP) 
which, in turn,  lies within our tile LMC 8\_8. 
In these surveys, a total of 207 T2CEPs were found (203 by OGLE~III and  
4 by OGLE~IV\footnote{\citet{sos12} also report the discovery of one
  yellow semiregular variable (SRd). Since this class of variables is
  not considered in this paper, we ignore this object in the present
  work.}), of which 65 are BL Her, 98 are W Vir and 44 are RV Tau pulsators.    

In this paper we present results for the T2CEPs included on 
13 ``tiles'' (1.5 deg$^2$) completely or nearly completely
observed, processed and catalogued by the VMC survey as of March 
2013 (and overlapping with the area investigated by OGLE~III), 
namely the tiles LMC 4\_6, 4\_8, 5\_3, 5\_5, 5\_7, 6\_4,  6\_5, 6\_6, 6\_8,
 7\_3, 7\_5, 7\_7 and 8\_8 (see Fig.~\ref{figMap} and Table~\ref{tabLog}). 
Tile LMC 6\_6 is centred on the well known  30 Dor star
forming region; tiles LMC 5\_5, 6\_4 and 6\_5 are placed on the bar of the LMC. 
The remaining tiles lie in less crowded regions of the galaxy. 

\begin{table*}
\scriptsize\tiny
\caption{Cross-identification and main characteristics of the T2CEPs in the 13 ``tiles'' analysed in this paper.  
The columns report: 1) OGLE identification; 2) right ascension (OGLE);
3) declination (OGLE); 4) variability class; 
5) intensity-averaged $I$ magnitude (OGLE); 6) intensity-averaged
$V$ magnitude (OGLE); 7) period (OGLE); 8) epoch of maximum light $-$2450000 d (OGLE);  
9) VMC identification as in the internal VSA release VMC v1.2/v1.3 (
August, 5 2013); 10) VMC Tile; 11) Number of $J$ and
$K_\mathrm{s}$ epochs, respectively; 12) Notes on individual stars.}
\label{tabData}
\begin{center}
\begin{tabular}{cccccccccccc}
\hline
\noalign{\smallskip}    
ID & RA  & DEC  & Type & $\langle I \rangle$ & $\langle V \rangle$ &  Period & Epoch & VMC-ID & Tile & N$_{\rm Epochs}$ & Notes\\
\noalign{\smallskip}   
   & J2000  & J2000  & &   mag & mag & d &  d & &  &  $J$,$K_\mathrm{s}$& \\                     
\noalign{\smallskip}    
(1)    & (2)  & (3) & (4) & (5)&(6)  & (7) &(8)  &(9)  &  (10)&(11)  &(12) \\                     
\noalign{\smallskip}
\hline
\noalign{\smallskip}   
OGLE-LMC-T2CEP-123   &    5:26:19.26   &   -70:15:34.7   &   BLHer   &    18.233   &    18.723   &   1.002626   &    454.80233   &    558361325273  & 5\_5  &    4,15    &        a);b)            \\
OGLE-LMC-T2CEP-069   &    5:14:56.77   &   -69:40:22.4   &   BLHer   &    18.372   &    18.919   &   1.021254   &    457.21815   &    558355522273  & 6\_4  &    4,14    &    a); b); c)         \\
OGLE-LMC-T2CEP-114   &    5:23:29.75   &   -68:19:07.2   &   BLHer   &    18.068   &    19.020   &   1.091089   &   2167.44939   &    558353567228  & 7\_5  &    4,14    &        b)            \\
OGLE-LMC-T2CEP-020   &    4:59:06.12   &   -67:45:24.6   &   BLHer   &    18.036   &    18.469   &   1.108126   &   2166.10854   &    558351437065  & 7\_3  &    4,16    &  a);b)              \\
OGLE-LMC-T2CEP-071   &    5:15:08.63   &   -68:54:53.5   &   BLHer   &    17.872   &    18.382   &   1.152164   &    457.43379   &    558354926512  & 6\_4  &    4,14    &                    \\
OGLE-LMC-T2CEP-089   &    5:18:35.72   &   -69:45:45.7   &   BLHer   &    18.032   &    18.492   &   1.167298   &    455.65166   &    558355569068  & 6\_4  &   11,23    &                    \\
OGLE-LMC-T2CEP-061   &    5:12:30.42   &   -69:07:16.2   &   BLHer   &    18.018   &    18.588   &   1.181512   &    457.30501   &    558355098130  & 6\_4  &    4,14    &                    \\
OGLE-LMC-T2CEP-107   &    5:22:05.79   &   -69:40:24.5   &   BLHer   &    17.684   &    18.482   &   1.209145   &    455.57377   &    558356704139  & 6\_5  &    7,9     &        d);e)            \\
OGLE-LMC-T2CEP-077   &    5:16:21.44   &   -69:36:59.2   &   BLHer   &    17.762   &    18.039   &   1.213802   &    456.99603   &    558355472930  & 6\_4  &    4,14    &                    \\
OGLE-LMC-T2CEP-165   &    5:38:15.29   &   -69:28:57.1   &   BLHer   &    18.761   &    19.723   &   1.240833   &   2187.68339   &    558357659836  & 6\_6  &    5,14    &                    \\
OGLE-LMC-T2CEP-102   &    5:21:19.67   &   -69:56:56.2   &   BLHer   &    17.758   &    18.231   &   1.266018   &    455.07285   &    558356982625  & 6\_5  &    7,9     &          d);e)          \\
OGLE-LMC-T2CEP-194   &    5:57:12.03   &   -72:17:13.3   &   BLHer   &    17.874   &    18.447   &   1.314467   &   2194.11008   &    558367367174  & 4\_8  &    5,10    &                    \\
OGLE-LMC-T2CEP-136   &    5:29:48.11   &   -69:35:32.1   &   BLHer   &    17.823   &    18.095   &   1.323038   &    454.37319   &    558356602471  & 6\_5  &    7,9     &         b)             \\
OGLE-LMC-T2CEP-138   &    5:30:10.87   &   -68:49:17.1   &   BLHer   &    18.059   &    18.827   &   1.393591   &   2167.52491   &    558356009909  & 6\_5  &    7,9     &        b);d)            \\
OGLE-LMC-T2CEP-109   &    5:22:12.83   &   -69:41:50.6   &   BLHer   &    19.559   &    21.212   &   1.414553   &    454.69580   &    558356727002  & 6\_5  &    7,9     &        c),d)            \\
OGLE-LMC-T2CEP-105   &    5:21:58.32   &   -70:16:35.1   &   BLHer   &    17.645   &    18.206   &   1.489298   &    830.77386   &    558361351217  & 5\_5  &    4,15    &                    \\
OGLE-LMC-T2CEP-122   &    5:25:48.19   &   -68:29:11.4   &   BLHer   &    18.241   &    19.028   &   1.538669   &   2167.45087   &    558353653819  & 7\_5  &    4,14    &                    \\
OGLE-LMC-T2CEP-171   &    5:39:40.96   &   -69:58:01.3   &   BLHer   &    17.824   &    18.512   &   1.554749   &    726.82805   &    558358012379  & 6\_6  &    5,14    &                    \\
OGLE-LMC-T2CEP-068   &    5:14:27.05   &   -68:58:02.0   &   BLHer   &    17.671   &    18.264   &   1.609301   &    456.51294   &    558354968904  & 6\_4  &    4,14    &                    \\
OGLE-LMC-T2CEP-124   &    5:26:55.80   &   -68:51:53.9   &   BLHer   &    17.889   &    18.614   &   1.734867   &   2167.63818   &    558356040530  & 6\_5  &    7,9     &                    \\
OGLE-LMC-T2CEP-008   &    4:51:11.51   &   -69:57:27.0   &   BLHer   &    17.842   &    18.585   &   1.746099   &   2165.20369   &    558358656758  & 5\_3  &    4,11    &   c);d);f)       \\
OGLE-LMC-T2CEP-142   &    5:30:34.92   &   -68:06:15.2   &   BLHer   &    17.580   &    18.458   &   1.760753   &   2167.01120   &    558353450542  & 7\_5  &    4,13    &        a);b);g)            \\
OGLE-LMC-T2CEP-084   &    5:17:07.50   &   -69:27:34.1   &   BLHer   &    17.512   &    17.841   &   1.770840   &    456.08800   &    558355348031  & 6\_4  &    1,8     &         a);b);g)           \\
OGLE-LMC-T2CEP-141   &    5:30:23.32   &   -71:39:00.6   &   BLHer   &    17.975   &    18.757   &   1.822954   &   2166.56437   &    558367767291  & 4\_6  &    6,14    &                    \\
OGLE-LMC-T2CEP-140   &    5:30:22.71   &   -69:15:38.6   &   BLHer   &    17.760   &    18.508   &   1.841144   &   2166.65700   &    558356311759  & 6\_5  &    7,9     &                    \\
OGLE-LMC-T2CEP-144   &    5:31:19.82   &   -68:51:54.9   &   BLHer   &    17.750   &    18.545   &   1.937450   &   2166.59387   &    558356035425  & 6\_5  &   10,20    &       a);b);d);f)             \\
OGLE-LMC-T2CEP-130   &    5:29:04.24   &   -70:41:37.9   &   BLHer   &    17.527   &    18.124   &   1.944694   &   2167.58469   &    558361658078  & 5\_5  &    4,15    &                    \\
OGLE-LMC-T2CEP-088   &    5:18:33.57   &   -70:50:19.2   &   BLHer   &    17.212   &    17.353   &   1.950749   &   2161.24295   &    558361779217  & 5\_5  &    4,15    &        c);d);e)        \\
OGLE-LMC-T2CEP-116   &    5:23:55.90   &   -69:25:30.1   &   BLHer   &    17.825   &    18.658   &   1.966679   &    445.61278   &    558356464708  & 6\_5  &    7,9     &                    \\
OGLE-LMC-T2CEP-121   &    5:25:42.79   &   -70:20:46.1   &   BLHer   &    17.713   &    18.430   &   2.061365   &   2166.37479   &    558361402653  & 5\_5  &    4,15    &                    \\
OGLE-LMC-T2CEP-166   &    5:38:29.09   &   -69:45:06.3   &   BLHer   &    16.927   &    17.696   &   2.110599   &   2186.16694   &    558357846207  & 6\_6  &    5,14    &       h)             \\
OGLE-LMC-T2CEP-064   &    5:13:55.87   &   -68:37:52.1   &   BLHer   &    17.514   &    18.151   &   2.127891   &   2167.00843   &    558354745198  & 6\_4  &    4,14    &                    \\
OGLE-LMC-T2CEP-167   &    5:39:02.56   &   -69:37:38.5   &   BLHer   &    17.781   &    18.597   &   2.311824   &   2187.14839   &    558357756388  & 6\_6  &    5,14    &                    \\
OGLE-LMC-T2CEP-092   &    5:19:23.63   &   -70:02:56.8   &   BLHer   &    17.401   &    18.143   &   2.616768   &   2122.71933   &    558357072491  & 6\_5  &   8,24    &                    \\
OGLE-LMC-T2CEP-148   &    5:31:52.26   &   -69:30:26.4   &   BLHer   &    17.442   &    18.194   &   2.671734   &    453.91138   &    558357678615  & 6\_6  &   12,23    &                    \\
OGLE-LMC-T2CEP-195   &    6:02:46.27   &   -72:12:47.0   &   BLHer   &    17.342   &    18.050   &   2.752929   &   2186.99000   &    558367354217  & 4\_8  &    5,10    &                    \\
OGLE-LMC-T2CEP-113   &    5:23:06.33   &   -69:32:20.5   &   BLHer   &    17.137   &    17.811   &   3.085460   &    455.01003   &    558356568619  & 6\_5  &    7,9     &          b);e)          \\
OGLE-LMC-T2CEP-049   &    5:09:21.88   &   -69:36:03.0   &   BLHer   &    17.130   &    17.703   &   3.235275   &    723.91243   &    558355501190  & 6\_4  &    4,14    &    b)               \\
OGLE-LMC-T2CEP-145   &    5:31:46.42   &   -68:58:44.0   &   BLHer   &    16.726   &    17.209   &   3.337302   &   2167.28023   &    558357363019  & 6\_6  &   12,23    &                    \\
OGLE-LMC-T2CEP-085   &    5:18:12.87   &   -71:17:15.4   &   BLHer   &    17.142   &    17.888   &   3.405095   &   2160.55457   &    558362047285  & 5\_5  &    4,15    &                    \\
OGLE-LMC-T2CEP-030   &    5:03:35.82   &   -68:10:16.2   &   BLHer   &    16.948   &    17.755   &   3.935369   &   2166.20673   &    558351663560  & 7\_3  &    4,16    &   a);b);g)              \\
OGLE-LMC-T2CEP-134   &    5:29:28.49   &   -69:48:00.4   &   pWVir   &    16.268   &    16.851   &   4.075726   &    454.54080   &    558356809300  & 6\_5  &    7,9     &                         \\          
OGLE-LMC-T2CEP-173   &    5:39:49.93   &   -69:50:52.9   &    WVir   &    18.416   &    20.149   &   4.147881   &    724.81727   &    558357918488  & 6\_6  &    5,14    &          a);b)              \\          
OGLE-LMC-T2CEP-120   &    5:25:29.55   &   -68:48:11.8   &    WVir   &    17.002   &    17.880   &   4.559053   &   2165.73588   &    558356005996  & 6\_5  &    7,9     &                         \\          
OGLE-LMC-T2CEP-052   &    5:09:59.34   &   -69:58:28.7   &   pWVir   &    16.395   &    16.861   &   4.687925   &   2164.81082   &    558355737497  & 6\_4  &    4,14    &                         \\          
OGLE-LMC-T2CEP-098   &    5:20:25.00   &   -70:11:08.7   &   pWVir   &    14.374   &    14.671   &   4.973737   &    829.46470   &    558361278143  & 5\_5  &    4,15    &                         \\          
OGLE-LMC-T2CEP-095   &    5:20:09.84   &   -68:18:35.3   &    WVir   &    17.009   &    17.873   &   5.000122   &   2121.24028   &    558353571684  & 7\_5  &    4,14    &          b);f);g);h)               \\          
OGLE-LMC-T2CEP-087   &    5:18:21.64   &   -69:40:45.2   &    WVir   &    16.887   &    17.770   &   5.184979   &    454.04523   &    558355510541  & 6\_4  &   11,23    &                         \\          
OGLE-LMC-T2CEP-023   &    5:00:13.00   &   -67:42:43.7   &   pWVir   &    15.511   &    16.101   &   5.234801   &   2163.87839   &    558351399660  & 7\_3  &    4,16    &                    \\        
OGLE-LMC-T2CEP-083   &    5:16:58.99   &   -69:51:19.3   &   pWVir   &    16.531   &    17.320   &   5.967650   &   2119.65683   &    558355634988  & 6\_4  &    4,14    &                         \\          
OGLE-LMC-T2CEP-062   &    5:13:19.12   &   -69:38:57.6   &    WVir   &    17.338   &    18.490   &   6.046676   &    453.31305   &    558355513592  & 6\_4  &    4,14    &           b);e)              \\          
OGLE-LMC-T2CEP-133   &    5:29:23.48   &   -70:24:28.5   &    WVir   &    16.671   &    17.497   &   6.281955   &   2162.68787   &    558361447993  & 5\_5  &    4,15    &                         \\          
OGLE-LMC-T2CEP-137   &    5:30:03.55   &   -69:38:02.8   &    WVir   &    16.728   &    17.633   &   6.362350   &    453.96088   &    558356644891  & 6\_5  &    7,9     &                         \\          
OGLE-LMC-T2CEP-183   &    5:44:32.99   &   -69:48:21.8   &    WVir   &    17.293   &    18.600   &   6.509627   &   2183.46556   &    558357893157  & 6\_6  &    5,13    &                    \\               
OGLE-LMC-T2CEP-043   &    5:06:00.44   &   -69:55:14.6   &    WVir   &    16.851   &    17.774   &   6.559427   &    462.41832   &    558355727258  & 6\_4  &    4,14    &         b);f);e);g);h)                \\          
OGLE-LMC-T2CEP-159   &    5:36:42.13   &   -69:31:11.7   &    WVir   &    16.805   &    17.769   &   6.625570   &   2182.53772   &    558357684253  & 6\_6  &    5,14    &                         \\          
OGLE-LMC-T2CEP-117   &    5:24:41.50   &   -71:06:44.6   &    WVir   &    16.640   &    17.539   &   6.629349   &   2165.52937   &    558361934091  & 5\_5  &    4,15    &                         \\          
OGLE-LMC-T2CEP-106   &    5:22:02.03   &   -69:27:25.3   &    WVir   &    16.612   &    17.493   &   6.706736   &    455.58483   &    558356498352  & 6\_5  &    7,9     &                         \\          
OGLE-LMC-T2CEP-078   &    5:16:29.09   &   -69:24:09.0   &   pWVir   &    16.308   &    17.206   &   6.716294   &    455.31768   &    558355301964  & 6\_4  &    4,14    &                     \\              
OGLE-LMC-T2CEP-063   &    5:13:43.86   &   -69:50:41.1   &    WVir   &    16.662   &    17.553   &   6.924580   &   2165.50032   &    558355642907  & 6\_4  &    4,14    &                         \\          
OGLE-LMC-T2CEP-110   &    5:22:19.48   &   -68:53:50.0   &    WVir   &    16.763   &    17.705   &   7.078468   &   2151.91051   &    558356071179  & 6\_5  &    7,9     &                         \\          
OGLE-LMC-T2CEP-181   &    5:43:37.42   &   -70:38:04.9   &   pWVir   &    16.193   &    16.972   &   7.212532   &    724.38026   &    558360373616  & 5\_7  &    4,8     &                    \\               
OGLE-LMC-T2CEP-047   &    5:07:46.53   &   -69:37:00.3   &    WVir   &    16.616   &    17.536   &   7.286212   &    723.50042   &    558355524174  & 6\_4  &    4,14    &                         \\          
OGLE-LMC-T2CEP-056   &    5:11:19.35   &   -69:34:32.3   &    WVir   &    16.677   &    17.654   &   7.289638   &    452.87968   &    558355469354  & 6\_4  &    4,14    &                         \\          
OGLE-LMC-T2CEP-100   &    5:21:14.64   &   -70:23:15.4   &    WVir   &    16.642   &    17.407   &   7.431095   &    825.70218   &    558361448406  & 5\_5  &    4,15    &                         \\          
OGLE-LMC-T2CEP-111   &    5:22:22.30   &   -70:52:46.8   &    WVir   &    16.542   &    17.440   &   7.495684   &    829.55773   &    558361794595  & 5\_5  &    4,15    &                           \\        
OGLE-LMC-T2CEP-170   &    5:39:38.12   &   -68:48:24.9   &    WVir   &    16.703   &   -99.990   &   7.682906   &   2181.19087   &    558357268116  & 6\_6  &    5,14    &         i)             \\          
OGLE-LMC-T2CEP-151   &    5:34:35.73   &   -69:59:14.9   &    WVir   &    16.479   &    17.384   &   7.887246   &    455.11756   &    558358035015  & 6\_6  &    5,14    &                         \\          
OGLE-LMC-T2CEP-179   &    5:43:04.02   &   -70:01:33.6   &    WVir   &    16.744   &    17.805   &   8.050065   &   2185.44813   &    558358064065  & 6\_6  &    4,14    &                    \\               
OGLE-LMC-T2CEP-182   &    5:43:46.89   &   -70:42:36.5   &    WVir   &    16.312   &    17.265   &   8.226419   &   2188.39082   &    558360430553  & 5\_7  &    4,8     &                    \\               
OGLE-LMC-T2CEP-094   &    5:19:53.20   &   -69:53:09.9   &    WVir   &    16.588   &    17.529   &   8.468490   &   2120.73841   &    558356923555  & 6\_5  &    7,9     &                         \\          
OGLE-LMC-T2CEP-019   &    4:58:49.42   &   -68:04:27.8   &   pWVir   &    15.989   &    16.853   &   8.674863   &   2162.74938   &    558351644677  & 7\_3  &    4,16    &                    \\
OGLE-LMC-T2CEP-039   &    5:05:11.31   &   -67:12:45.3   &    WVir   &    16.322   &    17.192   &   8.715837   &   2166.31977   &    558351083913  & 7\_3  &    4,16    &                         \\          
OGLE-LMC-T2CEP-028   &    5:03:00.85   &   -70:07:33.7   &   pWVir   &    15.543   &    16.045   &   8.784807   &   2168.94800   &    558358668771  & 5\_3  &    4,9     &                         \\          
OGLE-LMC-T2CEP-074   &    5:15:48.75   &   -68:48:48.1   &    WVir   &    16.070   &    16.892   &   8.988344   &   2123.38975   &    558354851839  & 6\_4  &    4,14    &                         \\          
OGLE-LMC-T2CEP-152   &    5:34:37.58   &   -70:01:08.5   &    WVir   &    16.453   &    17.323   &   9.314921   &    453.02663   &    558358053632  & 6\_6  &    5,14    &                         \\          
OGLE-LMC-T2CEP-021   &    4:59:34.97   &   -71:15:31.2   &   pWVir   &    15.884   &    16.580   &   9.759502   &   2161.10277   &    558359420632  & 5\_3  &    4,11    &                   \\
OGLE-LMC-T2CEP-132   &    5:29:08.23   &   -69:56:04.3   &   pWVir   &    15.818   &    16.548   &  10.017829   &    448.21817   &    558356939981  & 6\_5  &    7,9     &                         \\          
OGLE-LMC-T2CEP-146   &    5:31:48.01   &   -68:49:12.1   &    WVir   &    16.392   &    17.347   &  10.079593   &   2161.81703   &    558357277233  & 6\_6  &   12,23    &                         \\          
OGLE-LMC-T2CEP-097   &    5:20:20.58   &   -69:12:20.9   &    WVir   &    16.177   &    17.064   &  10.510167   &    446.10816   &    558356294442  & 6\_5  &    7,9     &                         \\          
OGLE-LMC-T2CEP-022   &    4:59:58.56   &   -70:34:27.8   &    WVir   &    16.271   &    17.179   &  10.716780   &   2157.78714   &    558359020369  & 5\_3  &    4,11    &                    \\
OGLE-LMC-T2CEP-201   &    5:15:12.67   &   -69:13:08.0   &   pWVir   &    14.611   &    15.152   &  11.007243   &    456.11301   &    558355159487  & 6\_4  &    4,14    &                    \\               
OGLE-LMC-T2CEP-101   &    5:21:18.87   &   -69:11:47.3   &    WVir   &    16.035   &    16.838   &  11.418560   &    444.88281   &    558356283672  & 6\_5  &    7,9     &                         \\          
OGLE-LMC-T2CEP-013   &    4:55:24.41   &   -69:55:43.4   &    WVir   &    16.184   &    17.119   &  11.544611   &   2157.45185   &    558358587418  & 5\_3  &    4,11    &                    \\
OGLE-LMC-T2CEP-178   &    5:42:19.01   &   -70:24:08.1   &    WVir   &    16.326   &    17.406   &  12.212367   &    726.43160   &    558360198448  & 5\_7  &    4,8     &                    \\               
OGLE-LMC-T2CEP-127   &    5:27:59.80   &   -69:23:27.5   &    WVir   &    16.120   &    17.092   &  12.669118   &    454.17111   &    558356420696  & 6\_5  &    7,9     &                         \\          
OGLE-LMC-T2CEP-118   &    5:25:15.05   &   -68:09:11.7   &    WVir   &    16.103   &    17.037   &  12.698580   &   2163.34477   &    558353477576  & 7\_5  &    4,14    &                         \\          
OGLE-LMC-T2CEP-103   &    5:21:35.27   &   -70:13:25.7   &    WVir   &    16.039   &    16.995   &  12.908278   &    824.38616   &    558361309970  & 5\_5  &    4,15    &                         \\          
OGLE-LMC-T2CEP-044   &    5:06:28.86   &   -69:43:58.8   &    WVir   &    16.099   &    17.108   &  13.270100   &    464.57726   &    558355611443  & 6\_4  &    4,14    &                         \\          
OGLE-LMC-T2CEP-026   &    5:02:11.56   &   -68:20:16.0   &    WVir   &    16.091   &    17.026   &  13.577869   &   2156.87252   &    558351786614  & 7\_3  &    4,16    &                    \\   
OGLE-LMC-T2CEP-096   &    5:20:10.42   &   -68:48:39.2   &    WVir   &    15.918   &    16.832   &  13.925722   &   2129.22374   &    558356025075  & 6\_5  &    7,9     &                         \\          
OGLE-LMC-T2CEP-157   &    5:36:02.60   &   -69:27:16.1   &    WVir   &    16.045   &    17.050   &  14.334647   &   2181.19312   &    558357639701  & 6\_6  &    5,14    &                         \\          
OGLE-LMC-T2CEP-017   &    4:56:16.02   &   -68:16:16.4   &    WVir   &    15.986   &    16.968   &  14.454754   &   2157.70744   &    558351791598  & 7\_3  &    4,16    &                    \\
OGLE-LMC-T2CEP-143   &    5:31:09.75   &   -69:15:48.9   &    WVir   &    15.806   &    16.701   &  14.570185   &   2166.57316   &    558356313034  & 6\_5  &   12,23    &                         \\          
\noalign{\smallskip}                                                                                                                                                                                      
\hline                                                                                                                                                                                                    
\noalign{\smallskip}                                                                                                                                                                                      
\end{tabular}                                                                                                                                                                                     
\end{center}                                                                                                                                                                                              
\end{table*}

\begin{table*}
\scriptsize\tiny
\begin{center}
\contcaption{}
\begin{tabular}{cccccccccccc}
\hline
\noalign{\smallskip}    
ID & RA  & DEC  & Type &  $\langle I \rangle$ & $\langle V \rangle$ & Period & Epoch & VMC-ID & Tile & N$_{\rm Epochs}$ & Notes\\
\noalign{\smallskip}   
   & J2000  & J2000  & &   mag & mag & d &  d & &  &  & \\                     
\noalign{\smallskip}    
(1)    & (2)  & (3) & (4) & (5)&(6)  & (7) &(8)  &(9)  &  (10)&(11)  &(12) \\                     
\noalign{\smallskip}
\hline
\noalign{\smallskip}   
OGLE-LMC-T2CEP-046   &    5:07:38.94   &   -68:20:05.9   &    WVir   &    15.547   &    16.415   &  14.743796   &   2162.69705   &    558351740940  & 7\_3  &    4,16    &      b);c);d);f)                   \\          
OGLE-LMC-T2CEP-139   &    5:30:22.56   &   -69:09:12.1   &    WVir   &    15.968   &    17.003   &  14.780410   &   2156.19900   &    558356235708  & 6\_5  &    7,9     &                         \\          
OGLE-LMC-T2CEP-177   &    5:40:36.54   &   -69:13:04.3   &    WVir   &    16.132   &    17.240   &  15.035903   &   2178.31837   &    558357492207  & 6\_6  &    5,14    &                    \\               
OGLE-LMC-T2CEP-099   &    5:20:44.48   &   -69:01:48.4   &    WVir   &    15.932   &    16.999   &  15.486788   &   2111.72112   &    558356167163  & 6\_5  &    7,9     &                         \\          
OGLE-LMC-T2CEP-086   &    5:18:17.80   &   -69:43:27.7   &    WVir   &    15.629   &    16.486   &  15.845500   &    452.84478   &    558355544575  & 6\_4  &   11,23    &                         \\          
OGLE-LMC-T2CEP-126   &    5:27:53.42   &   -70:51:30.9   &    WVir   &    16.210   &    17.436   &  16.326778   &   2167.50661   &    558361770086  & 5\_5  &    4,15    &                         \\          
OGLE-LMC-T2CEP-057   &    5:11:21.13   &   -68:40:13.3   &    WVir   &    15.749   &    16.707   &  16.632041   &   2159.16741   &    558354781673  & 6\_4  &    4,14    &                         \\          
OGLE-LMC-T2CEP-093   &    5:19:26.45   &   -69:51:51.0   &    WVir   &    15.130   &    15.861   &  17.593049   &    446.06633   &    558356904142  & 6\_5  &    7,9     &           j)              \\          
OGLE-LMC-T2CEP-128   &    5:28:43.81   &   -70:14:02.3   &    WVir   &    15.517   &    16.460   &  18.492694   &    453.20828   &    558361300181  & 5\_5  &    4,15    &                         \\          
OGLE-LMC-T2CEP-058   &    5:11:33.52   &   -68:35:53.7   &   RVTau   &    15.511   &    16.594   &  21.482951   &   2167.45398   &    558354737426  & 6\_4  &    4,14    &                \\                   
OGLE-LMC-T2CEP-104   &    5:21:49.10   &   -70:04:34.3   &   RVTau   &    14.937   &    15.830   &  24.879948   &    447.75745   &    558361170450  & 5\_5  &   11,24    &                    \\               
OGLE-LMC-T2CEP-115   &    5:23:43.53   &   -69:32:06.8   &   RVTau   &    15.593   &    16.651   &  24.966913   &   2145.84889   &    558356566155  & 6\_5  &    7,9     &                    \\               
OGLE-LMC-T2CEP-192   &    5:53:55.69   &   -70:17:11.4   &   RVTau   &    15.233   &    16.148   &  26.194001   &   2181.44982   &    558360150098  & 5\_7  &    4,8     &                    \\               
OGLE-LMC-T2CEP-135   &    5:29:38.50   &   -69:15:12.2   &   RVTau   &    15.194   &    16.162   &  26.522364   &   2144.30037   &    558356308540  & 6\_5  &    7,9     &                    \\               
OGLE-LMC-T2CEP-108   &    5:22:11.27   &   -68:11:31.3   &   RVTau   &    14.746   &    15.477   &  30.010843   &   2113.81336   &    558353504910  & 7\_5  &    4,14    &       k)             \\               
OGLE-LMC-T2CEP-162   &    5:37:44.95   &   -69:54:16.5   &   RVTau   &    15.112   &    16.200   &  30.394148   &    706.20990   &    558357961649  & 6\_6  &    5,14    &                    \\               
OGLE-LMC-T2CEP-180   &    5:43:12.87   &   -68:33:57.1   &   RVTau   &    14.502   &    15.303   &  30.996315   &   2178.20791   &    558352877374  & 7\_7  &    4,8     &                    \\               
OGLE-LMC-T2CEP-119   &    5:25:19.48   &   -70:54:10.0   &   RVTau   &    14.391   &    15.225   &  33.825094   &   2158.59349   &    558361803554  & 5\_5  &    4,15    &                    \\               
OGLE-LMC-T2CEP-050   &    5:09:26.15   &   -68:50:05.0   &   RVTau   &    14.964   &    15.661   &  34.748344   &    713.64755   &    558354903269  & 6\_4  &    4,14    &                    \\               
OGLE-LMC-T2CEP-200   &    5:13:56.43   &   -69:31:58.3   &   RVTau   &    15.092   &    16.124   &  34.916555   &    423.70670   &    558355423319  & 6\_4  &    4,14    &       k)             \\               
OGLE-LMC-T2CEP-065   &    5:14:00.75   &   -68:57:56.8   &   RVTau   &    14.699   &    15.611   &  35.054940   &    455.17514   &    558354970692  & 6\_4  &    4,14    &       k)             \\               
OGLE-LMC-T2CEP-091   &    5:18:45.48   &   -69:03:21.6   &   RVTau   &    14.203   &    14.899   &  35.749346   &    425.38622   &    558355015602  & 6\_4  &   11,23    &                    \\               
OGLE-LMC-T2CEP-203   &    5:22:33.79   &   -69:38:08.5   &   RVTau   &    15.395   &    16.723   &  37.126746   &    448.74961   &    558356665485  & 6\_5  &    7,9     &                    \\               
OGLE-LMC-T2CEP-202   &    5:21:49.09   &   -70:46:01.4   &   RVTau   &    15.167   &    16.359   &  38.135567   &    812.55923   &    558361722614  & 5\_5  &    4,15    &                    \\               
OGLE-LMC-T2CEP-112   &    5:22:58.36   &   -69:26:20.9   &   RVTau   &    14.065   &    14.749   &  39.397704   &    421.63429   &    558356478674  & 6\_5  &    7,9     &                    \\               
OGLE-LMC-T2CEP-051   &    5:09:41.93   &   -68:51:25.0   &   RVTau   &    14.569   &    15.440   &  40.606400   &    720.05675   &    558354917278  & 6\_4  &    4,14    &         k)           \\               
OGLE-LMC-T2CEP-080   &    5:16:47.43   &   -69:44:15.1   &   RVTau   &    14.341   &    15.175   &  40.916413   &    436.42111   &    558355560379  & 6\_4  &    4,14    &                    \\               
OGLE-LMC-T2CEP-149   &    5:32:54.46   &   -69:35:13.2   &   RVTau   &    14.151   &    14.868   &  42.480613   &   2149.99673   &    558357730269  & 6\_6  &    5,14    &                    \\               
OGLE-LMC-T2CEP-032   &    5:03:56.31   &   -67:27:24.6   &   RVTau   &    14.011   &    14.992   &  44.561195   &   2152.87623   &    558351226498  & 7\_3  &    4,16    &                    \\               
OGLE-LMC-T2CEP-147   &    5:31:51.00   &   -69:11:46.3   &   RVTau   &    13.678   &    14.391   &  46.795842   &   2135.14758   &    558357481187  & 6\_6  &   9,23    &                    \\               
OGLE-LMC-T2CEP-174   &    5:40:00.50   &   -69:42:14.7   &   RVTau   &    13.693   &    14.457   &  46.818956   &   2166.79927   &    558357814883  & 6\_6  &    5,14    &                      \\             
OGLE-LMC-T2CEP-067   &    5:14:18.11   &   -69:12:35.0   &   RVTau   &    13.825   &    14.627   &  48.231705   &    442.94273   &    558355160313  & 6\_4  &    4,14    &                    \\               
OGLE-LMC-T2CEP-075   &    5:16:16.06   &   -69:43:36.9   &   RVTau   &    14.568   &    15.728   &  50.186569   &    430.99079   &    558355554309  & 6\_4  &    4,14    &                    \\               
OGLE-LMC-T2CEP-014   &    4:55:35.40   &   -69:54:04.2   &   RVTau   &    14.312   &    15.103   &  61.875713   &   2161.68872   &    558358564467  & 5\_3  &    4,11    &       k)             \\               
OGLE-LMC-T2CEP-129   &    5:28:54.60   &   -69:52:41.1   &   RVTau   &    14.096   &    14.813   &  62.508947   &    397.72780   &    558356885794  & 6\_5  &    7,9     &                    \\               
OGLE-LMC-T2CEP-045   &    5:06:34.06   &   -69:30:03.7   &   RVTau   &    13.729   &    14.787   &  63.386339   &   2148.64483   &    558355447114  & 6\_4  &    4,14    &                    \\               
\noalign{\smallskip}                                                                                                                                                                                      
\hline                                                                                                                                                                                                    
\noalign{\smallskip}                                                                                                                                                                                      
\multicolumn{12}{l}{a) Large separation ($>$ 0.5 arcsec) between VMC and OGLE III star centroids likely due to crowding; b) blended object; c)  faint object; d) poor light curve;}\\                            
\multicolumn{12}{l}{ e) very low amplitude in the optical; f) source lies within a strip of the tile that has half the exposure of most of the tile \citep[see][]{Cross12}; } \\
\multicolumn{12}{l}{ g) poorly sampled or heavily dispersed light
  curve (due to e.g. blending, saturation); h)  source image comes partly from detector 16} \\
\multicolumn{12}{l}{(on the top half of detector 16, the quantum eﬃciency (QE)
varies on short timescales making flat-fields inaccurate; \citealt{Cross12});}  \\                                          
\multicolumn{12}{l}{ i) missing OGLE $V$ magnitude; j) light curve showing pulsation plus eclipse according to OGLE III; k) correction for saturation not effective}
\end{tabular}                                                                                                                                                                                     
\end{center}                                                                                                                                                                                              
\end{table*}

A detailed description of the general observing strategy of the VMC
survey can be found in \citet{Cioni11}. As for the variable stars, the
specific procedures adopted to study these objects were discussed in
\citet{Moretti14}.  Here we only briefly recall that the VMC
$K_\mathrm{s}$-band time series observations were scheduled in 12
separate epochs distributed over several consecutive months.  This
strategy allows us to obtain well sampled light curves for a variety
of variable types (including RR Lyrae variables and Cepheids of all
types). Concerning the $J$ and $Y$ bands, the average number of epochs is 3, 
as a result of the observing strategy in these bands  (i.e. monitoring
was not planned). Hence, some epochs could occur in the same
night and even one after the other. 
We note that in this paper we did not consider the
  $Y$-band data for several reasons: i) this filter is very rarely used 
in the context of distance scale; ii) its photometric zero point
is difficult to calibrate (no 2MASS measures); iii) because the $Y$-
band is bluer than the typical NIR bands, the $PL$, $PLC$ and $PW$ relations
in this filter are expected to be more dispersed \citep[see e.g.][]{Madore2012} and of lesser 
utility with respect to those in $J$ and $K_\mathrm{s}$.

The VMC data, processed through the pipeline
\citep{Irwin_etal04} of the VISTA Data Flow System
\citep[VDFS,][]{Emerson_etal04} are in the VISTA photometric system 
(Vegamag=0). 
The time series photometry used in this paper 
was retrieved from the VISTA Science
Archive\footnote{http://horus.roe.ac.uk/vsa/} \citep[VSA,][]{Cross12}.
For details about the data reduction we refer the
reader to the aforementioned papers. Nevertheless, we underline two
characteristics of the data reduction which we think may have
importance in the subsequent discussion. First, the pipeline is able to correct the
photometry of stars close to the saturation limit \citep{Irwin2009}. This is
relevant in the context of this paper because the RV Tau variables discussed here are
very bright objects $K_\mathrm{s} \sim 12-13$ mag, close to
the saturation limits of the VMC survey. The photometry of these stars
takes advantage of the VDFS ability to treat saturated images,
however, as we will see below, the corrections applied are not always 
sufficient to fully recover the data. Second, the data retrieved from
VSA include quality flags which are very useful to understand if the
images have problems. We shall use this information later in this paper. 

According to OGLE~III/IV, 130 T2CEPs are expected to lie in the 13 tiles
analysed in this paper. Note that no T2CEP from OGLE~III or OGLE~IV falls inside
our tiles 6\_8 or 8\_8, respectively. Hence, in the following we only use OGLE~III data.
Figure~\ref{figMap} and Table~\ref{tabLog} show the distribution of
such stars through the VMC tiles.

Table~\ref{tabData} lists the 130 T2CEPs analysed here, together with
their main properties as measured by OGLE~III and the information
about the VMC tile they belong to, as well as the number of epochs of
observations in the $J$- and $K_\mathrm{s}$-bands. In total, our sample is
composed of 41 BL Her, 62 W Vir (12 pW Vir) and 27 RV Tau variables,
corresponding to 63\%, 63\% (75\%) and 61\% of the known LMC populations
of the three different variable classes, respectively. 

The OGLE~III catalogues of  T2CEP variables were cross-correlated against the VMC catalogue  
to obtain the $J$ and $K_\mathrm{s}$ light curves for these variables.   
 All the 130 T2CEPs were found to have a counterpart in the VMC
catalogue within 2 arcsec from the OGLE III positions. 
The great majority of the objects showed separation in position
with respect to OGLE III less than 0.1 arcsec. 
However, 8 stars  (OGLE-LMC-T2CEP-020, 030, 069, 084, 123, 142, 144,
173) present separations significantly larger than average ($>$ 0.5
arcsec). Figure~\ref{figCharts} shows the OGLE III and VMC finding  
charts of 29 stars with some kind of identification or data problem,
within which we included the eight objects quoted above. It can be seen that all the  
stars lie in crowded regions or are clearly blended by other  
stars or diffuse objects (e.g. OGLE-LMC-T2CEP-142). We will discuss these objects further in the following sections.

\subsection{T2CEPs light curves}
\label{t2ceps}

The VMC time series $J$ and $K_\mathrm{s}$ photometry for the 130 objects is provided
in Table~\ref{sampleTimeSeries}, which is published in its entirety in
the on-line version of the paper.

Periods and epochs of maximum light available from the 
OGLE~III catalogue were used to  fold the
$J$- and $K_\mathrm{s}$-band light curves produced by the VMC
observations. 
Given the larger number of epochs in $K_\mathrm{s}$ with respect to
$J$, we discuss first the $K_\mathrm{s}$-band data.

The $K_\mathrm{s}$--band light curves for a sample of 120 T2CEPs with
useful light curves are shown in Fig.~\ref{figureK}. 
Apart from a few cases these light curves are generally well sampled and nicely shaped. 
Some clearly discrepant data points (empty circles in Fig.~\ref{figureK}) in 
the light curves were excluded from the fit but were plotted in the 
figure for completeness. Note that most of these ``bad'' data points 
belong to  observations collected during nights that did not strictly 
meet the VMC observing constraints \citep[see Table 2 in][]{Cioni11}. The final spline fit to the data is shown 
by a solid line in Fig. ~\ref{figureK}. 
Intensity-averaged $\langle K_\mathrm{s} \rangle $ magnitudes were derived from the light curves
using custom software written in {\sc C}, that performs
a spline interpolation to the data with no need of using templates.
The numerical model of the light-curve is thus obtained and then
integrated in intensity to obtain the mean intensity which is
eventually transformed to mean magnitude.
 
\begin{figure*}
\includegraphics[height=22cm]{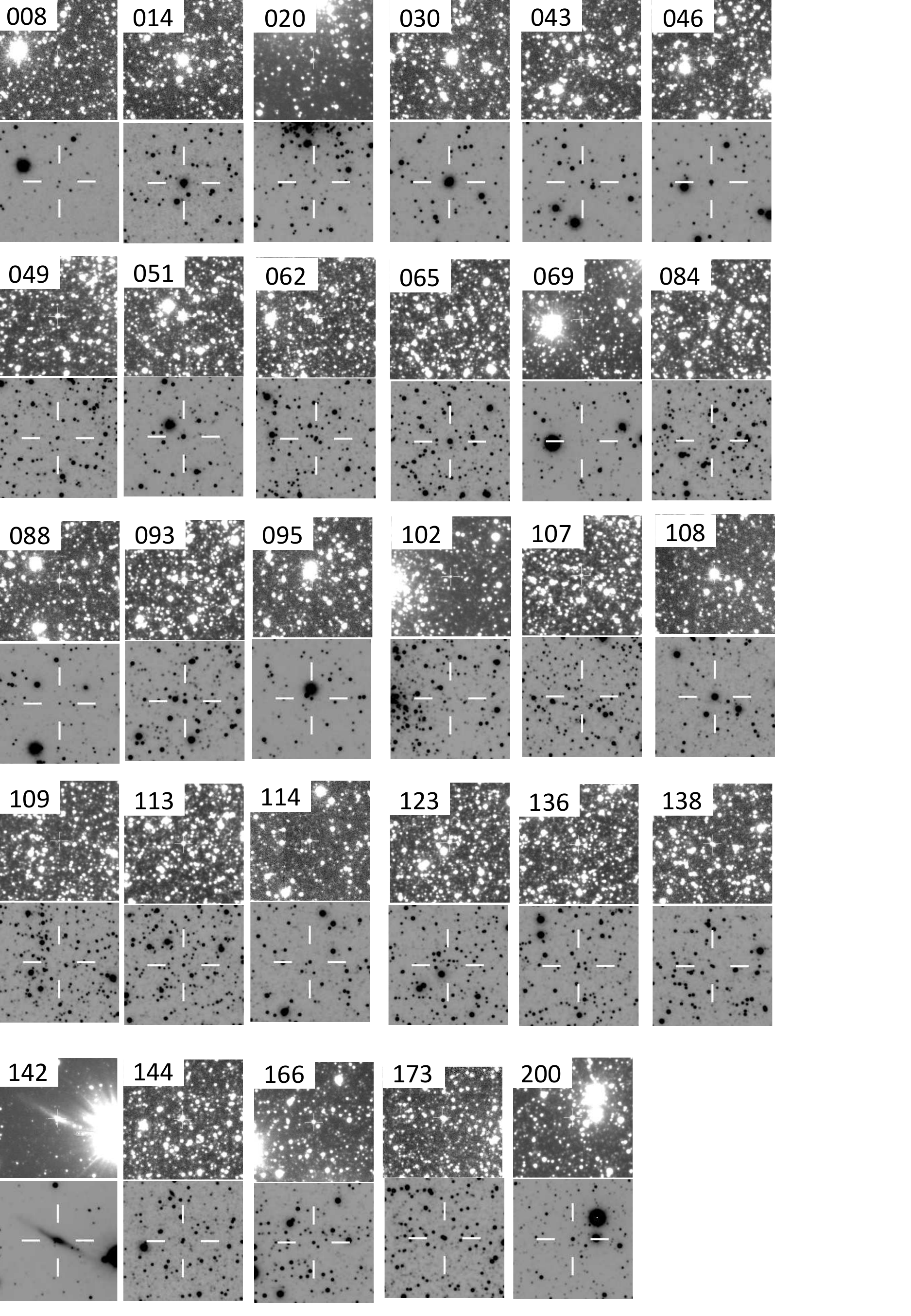}
\caption{Sky pictures for 29 problematic stars extracted from the VMC (bottom panels) 
 and the OGLE~III (top panels) archives. The target is identified 
 with the last three digits of the  OGLE III identification  
(i.e. without the prefix ``OGLE-LMC-T2CEP-'').} 
\label{figCharts}
\end{figure*}

Ten objects in our sample showed unusable light curves, namely:
OGLE-LMC-T2CEP-014, 030, 043, 051, 065, 084, 095, 108, 142, 200.
Their light curves are displayed  in Fig.~\ref{figureK_sfigate}, whereas their finding 
charts are shown in Fig.~\ref{figCharts}. A quick analysis of the finding charts 
reveals that all these stars have significant problems of 
crowding/blending. Three of the aforementioned objects
(OGLE-LMC-T2CEP-030, 084, 142) have centroids significantly
shifted with respect to OGLE's, thus confirming the presence of strong
blending. 

\begin{figure}
\includegraphics[width=8.5cm]{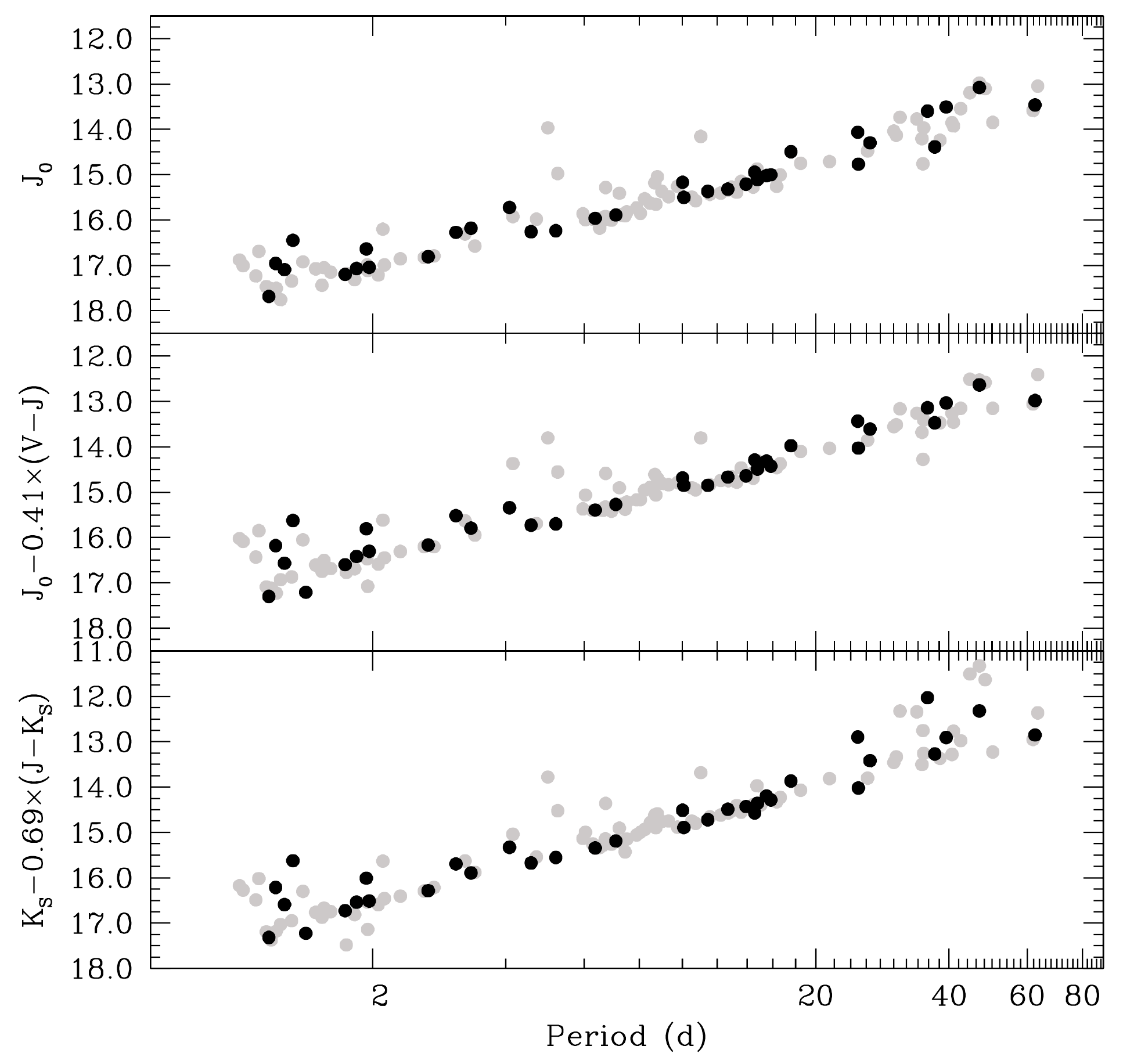}
\caption{From top to bottom: $PL$ in the $J$ band, $PW$ in ($J$,$V-J$) and $PW$ in 
  ($K_\mathrm{s}$,$J-K_\mathrm{s}$) for T2CEPs whose intensity-averaged 
  $\langle J \rangle$ magnitude was obtained on the basis of direct spline fit (black 
  filled circles) or template fit (grey filled circles). See the text for details. } 
\label{plj_comp}
\end{figure}

As for the $J$--band data, Fig.~\ref{figureJspline} shows the light
curves for the 34 stars that have sufficiently good data to allow an independent spline fit (solid line
in the figure). Figures~\ref{figureJtempl} and ~\ref{figureJ_sfigate}
show the light curves for the remaining 86 and 10 objects with small 
number of epochs ($\sim$4-5 on average) and dispersed light curves,
respectively. The latter variables show the same problems 
reported for the $K_\mathrm{s}$-band.
To estimate the intensity-averaged $J$ magnitude for the 86 stars
possessing only few epochs of observation, we decided to use the
spline fit curves in the $K_\mathrm{s}$-band as 
templates.\footnote{A comparison of Fig. A.1 ($K_\mathrm{s}$ light curves) and A.3 ($J$-light curves for stars possessing sufficient data points to be 
analysed independently from $K_\mathrm{s}$-band) show that at present
level of precision, the light curves in $J$ and $K_\mathrm{s}$ 
are sufficiently similar to allow us using the $K_\mathrm{s}$-spline fits as templates.} To this aim, for
each star we performed the following steps: 1) subtracted the average
$\langle K_\mathrm{s} \rangle$ magnitude from the $K_\mathrm{s}$
spline fit curve; 2)
adjusted by eye the data obtained in this way to fit the $J$ light curve by
i) adding a zero point; ii) multiplying the amplitude by a
proper factor; iii) shifting the light curve in phase.   
The factor needed for point ii) is the ratio
Amp($J$)/Amp($K_\mathrm{s}$). To estimate this number, we
used the 34 stars with independent $J$--band spline fit, obtaining a value
of 1.1$\pm$0.2. The uncertainty of $\sim$20\% may appear large, but it
does not actually  represent a problem since its contribution to the error on the
intensity-averaged $J$ is of the order of 0.5\%. In some favourable cases, the
few data points covered both maximum and minimum of the light curve and it was
then possible to constrain directly the amplitude ratio. 
The shift in phase (point iii above) varied from case to case, but was on average
close to 0.05-0.06. The final error on the  intensity-averaged $J$
magnitude was calculated by summing in quadrature the error on the
$K_\mathrm{s}$ magnitude, the uncertainty on the $J$ magnitude caused by
the error on the amplitude ratio, and an additional 1\% to take into
account the uncertainty on the phase shift.
The goodness of this procedure can be appreciated in
Fig.~\ref{plj_comp}, where we show in different colours the $PL$ and $PW$
relations (see next section for a detailed description of these
relations) for the stars
with intensity-averaged $J$ photometry obtained directly from spline
fits  (black points) and with the template fits (grey points). The figure
clearly shows that the results obtained on the basis of the
$K_\mathrm{s}$ templates are usable for scientific purposes.
The final $\langle J \rangle$,$\langle K_\mathrm{s} \rangle$
magnitudes with relative uncertainties, as well as, pulsational
amplitudes and adopted reddening values (see Sect.~\ref{section3}) are provided in Table~\ref{tabResults}.

\begin{table}
\scriptsize
\caption{$J$ and $K_\mathrm{s}$ time series photometry for the T2CEPs
investigated in this paper. The data below refer to the variable OGLE-LMC-T2CEP-123.}
\label{sampleTimeSeries}
\begin{center}
\begin{tabular}{ccc}
\hline
\noalign{\smallskip} 
HJD-2\,400\,000 & $J$  & $\sigma_{J}$  \\
\noalign{\smallskip}
\hline 
\noalign{\smallskip} 
   55487.77111 &   16.963  &   0.014 \\
   55487.80976 &   16.959  &   0.014 \\
   55497.79317 &   16.989  &   0.014 \\
   55497.86048 &   16.950  &   0.013 \\
\noalign{\smallskip}
\hline 
\noalign{\smallskip} 
HJD-2\,400\,000 & $K_\mathrm{s}$  & $\sigma_{K_\mathrm{s}}$  \\
\noalign{\smallskip}
\hline 
\noalign{\smallskip} 
   55495.82644  &  16.520  &   0.020 \\
   55497.75937  &  16.520  &   0.020 \\
   55497.81507  &  16.513  &   0.024 \\
   55499.82170  &  16.517  &   0.023 \\
   55511.74774  &  16.507  &   0.020 \\
   55516.77236  &  16.496  &   0.023 \\
   55526.78868  &  16.498  &   0.021 \\
   55539.82483  &  16.488  &   0.022 \\
   55557.73937  &  16.482  &   0.023 \\
   55563.71325  &  16.465  &   0.021 \\
   55587.65755  &  16.470  &   0.023 \\
   55844.79771  &  16.526  &   0.020 \\
   55865.82753  &  16.483  &   0.021 \\
   55887.74744  &  16.477  &   0.022 \\
   55937.67877  &  16.454  &   0.021 \\
\noalign{\smallskip}
\hline
\noalign{\smallskip}
\end{tabular}
\end{center}
Table~\ref{sampleTimeSeries} is published in its entirety only in the
electronic edition of the journal. 
A portion is shown here for guidance regarding its form and content.
\end{table}

We recall that the $J$,$K_\mathrm{s}$ photometry
presented in this paper is set in the VISTA system. A consistent comparison
between our results and those in the widely used 2MASS system
\citep[Two Micron All Sky Survey][]{2mass} can be performed after
applying proper system transformations as for instance those provided by the Cambridge Astronomy Survey
Unit (CASU)\footnote{http://casu.ast.cam.ac.uk/surveys-projects/vista/technical/photometric-properties}: 
($J$$-$$K_\mathrm{s}$)(2MASS)=1.081($J$-$K_\mathrm{s}$)(VISTA);   
$J$(2MASS)=$J$(VISTA)+0.07($J$-$K_\mathrm{s}$)(VISTA)  
and $K_\mathrm{s}$(2MASS)=$K_\mathrm{s}$(VISTA)$-$0.011($J$-$K_\mathrm{s}$)(VISTA). 

\begin{figure}
\includegraphics[width=8.5cm]{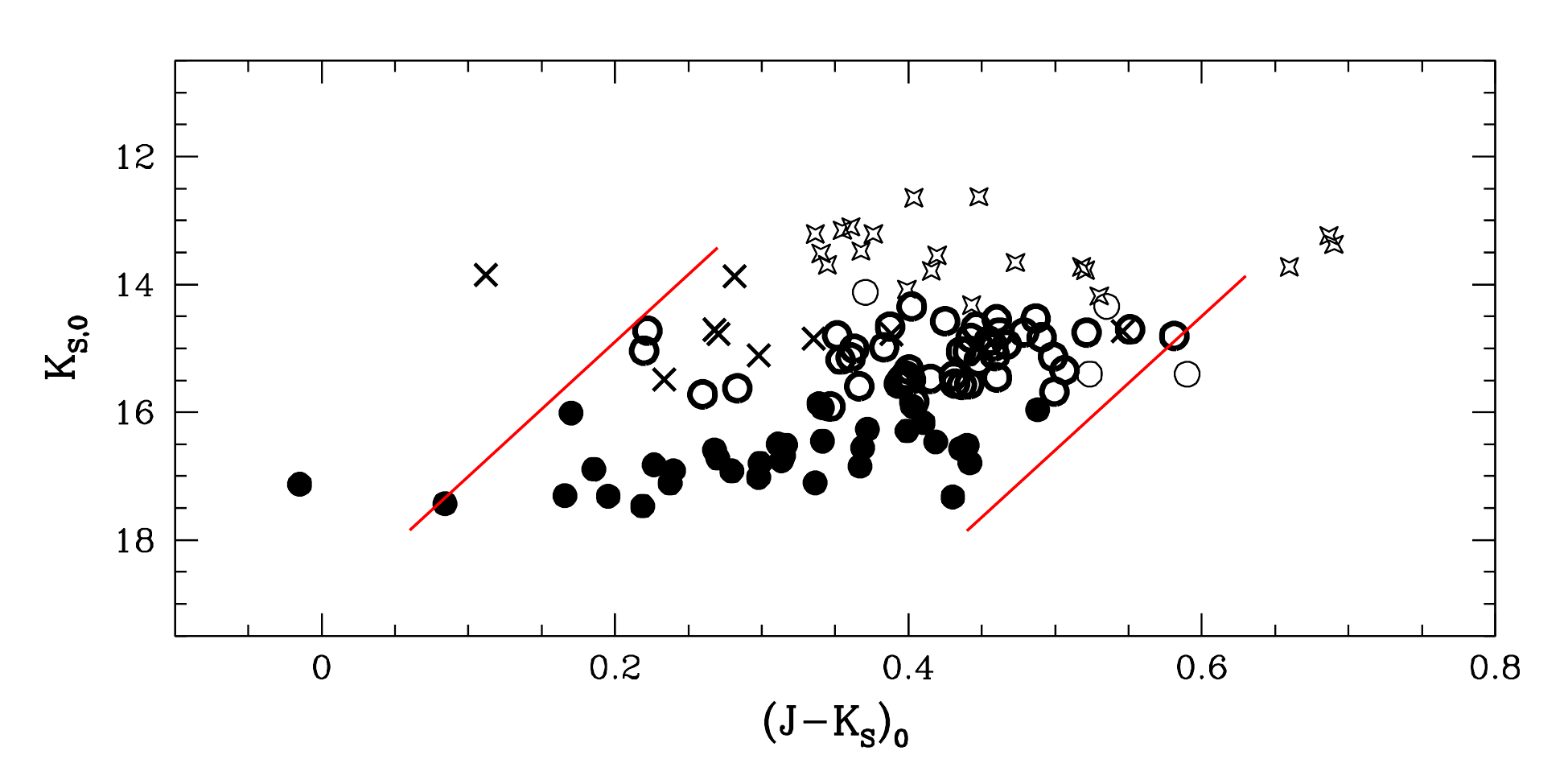}
\caption{Observed instability strip in the plane $K_\mathrm{s,0}$ vs 
  $(J-K_\mathrm{s})_0$. Filled circles, open circles, crosses and
  stars show BL Her, W Vir, pW Vir and RV Tau variables, respectively. The 
solid lines show the approximate borders of the BL Her/W Vir 
instability strip. Blue and red edges are described by the 
following equations: 
$K_\mathrm{s,0}$=19.1$-$21$(J-K_\mathrm{s})_0$ 
(0.06$<(J-K_\mathrm{s})_0<$0.27 mag) and  $K_\mathrm{s,0}$=27.1$-$21$(J-K_\mathrm{s})_0$ 
(0.44$<(J-K_\mathrm{s})_0<$0.63 mag), respectively.} 
\label{isjk}
\end{figure}

Since the $\langle$$J$ $\rangle$$-$$\langle$ $K_\mathrm{s}$$
\rangle$ colour of our T2CEP sample typically ranges from 0.1 to 0.6
mag, the VISTA and 2MASS $K_\mathrm{s}$ can be considered equivalent
for T2Ceps (see Fig.~\ref{isjk})  and for CCs
\citep[see][]{Ripepi12b}, 
to a very good approximation (better than $\sim$5 mmag).

\section{$J$, $K_\mathrm{s}$-band Period-Luminosity,
  Period-Luminosity-Colour
 and Period-Wesenheit relations}
\label{section3}

The data reported in Table~\ref{tabResults} allow us to calculate
different useful relationships adopting various combinations of
magnitudes and colours. In particular, we derived $PL$ relations in $J$
and $K_\mathrm{s}$ as well as $PW$ and $PLC$ relations for the following
combinations: ($J,V-J$); ($K_\mathrm{s}$,$V-K_\mathrm{s}$);
($K_\mathrm{s}$,$J-K_\mathrm{s}$).  

We first corrected magnitudes and colours for  reddening using the
recent extinction maps by \citet{haschke}. Individual
$E$($V$$-$$I$) reddening values  for the
120 T2CEPs with useful VMC data are reported in column 10 of
Table~\ref{tabResults}. The reliability of this reddening correction
can be questioned by observing that it has been derived from the
analysis of the Red Clump (RC) stars, which trace the intermediate-age 
population (2-9 Gyr) instead of the old one to whom BL Her and
W Vir belong to. However, we recall that in the NIR bands the interstellar absorption is very
low: $A_J \sim 0.25 A_V$ and $A_{K_\mathrm{s}} \sim 0.1 A_V$, where
$A_V$ is the absorption in the visible. 
Hence, even in the unlikely case of a  10\% large error in our $A_V$ 
estimates, this would introduce an amount of uncertainties of 
only $\sim$2.5\%  and $\sim$1\% in $J$ and $K_\mathrm{s}$, respectively.
An a posteriori indication about the global correctness of the
adopted reddening correction is represented by the concordance of
results provided by the $PL$ (reddening dependent) and $PW$ (reddening
independent) relations (see Sect.~\ref{section4} and ~\ref{discussion}). 
The reddening values were converted using  
the following equations: $E$($V$$-$$J$)=1.80$E$($V$$-$$I$);
$E$($V$$-$$K_\mathrm{s}$)=2.24$E$($V$$-$$I$)
$E$($J$$-$$ K_\mathrm{s}$)=0.43$E$($V$$-$$I$)
\citep{Cardelli1989,Gao2013}.\footnote{The coefficients we used are
  suited for the 2MASS system, to which the VISTA system is tied  
(see Sec. 2.1).} The coefficients of the $PW$
relations were calculated in a similar way.

\begin{table*}
\scriptsize\tiny
\caption{Results for the 120 T2CEPs with useful NIR light curves analysed in this paper.  
The columns report: 1) OGLE identification; 2) variability class ; 3) period (OGLE); 4) intensity--averaged $J$ magnitude; 5) uncertainty on the $\langle J \rangle$ 6) intensity--averaged $K_\mathrm{s}$ magnitude; 7) uncertainty on the $\langle K_\mathrm{s} \rangle$;  
8) peak--to--peak amplitude in $J$; 9) peak--to--peak amplitude in $ K_\mathrm{s}$; 10) adopted reddening; 11) T=results in $J$ obtained on the basis of the 
$K_\mathrm{s}$ template, S=results in $J$ obtained on the basis of direct spline fitting to the data.}
\label{tabResults}
\begin{center}
\begin{tabular}{ccccccccccc}
\hline
\noalign{\smallskip}   
ID & Var. Class & Period & $\langle J \rangle$ & $\sigma_{\langle J \rangle} $ &  $\langle K_\mathrm{s} \rangle$ & $\sigma_{\langle K_\mathrm{s} \rangle}$ & $Amp(J)$ & $Amp(K_\mathrm{s})$ & $E($V$$-$$I$)$ & Note \\ 
   & & d & mag & mag &mag & mag & mag & mag & mag &\\                     
(1)    & (2)  & (3) & (4) & (5) &(6)  & (7) &(8) & (9) & (10) & (11) \\                     
\noalign{\smallskip}
\hline
\noalign{\smallskip}    
OGLE-LMC-T2CEP-123   &  BLHer    &     1.0026263    &     16.939   &        0.021   &      16.486   &       0.013   &     0.05   &     0.05   &       0.080  &  T              \\
OGLE-LMC-T2CEP-069   &  BLHer    &     1.0212542    &     17.042   &        0.033   &      16.585   &       0.021   &     0.10   &     0.10   &       0.050  &  T              \\
OGLE-LMC-T2CEP-114   &  BLHer    &     1.0910886    &     17.329   &        0.069   &      16.831   &       0.019   &     0.17   &     0.16   &       0.130  &  T              \\
OGLE-LMC-T2CEP-020   &  BLHer    &     1.1081258    &     16.735   &        0.043   &      16.310   &       0.022   &     0.09   &     0.07   &       0.060  &  T              \\
OGLE-LMC-T2CEP-071   &  BLHer    &     1.1521638    &     17.522   &        0.022   &      17.326   &       0.026   &     0.40   &     0.38   &       0.070  &  T              \\
OGLE-LMC-T2CEP-089   &  BLHer    &     1.1672977    &     17.715   &        0.018   &      17.479   &       0.043   &     0.40   &     0.32   &       0.040  &  S              \\
OGLE-LMC-T2CEP-061   &  BLHer    &     1.1815124    &     17.581   &        0.037   &      17.458   &       0.031   &     0.38   &     0.19   &       0.090  &  T              \\
OGLE-LMC-T2CEP-107   &  BLHer    &     1.2091451    &     16.979   &        0.005   &      16.526   &       0.016   &     0.19   &     0.13   &       0.030  &  S              \\
OGLE-LMC-T2CEP-077   &  BLHer    &     1.2138023    &     17.521   &        0.045   &      17.317   &       0.025   &     0.18   &     0.17   &       0.020  &  T              \\
OGLE-LMC-T2CEP-165   &  BLHer    &     1.2408330    &     17.889   &        0.049   &      17.381   &       0.024   &     0.34   &     0.32   &       0.180  &  T              \\
OGLE-LMC-T2CEP-102   &  BLHer    &     1.2660176    &     17.146   &        0.010   &      16.817   &       0.020   &     0.20   &     0.13   &       0.070  &  S              \\
OGLE-LMC-T2CEP-194   &  BLHer    &     1.3144675    &     17.406   &        0.017   &      17.134   &       0.018   &     0.38   &     0.24   &       0.080  &  T              \\
OGLE-LMC-T2CEP-136   &  BLHer    &     1.3230384    &     16.492   &        0.011   &      15.978   &       0.006   &     0.31   &     0.08   &       0.060  &  S              \\
OGLE-LMC-T2CEP-138   &  BLHer    &     1.3935906    &     16.975   &        0.043   &      16.576   &       0.017   &     0.07   &     0.07   &       0.070  &  T              \\
OGLE-LMC-T2CEP-109   &  BLHer    &     1.4145528    &     18.610   &        0.056   &      17.790   &       0.038   &     0.43   &     0.38   &       0.030  &  S              \\
OGLE-LMC-T2CEP-105   &  BLHer    &     1.4892979    &     17.134   &        0.012   &      16.914   &       0.021   &     0.41   &     0.27   &       0.080  &  T              \\
OGLE-LMC-T2CEP-122   &  BLHer    &     1.5386690    &     17.520   &        0.034   &      17.136   &       0.018   &     0.24   &     0.23   &       0.110  &  T              \\
OGLE-LMC-T2CEP-171   &  BLHer    &     1.5547492    &     17.175   &        0.017   &      16.875   &       0.017   &     0.18   &     0.17   &       0.170  &  T              \\
OGLE-LMC-T2CEP-068   &  BLHer    &     1.6093007    &     17.225   &        0.028   &      16.942   &       0.018   &     0.27   &     0.26   &       0.100  &  T              \\
OGLE-LMC-T2CEP-124   &  BLHer    &     1.7348666    &     17.280   &        0.009   &      16.953   &       0.030   &     0.30   &     0.30   &       0.110  &  S              \\
OGLE-LMC-T2CEP-008   &  BLHer    &     1.7460989    &     17.257   &        0.023   &      17.389   &       0.028   &     0.08   &     0.08   &       0.100  &  T              \\
OGLE-LMC-T2CEP-141   &  BLHer    &     1.8229539    &     17.389   &        0.023   &      17.048   &       0.021   &     0.36   &     0.40   &       0.100  &  T              \\
OGLE-LMC-T2CEP-140   &  BLHer    &     1.8411435    &     17.127   &        0.012   &      16.779   &       0.014   &     0.21   &     0.27   &       0.080  &  S              \\
OGLE-LMC-T2CEP-144   &  BLHer    &     1.9374502    &     16.726   &        0.017   &      16.302   &       0.011   &     0.22   &     0.20   &       0.120  &  S              \\
OGLE-LMC-T2CEP-130   &  BLHer    &     1.9446935    &     17.036   &        0.016   &      16.740   &       0.021   &     0.36   &     0.34   &       0.060  &  T              \\
OGLE-LMC-T2CEP-088   &  BLHer    &     1.9507490    &     17.158   &        0.012   &      17.147   &       0.028   &     0.09   &     0.09   &       0.060  &  T              \\
OGLE-LMC-T2CEP-116   &  BLHer    &     1.9666793    &     17.086   &        0.038   &      16.746   &       0.007   &     0.23   &     0.32   &       0.060  &  S              \\
OGLE-LMC-T2CEP-121   &  BLHer    &     2.0613655    &     17.234   &        0.033   &      16.854   &       0.014   &     0.45   &     0.43   &       0.030  &  T              \\
OGLE-LMC-T2CEP-166   &  BLHer    &     2.1105987    &     16.343   &        0.015   &      15.922   &       0.006   &     0.23   &     0.22   &       0.190  &  T              \\
OGLE-LMC-T2CEP-064   &  BLHer    &     2.1278906    &     17.043   &        0.019   &      16.698   &       0.025   &     0.47   &     0.45   &       0.070  &  T              \\
OGLE-LMC-T2CEP-167   &  BLHer    &     2.3118238    &     17.091   &        0.045   &      16.685   &       0.010   &     0.50   &     0.48   &       0.320  &  T              \\
OGLE-LMC-T2CEP-092   &  BLHer    &     2.6167684    &     16.864   &        0.097   &      16.526   &       0.066   &     0.69   &     0.66   &       0.050  &  T              \\
OGLE-LMC-T2CEP-148   &  BLHer    &     2.6717338    &     16.853   &        0.011   &      16.516   &       0.015   &     0.43   &     0.56   &       0.060  &  S              \\
OGLE-LMC-T2CEP-195   &  BLHer    &     2.7529292    &     16.850   &        0.021   &      16.474   &       0.008   &     0.55   &     0.46   &       0.080  &  T              \\
OGLE-LMC-T2CEP-113   &  BLHer    &     3.0854602    &     16.285   &        0.002   &      15.935   &       0.008   &     0.10   &     0.06   &       0.020  &  S              \\
OGLE-LMC-T2CEP-049   &  BLHer    &     3.2352751    &     16.359   &        0.015   &      15.926   &       0.010   &     0.25   &     0.24   &       0.070  &  T              \\
OGLE-LMC-T2CEP-145   &  BLHer    &     3.3373019    &     16.269   &        0.008   &      16.047   &       0.015   &     0.11   &     0.08   &       0.120  &  S              \\
OGLE-LMC-T2CEP-085   &  BLHer    &     3.4050955    &     16.640   &        0.017   &      16.191   &       0.011   &     0.47   &     0.45   &       0.090  &  T              \\
OGLE-LMC-T2CEP-134   &  pWVir    &     4.0757258    &     15.782   &        0.009   &      15.514   &       0.007   &     0.31   &     0.36   &       0.080  &  S              \\
OGLE-LMC-T2CEP-173   &  WVir    &     4.1478811    &     16.049   &        0.018   &      15.452   &       0.005   &     0.12   &     0.11   &       0.170  &  T               \\
OGLE-LMC-T2CEP-120   &  WVir    &     4.5590530    &     16.354   &        0.007   &      15.951   &       0.009   &     0.38   &     0.38   &       0.130  &  S               \\
OGLE-LMC-T2CEP-052   &  pWVir    &     4.6879253    &     16.031   &        0.018   &      15.741   &       0.022   &     0.14   &     0.13   &       0.070  &  T              \\
OGLE-LMC-T2CEP-098   &  pWVir    &     4.9737372    &     14.056   &        0.014   &      13.892   &       0.005   &     0.15   &     0.14   &       0.120  &  T              \\
OGLE-LMC-T2CEP-087   &  WVir    &     5.1849790    &     16.302   &        0.013   &      15.859   &       0.015   &     0.30   &     0.31   &       0.090  &  S               \\
OGLE-LMC-T2CEP-023   &  pWVir    &     5.2348007    &     15.005   &        0.043   &      14.720   &       0.013   &     0.36   &     0.34   &       0.040  &  T              \\
OGLE-LMC-T2CEP-083   &  pWVir    &     5.9676496    &     15.936   &        0.054   &      15.462   &       0.011   &     0.48   &     0.46   &       0.100  &  T              \\
OGLE-LMC-T2CEP-062   &  WVir    &     6.0466764    &     16.060   &        0.019   &      15.431   &       0.003   &     0.05   &     0.05   &       0.090  &  T               \\
OGLE-LMC-T2CEP-133   &  WVir    &     6.2819551    &     16.013   &        0.010   &      15.564   &       0.013   &     0.09   &     0.09   &       0.040  &  T               \\
OGLE-LMC-T2CEP-137   &  WVir    &     6.3623499    &     16.044   &        0.004   &      15.630   &       0.010   &     0.11   &     0.11   &       0.110  &  S               \\
OGLE-LMC-T2CEP-183   &  WVir    &     6.5096275    &     16.325   &        0.016   &      15.739   &       0.016   &     0.15   &     0.14   &       0.200  &  T               \\
OGLE-LMC-T2CEP-159   &  WVir    &     6.6255696    &     16.089   &        0.015   &      15.605   &       0.010   &     0.09   &     0.09   &       0.110  &  T               \\
OGLE-LMC-T2CEP-117   &  WVir    &     6.6293487    &     16.007   &        0.012   &      15.579   &       0.005   &     0.12   &     0.11   &       0.080  &  T               \\
OGLE-LMC-T2CEP-106   &  WVir    &     6.7067363    &     15.956   &        0.055   &      15.474   &       0.010   &     0.16   &     0.15   &       0.050  &  T               \\
OGLE-LMC-T2CEP-078   &  pWVir    &     6.7162943    &     15.349   &        0.016   &      14.764   &       0.011   &     0.15   &     0.14   &       0.090  &  T              \\
OGLE-LMC-T2CEP-063   &  WVir    &     6.9245800    &     16.040   &        0.023   &      15.577   &       0.016   &     0.14   &     0.13   &       0.050  &  T               \\
OGLE-LMC-T2CEP-110   &  WVir    &     7.0784684    &     15.978   &        0.008   &      15.511   &       0.017   &     0.16   &     0.15   &       0.120  &  S               \\
OGLE-LMC-T2CEP-181   &  pWVir    &     7.2125323    &     15.505   &        0.013   &      15.151   &       0.005   &     0.07   &     0.07   &       0.130  &  T              \\
OGLE-LMC-T2CEP-047   &  WVir    &     7.2862123    &     15.943   &        0.018   &      15.511   &       0.011   &     0.14   &     0.13   &       0.070  &  T               \\
OGLE-LMC-T2CEP-056   &  WVir    &     7.2896382    &     15.965   &        0.017   &      15.522   &       0.004   &     0.16   &     0.15   &       0.110  &  T               \\
OGLE-LMC-T2CEP-100   &  WVir    &     7.4310950    &     15.965   &        0.012   &      15.647   &       0.020   &     0.29   &     0.28   &       0.080  &  T               \\
OGLE-LMC-T2CEP-111   &  WVir    &     7.4956838    &     15.865   &        0.011   &      15.441   &       0.006   &     0.19   &     0.18   &       0.060  &  T               \\
OGLE-LMC-T2CEP-170   &  WVir    &     7.6829062    &     15.926   &        0.018   &      15.423   &       0.004   &     0.16   &     0.15   &       0.180  &  T               \\
OGLE-LMC-T2CEP-151   &  WVir    &     7.8872458    &     15.814   &        0.016   &      15.366   &       0.009   &     0.14   &     0.13   &       0.110  &  T               \\
OGLE-LMC-T2CEP-179   &  WVir    &     8.0500650    &     15.932   &        0.014   &      15.378   &       0.005   &     0.14   &     0.13   &       0.110  &  T               \\
OGLE-LMC-T2CEP-182   &  WVir    &     8.2264194    &     15.628   &        0.035   &      15.218   &       0.007   &     0.37   &     0.35   &       0.130  &  T               \\
OGLE-LMC-T2CEP-094   &  WVir    &     8.4684897    &     15.659   &        0.048   &      15.143   &       0.006   &     0.10   &     0.10   &       0.040  &  T               \\
OGLE-LMC-T2CEP-019   &  pWVir    &     8.6748634    &     15.263   &        0.024   &      14.880   &       0.015   &     0.33   &     0.31   &       0.110  &  T              \\
OGLE-LMC-T2CEP-039   &  WVir    &     8.7158373    &     15.682   &        0.018   &      15.217   &       0.009   &     0.19   &     0.18   &       0.040  &  T               \\
OGLE-LMC-T2CEP-028   &  pWVir    &     8.7848073    &     15.083   &        0.016   &      14.791   &       0.006   &     0.32   &     0.30   &       0.050  &  T              \\
OGLE-LMC-T2CEP-074   &  WVir    &     8.9883439    &     15.414   &        0.019   &      15.025   &       0.025   &     0.22   &     0.21   &       0.060  &  T               \\
OGLE-LMC-T2CEP-152   &  WVir    &     9.3149211    &     15.559   &        0.013   &      15.080   &       0.004   &     0.39   &     0.37   &       0.100  &  T               \\
OGLE-LMC-T2CEP-021   &  pWVir    &     9.7595024    &     15.309   &        0.046   &      15.059   &       0.018   &     0.16   &     0.15   &       0.070  &  T              \\
OGLE-LMC-T2CEP-132   &  pWVir    &    10.0178287    &     15.227   &        0.015   &      14.804   &       0.005   &     0.22   &     0.09   &       0.080  &  S              \\
OGLE-LMC-T2CEP-146   &  WVir    &    10.0795925    &     15.576   &        0.026   &      15.172   &       0.021   &     0.37   &     0.29   &       0.100  &  S               \\
OGLE-LMC-T2CEP-097   &  WVir    &    10.5101666    &     15.530   &        0.062   &      15.068   &       0.006   &     0.28   &     0.27   &       0.050  &  T               \\
OGLE-LMC-T2CEP-022   &  WVir    &    10.7167800    &     15.598   &        0.011   &      15.126   &       0.015   &     0.35   &     0.33   &       0.030  &  T               \\
OGLE-LMC-T2CEP-201   &  pWVir    &    11.0072431    &     14.195   &        0.018   &      13.892   &       0.007   &     0.06   &     0.06   &       0.050  &  T              \\
OGLE-LMC-T2CEP-101   &  WVir    &    11.4185596    &     15.427   &        0.009   &      15.009   &       0.007   &     0.45   &     0.40   &       0.080  &  S               \\
OGLE-LMC-T2CEP-013   &  WVir    &    11.5446113    &     15.498   &        0.014   &      15.001   &       0.013   &     0.22   &     0.21   &       0.090  &  T               \\
OGLE-LMC-T2CEP-178   &  WVir    &    12.2123667    &     15.517   &        0.020   &      14.985   &       0.008   &     0.33   &     0.31   &       0.150  &  T               \\
OGLE-LMC-T2CEP-127   &  WVir    &    12.6691185    &     15.372   &        0.022   &      14.851   &       0.011   &     0.48   &     0.37   &       0.070  &  S               \\
OGLE-LMC-T2CEP-118   &  WVir    &    12.6985804    &     15.412   &        0.038   &      14.914   &       0.007   &     0.72   &     0.69   &       0.100  &  T               \\
OGLE-LMC-T2CEP-103   &  WVir    &    12.9082775    &     15.336   &        0.011   &      14.859   &       0.019   &     0.40   &     0.38   &       0.080  &  T               \\
OGLE-LMC-T2CEP-044   &  WVir    &    13.2701004    &     15.455   &        0.030   &      14.835   &       0.013   &     0.30   &     0.29   &       0.090  &  T               \\
OGLE-LMC-T2CEP-026   &  WVir    &    13.5778689    &     15.209   &        0.089   &      14.823   &       0.012   &     0.39   &     0.37   &       0.080  &  T               \\
OGLE-LMC-T2CEP-096   &  WVir    &    13.9257224    &     15.277   &        0.056   &      14.776   &       0.006   &     0.81   &     0.75   &       0.090  &  S               \\
OGLE-LMC-T2CEP-157   &  WVir    &    14.3346466    &     15.304   &        0.045   &      14.782   &       0.043   &     0.66   &     0.63   &       0.100  &  T               \\
OGLE-LMC-T2CEP-017   &  WVir    &    14.4547544    &     15.354   &        0.056   &      14.785   &       0.021   &     0.81   &     0.77   &       0.110  &  T               \\
OGLE-LMC-T2CEP-143   &  WVir    &    14.5701846    &     14.991   &        0.075   &      14.743   &       0.068   &     1.05   &     0.72   &       0.060  &  S               \\
\noalign{\smallskip}                                                                                                                                                                    \hline                                                                                                                                                                                  \noalign{\smallskip}                                                                                                                                                                    \end{tabular}                                                                                                                                                                          
\end{center}
\end{table*}

\begin{table*}
\scriptsize\tiny
\contcaption{}
\begin{center}
\begin{tabular}{ccccccccccc}
\hline
\noalign{\smallskip}   
ID & Var. Class & Period & $\langle J \rangle$ & $\sigma_{\langle J \rangle}$ &  $\langle K_\mathrm{s} \rangle$ & $\sigma_{\langle K_\mathrm{s} \rangle}$ & $Amp(J)$ & $Amp(K_\mathrm{s})$ & $E($V$$-$$I$)$ & Note \\ 
   & & d & mag & mag &mag & mag & mag & mag & mag &\\                     
(1)    & (2)  & (3) & (4) & (5) &(6)  & (7) &(8) & (9) & (10) & (11) \\                     
\noalign{\smallskip}
\hline
\noalign{\smallskip}    
OGLE-LMC-T2CEP-046   &  WVir    &    14.7437956    &     14.921   &        0.058   &      14.360   &       0.021   &     0.62   &     0.59   &       0.060  &  T                 \\
OGLE-LMC-T2CEP-139   &  WVir    &    14.7804104    &     15.220   &        0.014   &      14.709   &       0.005   &     0.50   &     0.51   &       0.150  &  S                 \\
OGLE-LMC-T2CEP-177   &  WVir    &    15.0359027    &     15.245   &        0.024   &      14.741   &       0.007   &     0.69   &     0.66   &       0.270  &  T                 \\
OGLE-LMC-T2CEP-099   &  WVir    &    15.4867877    &     15.094   &        0.003   &      14.564   &       0.005   &     0.51   &     0.52   &       0.100  &  S                 \\
OGLE-LMC-T2CEP-086   &  WVir    &    15.8455000    &     15.024   &        0.011   &      14.586   &       0.017   &     0.79   &     0.80   &       0.030  &  S                 \\
OGLE-LMC-T2CEP-126   &  WVir    &    16.3267785    &     15.323   &        0.023   &      14.733   &       0.013   &     0.77   &     0.73   &       0.090  &  T                 \\
OGLE-LMC-T2CEP-057   &  WVir    &    16.6320415    &     15.052   &        0.021   &      14.566   &       0.013   &     0.82   &     0.78   &       0.060  &  T                 \\
OGLE-LMC-T2CEP-093   &  WVir    &    17.5930492    &     14.524   &        0.021   &      14.136   &       0.019   &     0.61   &     0.47   &       0.040  &  S                 \\
OGLE-LMC-T2CEP-128   &  WVir    &    18.4926938    &     14.787   &        0.023   &      14.363   &       0.054   &     0.71   &     0.68   &       0.050  &  T                 \\
OGLE-LMC-T2CEP-058   &  RVTau    &    21.4829509    &     14.777   &        0.017   &      14.208   &       0.014   &     0.75   &     0.71   &       0.090  &  T                \\
OGLE-LMC-T2CEP-104   &  RVTau    &    24.8799480    &     14.131   &        0.020   &      13.402   &       0.043   &     0.32   &     0.61   &       0.090  &  S                \\
OGLE-LMC-T2CEP-115   &  RVTau    &    24.9669126    &     14.790   &        0.002   &      14.334   &       0.013   &     0.66   &     0.63   &       0.030  &  S                \\
OGLE-LMC-T2CEP-192   &  RVTau    &    26.1940011    &     14.521   &        0.033   &      14.096   &       0.008   &     1.09   &     1.04   &       0.060  &  T                \\
OGLE-LMC-T2CEP-135   &  RVTau    &    26.5223638    &     14.350   &        0.016   &      13.799   &       0.015   &     1.09   &     0.76   &       0.070  &  S                \\
OGLE-LMC-T2CEP-162   &  RVTau    &    30.3941483    &     14.294   &        0.043   &      13.726   &       0.043   &     0.57   &     0.41   &       0.220  &  T                \\
OGLE-LMC-T2CEP-180   &  RVTau    &    30.9963145    &     13.785   &        0.068   &      12.921   &       0.033   &     0.42   &     0.40   &       0.070  &  T                \\
OGLE-LMC-T2CEP-119   &  RVTau    &    33.8250938    &     13.832   &        0.021   &      12.951   &       0.064   &     0.89   &     0.85   &       0.080  &  T                \\
OGLE-LMC-T2CEP-050   &  RVTau    &    34.7483438    &     14.257   &        0.030   &      13.811   &       0.014   &     0.19   &     0.18   &       0.070  &  T                \\
OGLE-LMC-T2CEP-091   &  RVTau    &    35.7493456    &     13.652   &        0.045   &      12.693   &       0.055   &     0.62   &     0.64   &       0.070  &  S                \\
OGLE-LMC-T2CEP-203   &  RVTau    &    37.1267463    &     14.416   &        0.007   &      13.739   &       0.004   &     0.61   &     0.39   &       0.040  &  S                \\
OGLE-LMC-T2CEP-202   &  RVTau    &    38.1355674    &     14.310   &        0.013   &      13.753   &       0.015   &     0.07   &     0.07   &       0.090  &  T                \\
OGLE-LMC-T2CEP-112   &  RVTau    &    39.3977037    &     13.531   &        0.021   &      13.163   &       0.009   &     0.27   &     0.24   &       0.030  &  S                \\
OGLE-LMC-T2CEP-080   &  RVTau    &    40.9164131    &     13.957   &        0.027   &      13.253   &       0.047   &     0.44   &     0.42   &       0.040  &  T                \\
OGLE-LMC-T2CEP-149   &  RVTau    &    42.4806129    &     13.649   &        0.039   &      13.252   &       0.007   &     0.13   &     0.12   &       0.140  &  T                \\
OGLE-LMC-T2CEP-032   &  RVTau    &    44.5611948    &     13.232   &        0.030   &      12.212   &       0.090   &     0.36   &     0.34   &       0.050  &  T                \\
OGLE-LMC-T2CEP-147   &  RVTau    &    46.7958419    &     13.145   &        0.017   &      12.658   &       0.013   &     0.06   &     0.06   &       0.090  &  T                \\
OGLE-LMC-T2CEP-174   &  RVTau    &    46.8189562    &     13.089   &        0.016   &      12.048   &       0.030   &     0.46   &     0.44   &       0.150  &  T                \\
OGLE-LMC-T2CEP-067   &  RVTau    &    48.2317051    &     13.176   &        0.022   &      12.263   &       0.052   &     0.20   &     0.19   &       0.100  &  T                \\
OGLE-LMC-T2CEP-075   &  RVTau    &    50.1865686    &     13.900   &        0.110   &      13.502   &       0.033   &     0.78   &     0.74   &       0.070  &  T                \\
OGLE-LMC-T2CEP-129   &  RVTau    &    62.5089466    &     13.514   &        0.035   &      13.123   &       0.013   &     0.16   &     0.14   &       0.070  &  S                \\
OGLE-LMC-T2CEP-045   &  RVTau    &    63.3863391    &     13.098   &        0.024   &      12.664   &       0.021   &     0.16   &     0.15   &       0.070  &  T                \\
\noalign{\smallskip}
\hline
\noalign{\smallskip}
\end{tabular}
\end{center}
\end{table*}

In principle, an additional preliminary step would be required,
i.e. the correction for the inclination of the LMC disc-like structure
 by de-projecting each T2CEP with respect to the LMC centre. To do
 this we followed the procedure suggested in \citet{marel01} and adopted their values 
of the LMC centre, inclination, and position angle of the line of
nodes. However, we have a posteriori verified that the introduction of
this correction leads to worse results, i.e. larger dispersion in the
various relationships mentioned above. To verify if different choices
about the inclined disc parameters could improve the results, we have 
carried out the de-projection using several results present in the
literature \citep[see][and references
therein]{Haschke2012,Rubele2012,Subramanian2013}. Under no
circumstances the dispersion of the $PWs$ decreased (we used $PWs$ as
reference because they are reddening-free). 
To explain this occurrence we can reasonably hypothesise that the
T2CEPs (actually BL Her and W Vir), being old (age$>$ 10 Gyr) objects, 
are not preferentially distributed along the main disc-like structure of the LMC. 
Alternatively, the adopted parameters for the de-projection are not
 accurate enough, although this conclusion may be influenced by the 
relatively small number of objects. Subsequent studies using a larger 
number of objects observed in the VMC context  sampling different
populations (CCs, T2CEPs, RR Lyrae stars) will clarify the issue.
In any case, in the following analysis we did not apply any magnitude
correction accounting for the LMC disk structure.


\begin{figure*}
\hbox{
\includegraphics[width=8.5cm]{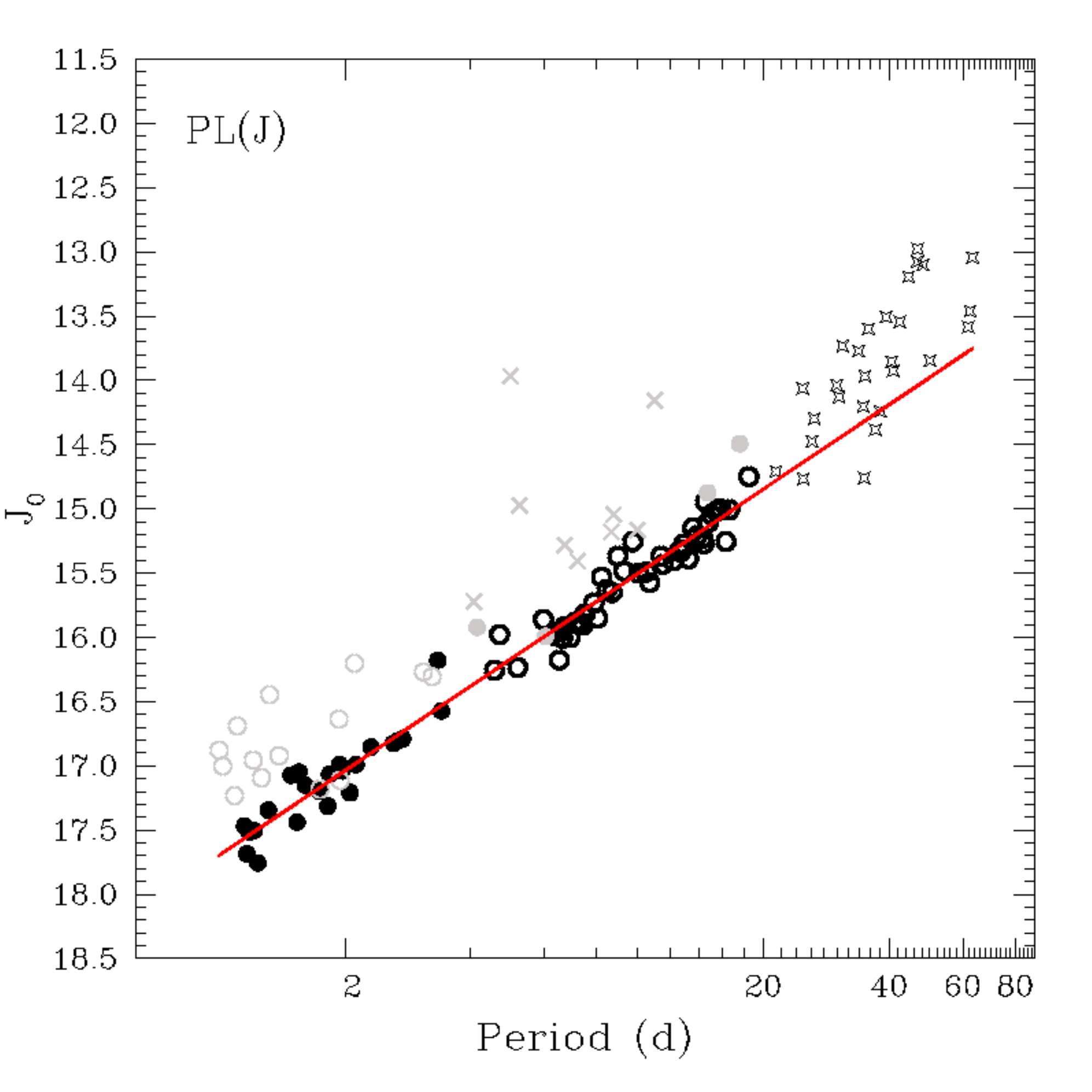}
\includegraphics[width=8.5cm]{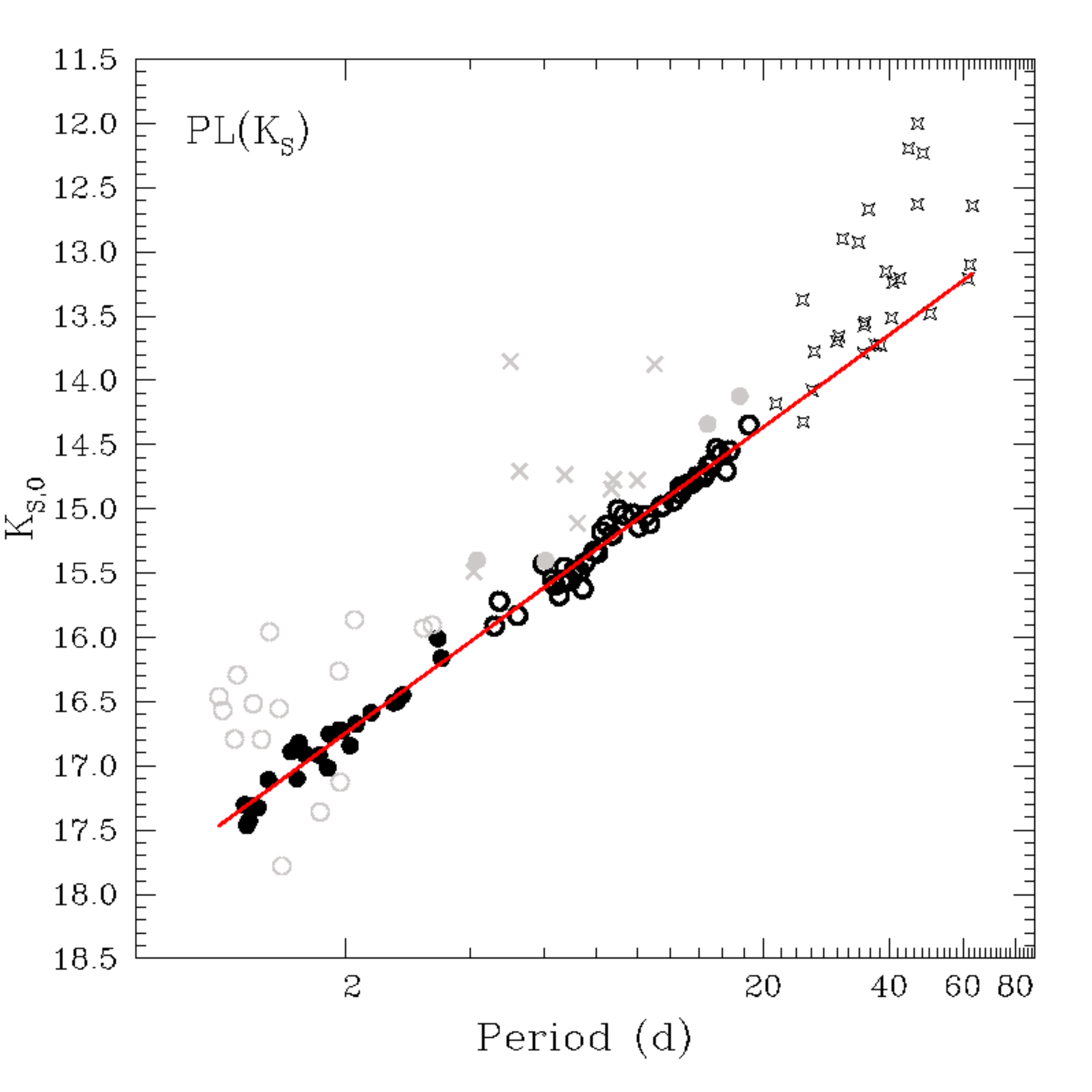}
}
\caption{$PL(J)$ and $PL(K_\mathrm{s})$ relations for the T2CEPs investigated in
  this paper. The meaning of the symbols is the following: black filled
  and empty circles are the BL Her and W Vir
  variables used in the derivation of the $PL$, $PW$ and $PLC$
  relationships, respectively. Grey empty and filled circles are the
  BL Her and W Vir variables discarded because of problems in the
  photometry (see text). Grey crosses are the peculiar W Vir stars. The starred 
symbols represent the RV Tau variables. The size of the symbols is
generally representative of the measurement errors. The solid lines represent the
least-squares fit to the data shown in Table~\ref{NIR}. We recall that
RV Tau stars were not used in the calculation of the least-squares
fits (see text).} 
\label{pl}
\end{figure*}

\begin{figure*}
\hbox{
\includegraphics[width=8.5cm]{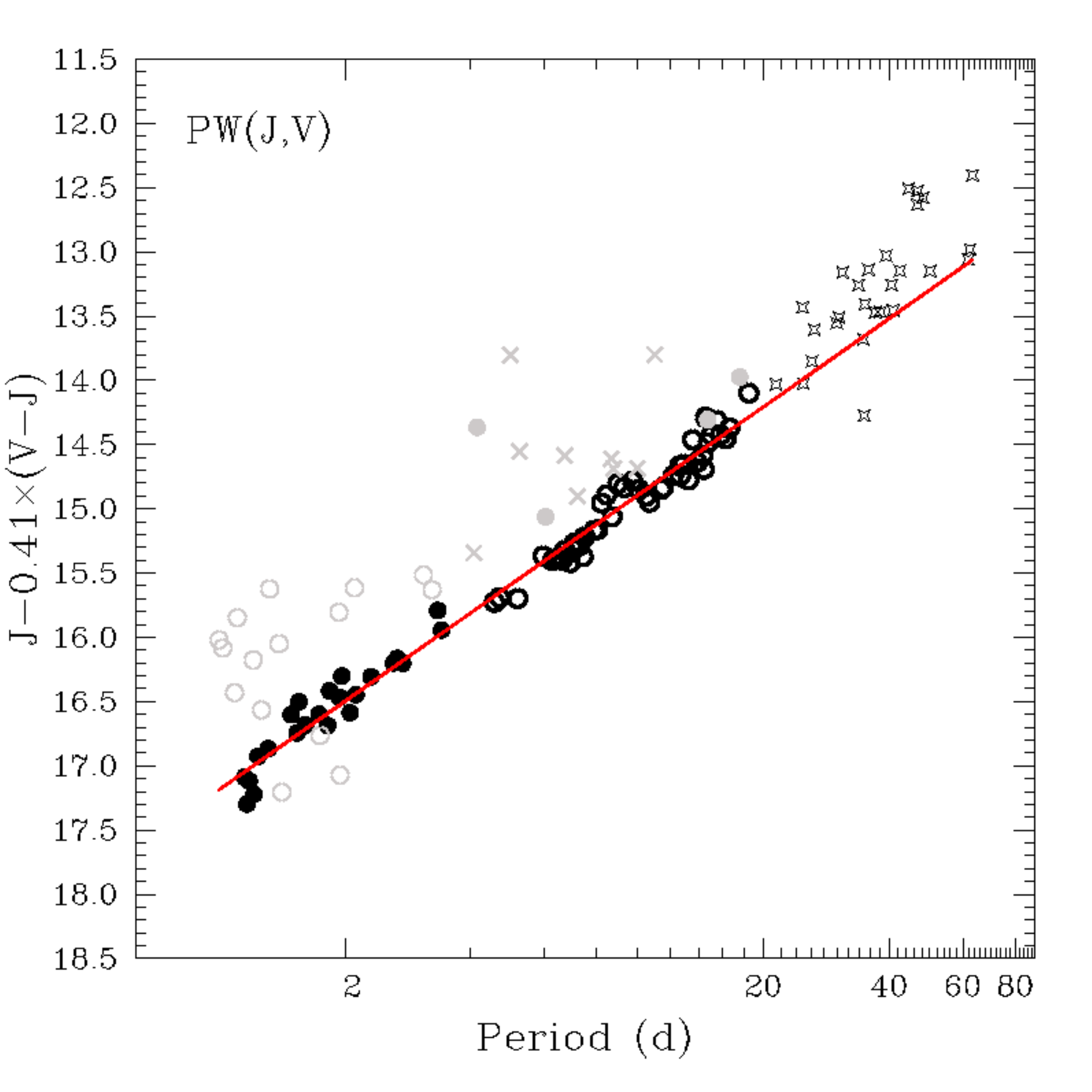}
\includegraphics[width=8.5cm]{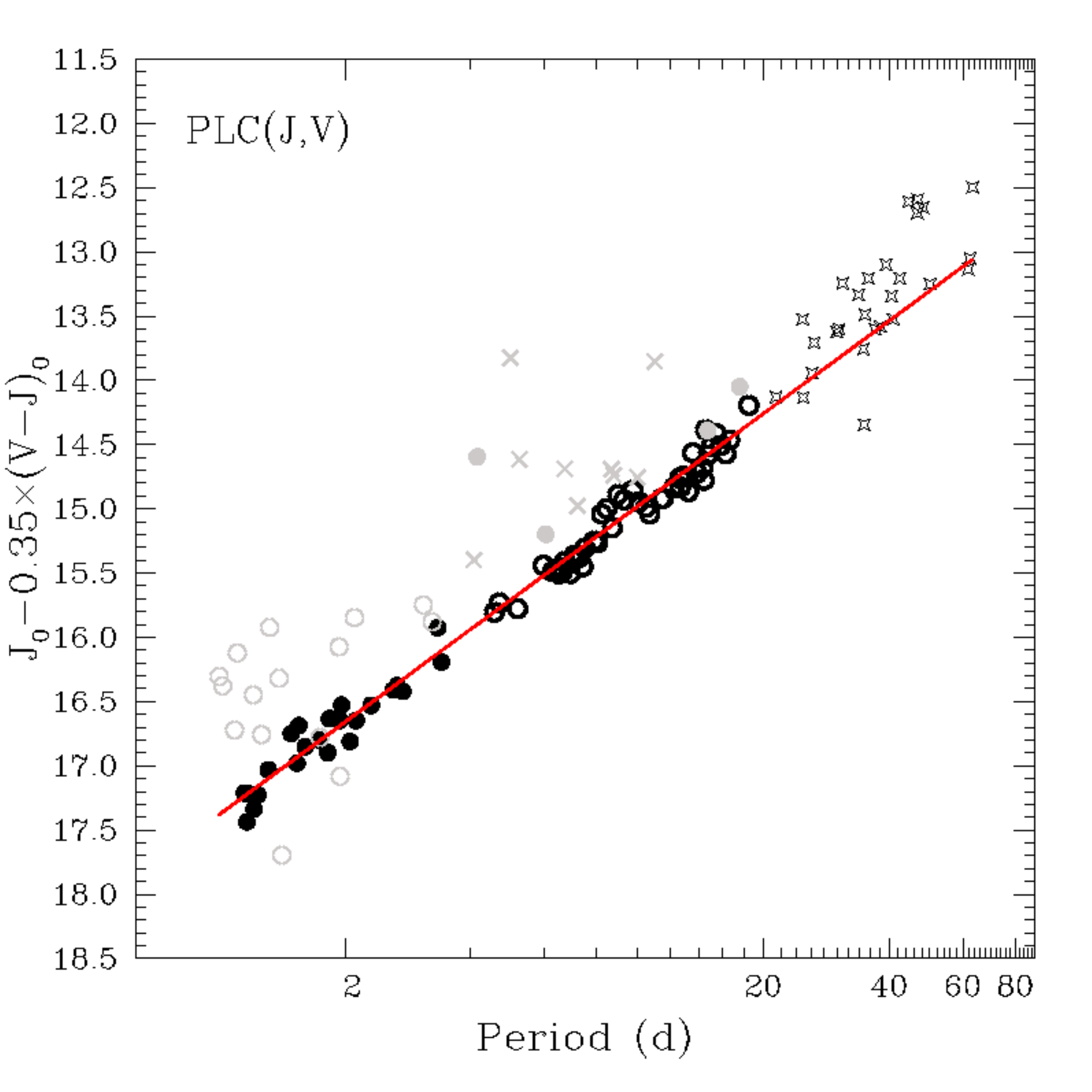}
}
\caption{$PW(J,V)$ and $PLC(J,V)$ for the T2CEPs investigated in
  this paper. Symbols are as in Fig.~\ref{pl}.} 
\label{vj}
\end{figure*}

\begin{figure*}
\hbox{
\includegraphics[width=8.5cm]{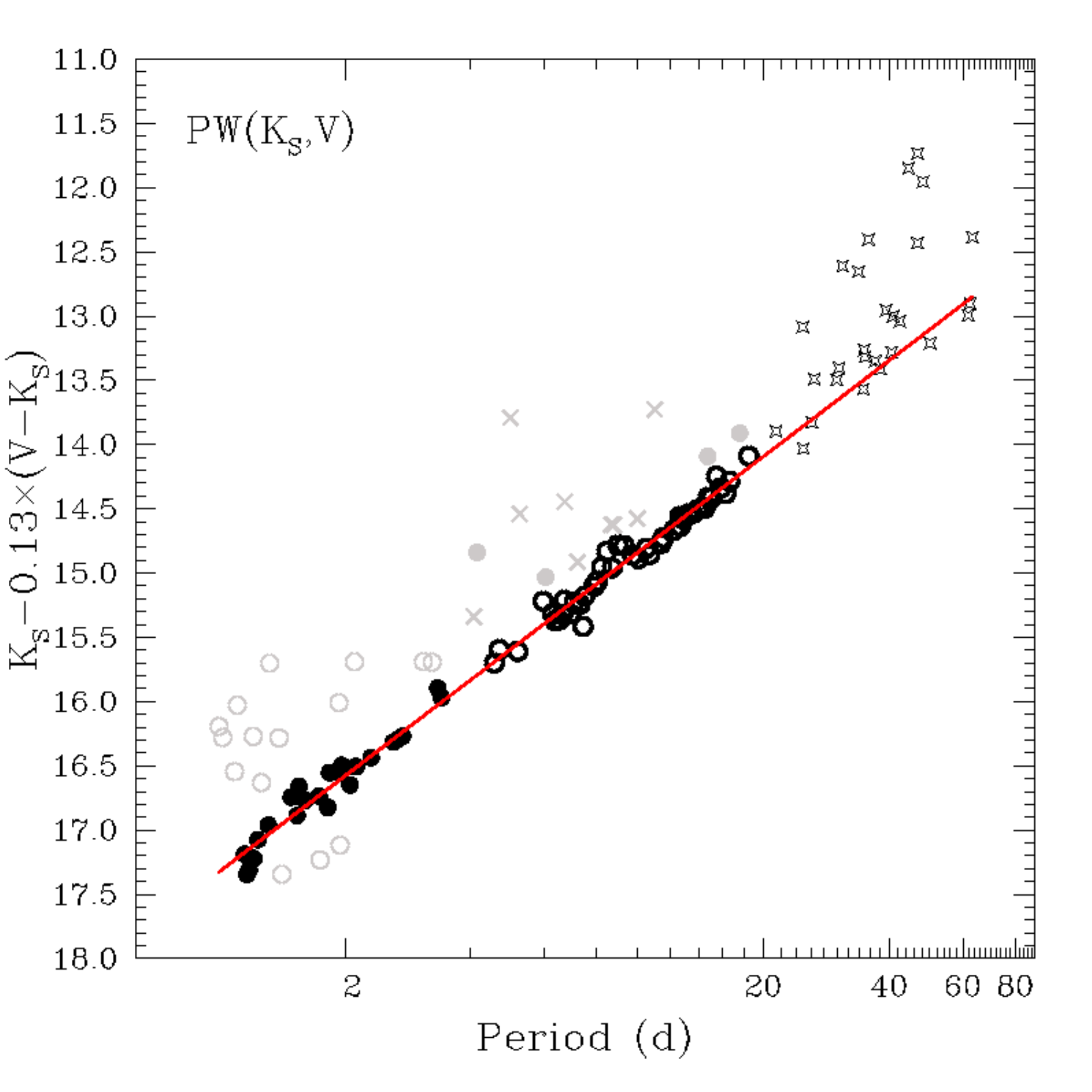}
\includegraphics[width=8.5cm]{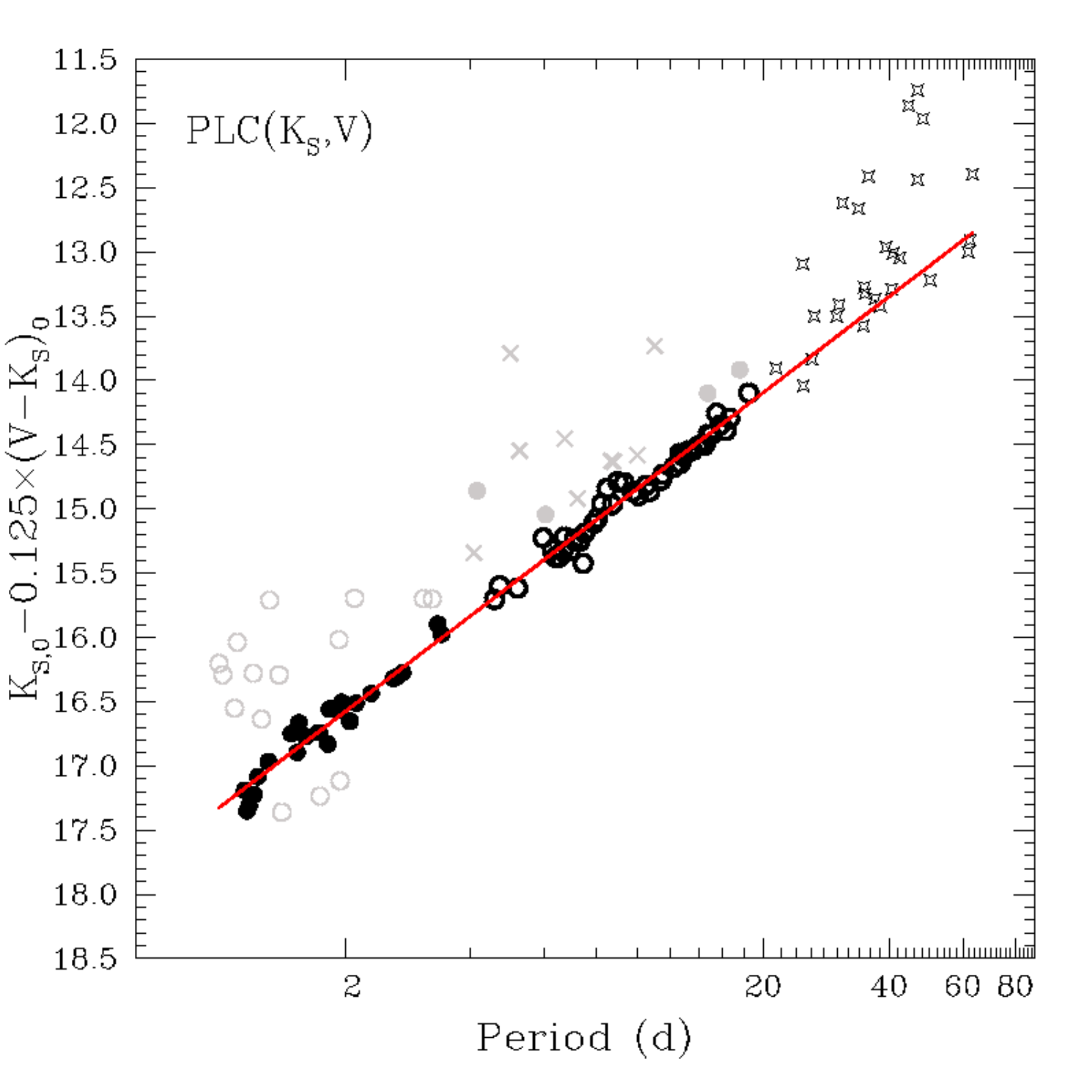}
}
\caption{$PW(K_\mathrm{s},V)$ and $PLC(K_\mathrm{s},V)$ for the T2CEPs investigated in
  this paper. Symbols are as in Fig.~\ref{pl}.} 
\label{vk}
\end{figure*}

\begin{figure*}
\hbox{
\includegraphics[width=8.5cm]{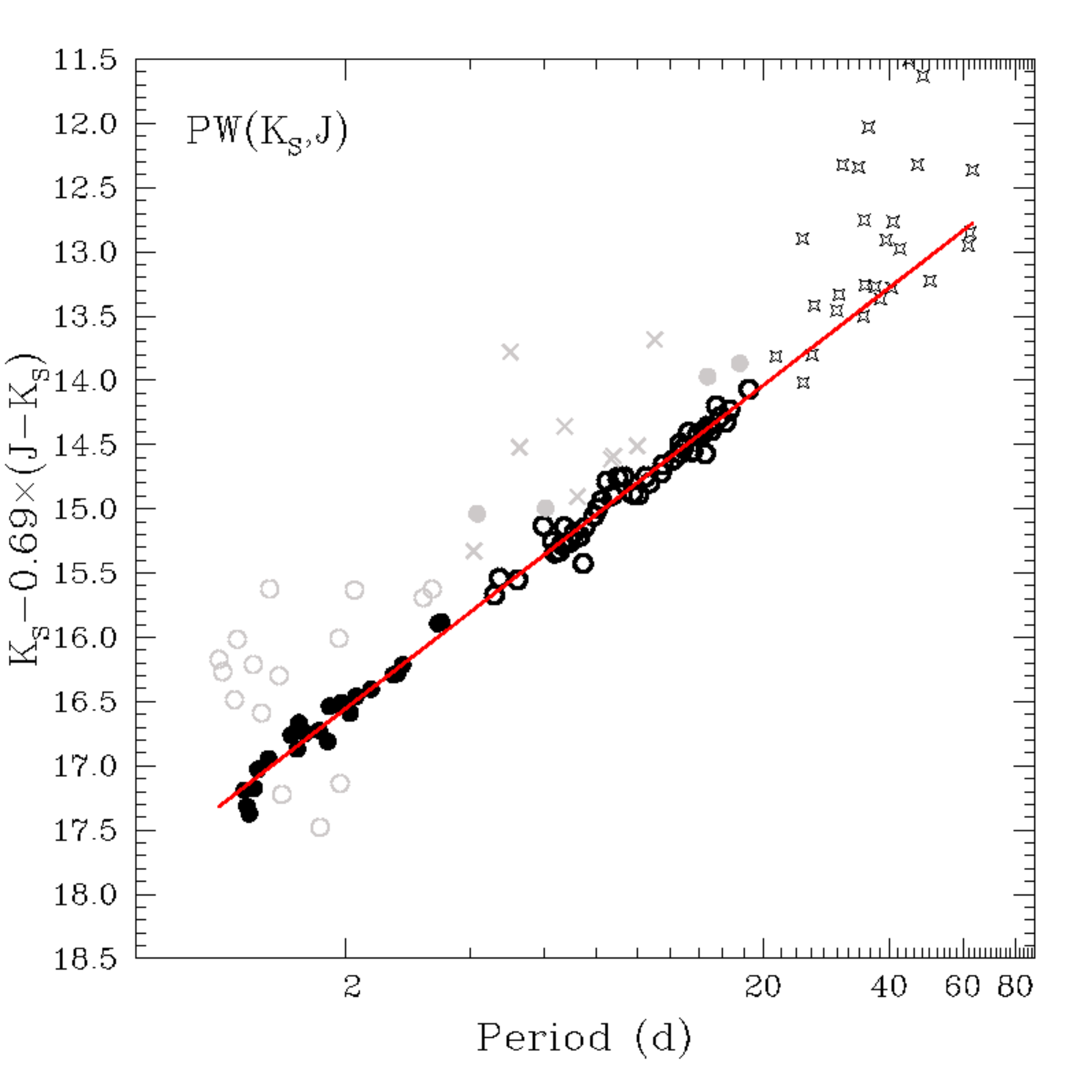}
\includegraphics[width=8.5cm]{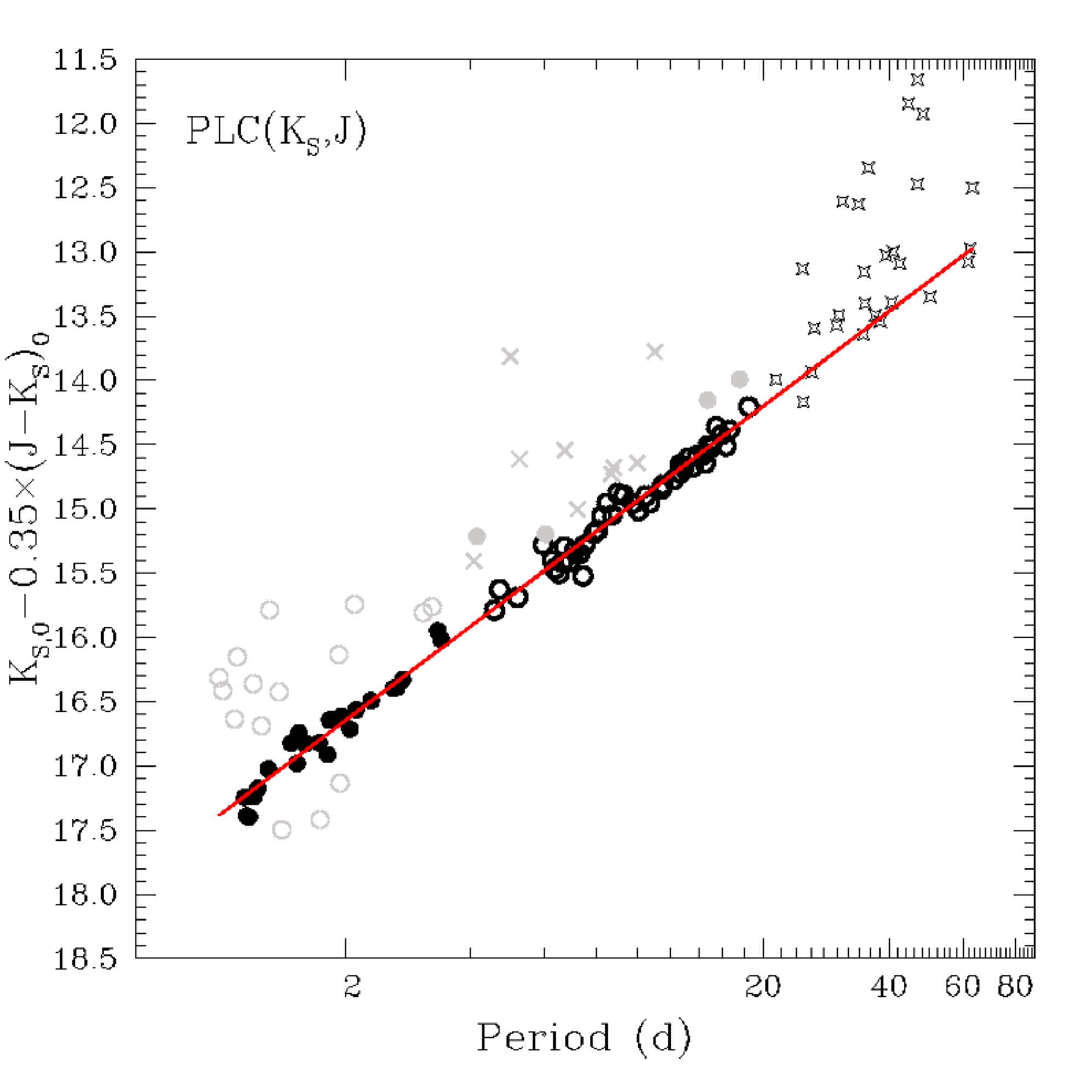}
}
\caption{$PW(K_\mathrm{s},J)$ and $PLC(K_\mathrm{s},J)$ for the T2CEPs investigated in
  this paper. Symbols are as in Fig.~\ref{pl}.} 
\label{jk}
\end{figure*}

\begin{figure}
\includegraphics[width=8.5cm]{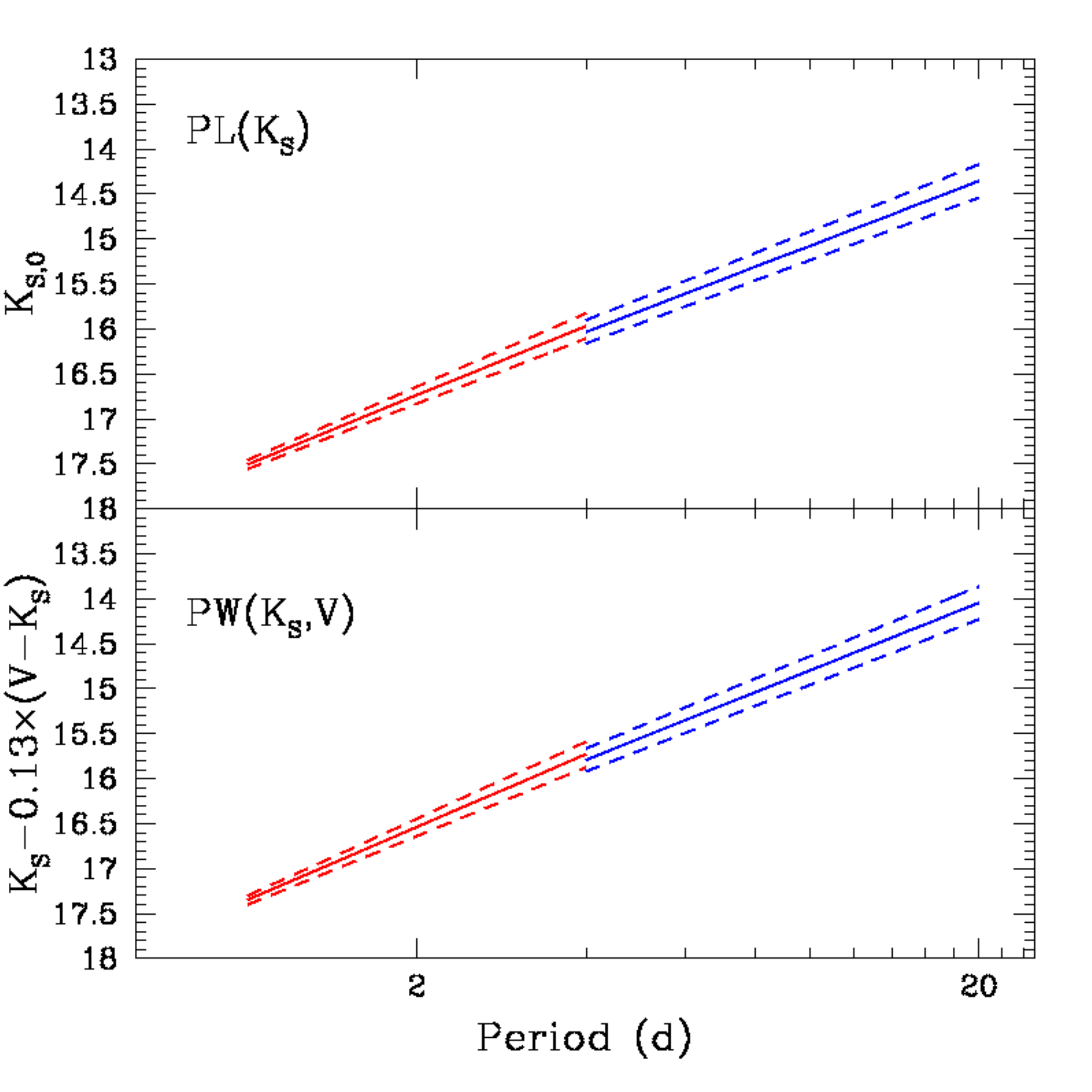}
\caption{Top panel: $PL(K_\mathrm{s})$ relation calculated separately for 
for BL Her (red) and W Vir (blue) variables. The solid and dashed 
lines show the best-fit $\pm$1 $\sigma$ error (both for slope and $ZP$), respectively. Bottom 
panel: as above but for the $PW(K_\mathrm{s},J-K_\mathrm{s})$ relation.  
}
\label{compRelations}
\end{figure}

Figures~\ref{pl},~\ref{vj},~\ref{vk} and ~\ref{jk} show all the  
relationships investigated here.   
An inspection of these figures confirms the findings by
\citet{Matsunaga2009} that BL Her and W Vir star follow a common $PL$
relation, whereas RV Tau show a different and more dispersed
relation (the dispersion is less severe in the $J$ than in
the $K_\mathrm{s}$-band). 
In our case the dispersion among RV Tau stars can in part be
due to the proximity of several bright variables to the saturation
limit. As a consequence, we decided to exclude these stars from the
calculation of the $PL$, $PW$ and $PLC$ relations.
To check if BL Her and W Vir stars can actually be fitted with an
unique relation we performed an independent test by fitting separately
the $PL(K_\mathrm{s},J)$ and $PW(K_\mathrm{s},V)$ relations for 
each class of variables. The result of this exercise is shown in
Fig.~\ref{compRelations}: for both relations, the two variable classes 
seem to show results that agree with each other well within 1
$\sigma$, thus confirming that we can use BL Her and W Vir variables 
together.

For each combination of periods, magnitudes and colours, we 
performed independent least-squares fits to the data, adopting 
equations of the form reported in Table~\ref{NIR}. The results of the
fitting procedure are shown in the same table as well as in  
Figures~\ref{pl},~\ref{vj},~\ref{vk} and ~\ref{jk} with a solid line. 
Note that the equations listed in Table~\ref{NIR} are given in terms
of absolute magnitudes since we subtracted the 
dereddened distance modulus ($DM_{0,LMC}$) of the LMC from each
equation. Thus, the absolute zero point (ZP) of the relations in
Table~\ref{NIR} can be simply obtained by using the preferred value
for the $DM_{0,LMC}$ value.

In deriving the equations of
Table~\ref{NIR}, we have implicitly neglected any dependence of both
$PL$ and $PW$ relations on the metallicity of the pulsators. This is in
agreement with \citet{Matsunaga2006}, who found a hardly significant
dependence of the $PL$ relations on metallicity (0.1$\pm$0.06 mag/dex), whereas
the theoretical models by \citet{Dicriscienzo2007} predict a very mild
metallicity dependence $\Delta$Mag/$\Delta$ [Fe/H]$\sim0.04-0.06$
mag/dex for both the $PL$ and $PW$ relations in the magnitudes and colours of interest.   
In any case, the very low dispersions of our $PL$ and $PW$ relations listed in
Table~\ref{NIR}, seems to suggest that the metallicity dependence, if
any, should be very small. Alternatively, a small dispersion in
metallicity among our sample could explain the results as
well. However, since the low metallicity dependence found by 
\citet{Matsunaga2006} is based on T2CEPs spanning a wide range of
[Fe/H], the latter explanation is less likely.

\begin{table*}
\small 
\caption{Relevant relationships derived in this work. Note that all the results are in 
the VISTA photometric system. $DM_{LMC,0}$ stands for the LMC dereddened distance modulus.}
\label{NIR}
\begin{center}
\begin{tabular}{llc}
\hline 
\noalign{\smallskip}
method &  Relation & $r.m.s$ (mag) \\
\noalign{\smallskip}
\hline 
\noalign{\smallskip}
$PL(J)$  & $M_{J,0}=(-2.19\pm0.04)\log P+(17.700\pm0.035)-DM_{LMC,0}$ & 0.13 \\
\noalign{\smallskip} 
$PL(K_\mathrm{s})$ & $M_{K_\mathrm{s,0}}=(-2.385\pm0.03)\log P+(17.47\pm0.02) -DM_{LMC,0}$&0.09\\
\noalign{\smallskip} 
$PW(J,V)$ & $M_J-0.41(V-J)=(-2.290\pm0.035)\log P+(17.19\pm0.03) -DM_{LMC,0}$&0.11 \\
\noalign{\smallskip} 
$PLC(J,V)$ & $M_{J,0}=(-2.40\pm0.05)\log P+(0.35\pm0.07)(V-J)_0+(17.385\pm0.065) -DM_{LMC,0}$ & 0.11 \\
\noalign{\smallskip} 
$PW(K_\mathrm{s},V)$ &
$M_{K_\mathrm{s}}-0.13(V-K_\mathrm{s})=(-2.49\pm0.03)\log P+(17.33\pm0.02)-DM_{LMC,0}$&0.08 \\
\noalign{\smallskip} 
$PLC(K_\mathrm{s},V)$ & $M_{K_\mathrm{s,0}}=(-2.48\pm0.04)\log P+(0.125\pm0.040)(V-K_\mathrm{s})_0+(17.33\pm0.05) -DM_{LMC,0}$&0.08 \\
\noalign{\smallskip} 
$PW(K_\mathrm{s},J)$ &$M_{K_\mathrm{s}}-0.69(J-K_\mathrm{s})= (-2.52\pm0.03)\log P+(17.320\pm0.025)-DM_{LMC,0}$&0.085\\
\noalign{\smallskip} 
$PLC(K_\mathrm{s},J)$ & $M_{K_\mathrm{s,0}}=(-2.45\pm0.04)\log P +(0.35\pm0.14)(J-K_\mathrm{s})_0+(17.39\pm0.04)-DM_{LMC,0}$&0.085\\
\noalign{\smallskip}
\hline 
\noalign{\smallskip}
\end{tabular}
\end{center}
\end{table*}

In each figure, a number of stars are shown with grey symbols. They
significantly deviate from almost all relationships discussed
above. The crosses represent the stars
classified by \citet{sos08} as peculiar W Vir (see column 4 in 
Table~\ref{tabData}), i.e. suspected binaries that do not follow the
optical $PL$ and $PW$ relations. We note that three of these peculiar W Vir
stars, namely OGLE-LMC-T2CEP-021, 052 and 083  do not show any difference
with respect to the normal W Vir stars in our $PL$, $PW$ and $PLC$ planes,
and were hence included in the calculations. 
As for BL Her and W Vir, 15 and 4 stars of the two classes were
not used in the least-squares fits because, with few exceptions, they show large
scattering in almost all the relationships calculated here, and, in
particular in the most reliable ones, namely the $PWs$ and $PLCs$
based on 
the $K_\mathrm{s}$-band photometry. The finding charts for all these
stars are displayed in Fig.~\ref{figCharts}, whereas the notes in 
Table~\ref{tabData} explain in detail the causes that led us to
exclude these objects, with blending by close companions being the most common cause.

Table~\ref{NIR} deserves some discussion: i) the dispersion of the
$PL(J)$ relation is, as expected, larger than for the $PL(K_\mathrm{s})$; ii)
for any combination of magnitude and colour, the dispersions of $PW$ and
$PLC$ are equal (this reflects the correctness of the reddening
correction applied in this paper); iii) the $PW(J,V)$ and $PLC(J,V)$ 
are significantly more dispersed than the $PW(K_\mathrm{s},V)$-$PLC(K_\mathrm{s},V)$ 
and $PW(K_\mathrm{s},J)$-$PLC(K_\mathrm{s},J)$ couples; iv) the best
combination of magnitude and colour (lower dispersion) appears to be the
$K_\mathrm{s}$,$V$; v) the color coefficients of the
$PW(K_\mathrm{s},V)$ and  $PLC(K_\mathrm{s},V)$ relations are
very similar and the two relations are coincident. Similarly, for 
$PW(J,V)$ and $PLC(J,V)$ the colour coefficients are the same within the
errors, whereas this is not true for the couple $PW(K_\mathrm{s},J)$;
$PLC(K_\mathrm{s},J)$. 

\begin{table*}
\small 
\caption{$PL(K_\mathrm{s})$ and $PW(K_\mathrm{s})$ relations for 
 LMC T2CEPs with the $ZP$ calibrated as follows: (2) by imposing a
  $DM_{LMC,0}$=18.46$\pm$0.03 mag \citep[from CCs in the
  LMC][]{Ripepi12b} in Table~\ref{NIR};  (3) by adopting Galactic
  T2CEP ($\kappa$ Pav) and RR Lyrae variables with $HST$ parallaxes
  \citep{Benedict2011} and T2CEPs with BW distance estimates
  \citep{Feast2008}; (4) by adopting only calibrators with the quoted
  $HST$ parallaxes. See text for  additional details.}
\label{ZP}
\begin{center}
\begin{tabular}{cccc}
\hline 
\noalign{\smallskip}
Relation& $ZP_{CC}$ & $ZP_{\pi+BW}$ & $ZP_{\pi}$ \\
\noalign{\smallskip}
(1) & (2) & (3) & (4) \\ 
\hline 
\noalign{\smallskip}
$K_\mathrm{s,0}=(-2.385\pm0.03)\log P+ZP$ & $-0.99\pm0.04$ &$-1.09\pm0.10$  &  $-1.11\pm0.10$ \\
\noalign{\smallskip}
$K_\mathrm{s}-0.13(V-K_\mathrm{s})=(-2.49\pm0.03)\log P 
+ZP$ & $-1.13\pm0.04$ &  $-1.24 \pm0.10$ & $-1.26\pm0.10$ \\
\noalign{\smallskip}
\hline 
\noalign{\smallskip}
\end{tabular}
\end{center}
\end{table*}

\section{Absolute calibration of $PL$, $PLC$ and $PW$ relations}
\label{section4}

\begin{figure}
\includegraphics[width=8.5cm]{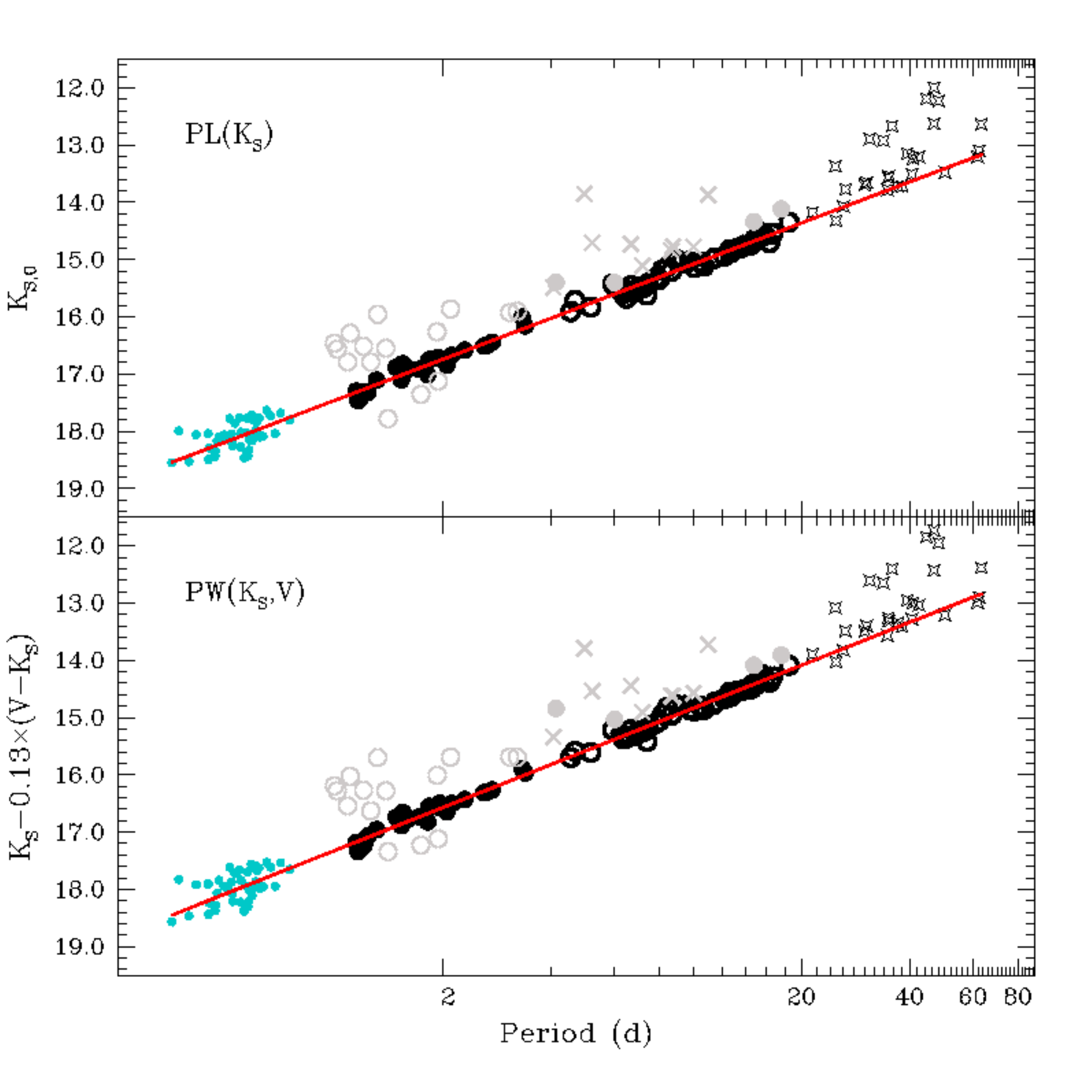}
\caption{$PL(K_\mathrm{s})$ and $PW(K_\mathrm{s},V)$ relations for the T2CEPs 
  analysed in this paper (symbols as in Fig.~\ref{pl}) and for the 
  sample of RR Lyrae stars in the LMC observed by 
  \citet{Borissova2009} (light blue filled circles). The red lines show 
  the relationships listed in Table~\ref{NIR} extended till the 
  periods of the RR Lyrae stars.} 
\label{borissova}
\end{figure}

\begin{figure*}
\includegraphics[width=12cm]{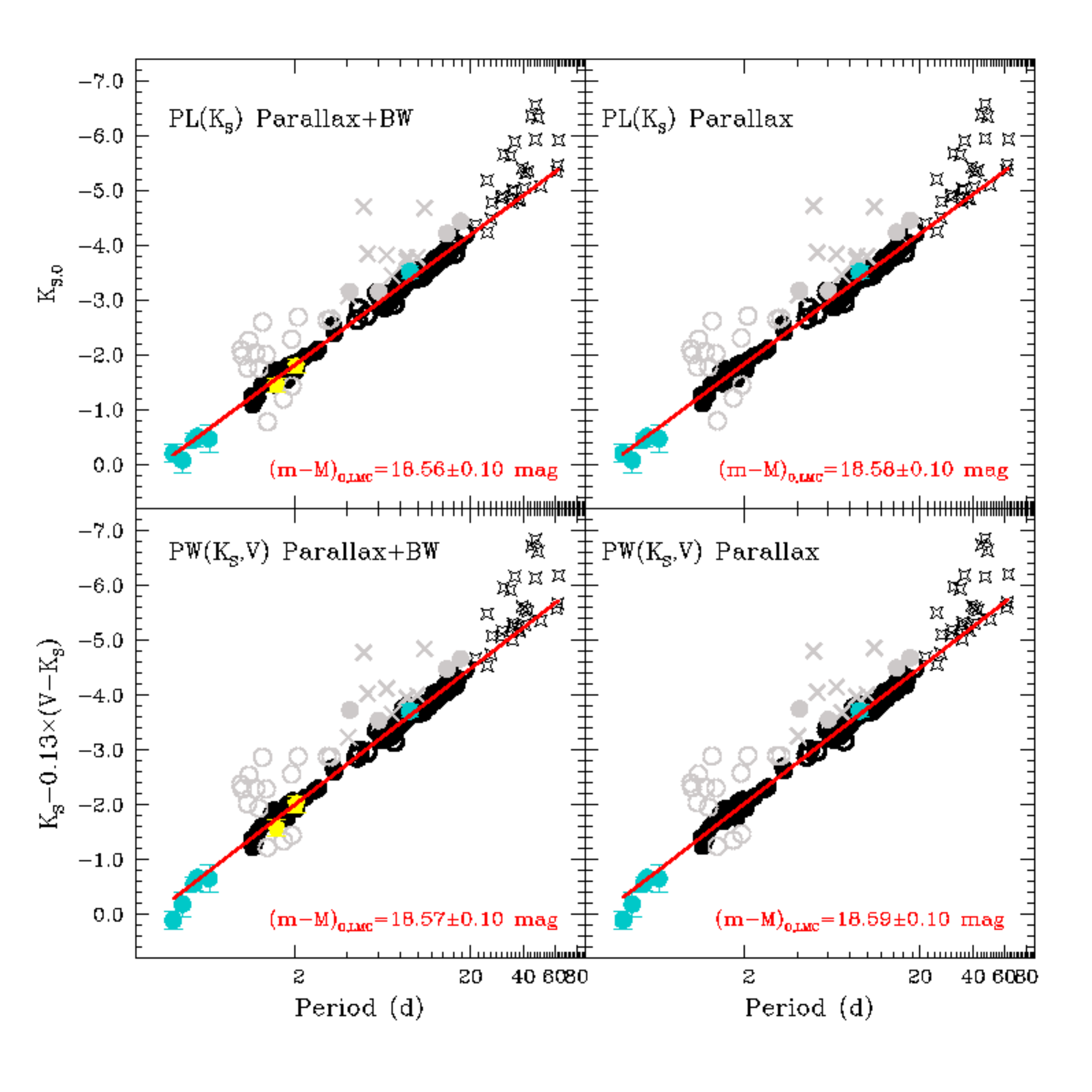}
\caption{Absolute $PL(K_\mathrm{s})$ and $PW(K_\mathrm{s},V)$
  relations for the T2CEPs 
  analysed in this paper (symbols as in Fig.~\ref{pl}). Light blue and 
  yellow filled circles show the objects whose distances were measured 
  through $HST$ parallaxes \citep{Benedict2011} or through 
  BW analysis \citep[][]{Feast2008}, respectively. 
 The red line shows the best-fit line to the data adopting the slope 
 from Table~\ref{NIR}, while $ZPs$ were calculated using the 
 objects with $HST$ parallaxes alone (right panels), and by adding to them the objects
 with BW analysis (left panels). 
The true $DMs$ estimated in each case for the LMC are also labelled
(see Sect.~\ref{discussion}).} 
\label{calibPL}
\end{figure*}

In Table~\ref{NIR}  we provided the absolute $ZP$ for the 
relevant $PL$, $PLC$ and $PW$ relations as a
function of the $DM_{0,LMC}$. 
However, it is of considerable astrophysical 
interest to obtain an independent absolute calibration for at least 
some of these relations. Indeed, this would allow us to 
obtain an independent measure of the distance to the LMC and to the GGCs 
hosting T2CEP variables.  To this aim, we can only rely on calibrators 
located close enough to the Sun to have a measurable 
parallax or whose distances have been estimated by Baade-Wesselink (BW) 
techniques \citep[see][for a review on this method]{Gautschy1987}. 
There are only two T2CEPs whose parallaxes were measured with 
reasonable accuracy with the $Hubble ~ Space ~ Telescope$
\citep[$HST$; ][]{Benedict2011}, namely $\kappa$ Pav (W Vir) and VY Pyx 
(BL Her). For two additional BL Her variables, SW Tau and V533 Cen, as 
well as for $\kappa$ Pav, a BW-based distance is 
also available \citep{Feast2008}.  However, VY Pyx turned out to be a 
peculiar star, unusable as calibrator \citep[see discussion 
in][]{Benedict2011}. 
As for $\kappa$ Pav, the pulsational parallax estimated by 
\citet{Feast2008} through BW analysis is about 2$\sigma$ smaller 
than the trigonometric parallax 
measured by $HST$ and adopted here ($\Delta \pi = 0.67 \pm 0.33$ mas). 
\citet{Feast2008} investigated the possible causes of the
discrepancy with respect to the Hipparcos parallax
\citep{vanLeeuwen2007}, which was even larger than the $HST$ one, 
but did not find any definitive explanation. A well known potential 
problem related with the application of the BW technique is 
the uncertainty on the projection factor {\it p} \citep[see, e.g.][and 
references therein]{Molinaro2012,Nardetto2014}. In their analysis 
\citet{Feast2008} derived and adopted a fixed {\it p}-factor =
1.23$\pm$ 0.03. However, several researchers suggested that the {\it p}-factor 
actually does depend on the period of the pulsator \citep[see e.g.][and
references therein]{Barnes2009,Laney2009,Storm2011a,Nardetto2014},
hence, for example, different {\it p}-factor values should be used for BL Her
and W Wir stars. 
Given the uncertainties on the projection factor discussed above, in the
following we will adopt the $HST$-based distance for $\kappa$ Pav,
and  the zero point of the different $PL, PW$ and $PLC$ relations will
be estimated including or not the BW-based distances for SW Tau and V533 Cen.
Finally, we note that $[Fe/H]$($\kappa$ Pav)$\approx$+0.0 dex \citep{Feast2008},
  i.e. at least 1 dex more metal rich than expected for typical T2CEPs.
Hence, some additional uncertainty when using this object as a distance indicator 
  can be caused by a possible metallicity effect. However, as
  discussed in Sect.~\ref{section3}, the metal
  dependence of the T2CEP $PLs$, if any, should be very small, and we do not
  expect the high metallicity of  $\kappa$ Pav to be an issue for our purposes.
To enlarge the number of reliable calibrators, a 
possibility is to use the 5 RR Lyrae stars whose parallax were measured 
with $HST$ by \citet{Benedict2011}. Indeed, as already hypothesised by 
\citet{Sollima2006} and  \citet{Feast2008}, RR Lyrae and T2CEPs follow 
the same $PL(K_\mathrm{s})$ relation \citep[][found similar results in 
the optical bands]{Caputo2004}. To further test this 
possibility, we draw in Fig.~\ref{borissova} the $PL(K_\mathrm{s})$ 
and $PW(K_\mathrm{s})$ relations for the T2CEPs analysed in this
paper, in comparison with the location occupied in the same planes 
by the RR Lyrae stars in the LMC \citep[light blue filled circles,
after ][]{Borissova2009}. 
The periods of c-type RR Lyrae stars were 
fundamentalised by adding $\delta$logP=0.127 \citep{articoloBlu} and the magnitudes 
have been corrected for the metallicity term devised by 
\citet{Sollima2006}, using the individual metallicity measurement compiled by 
\citet{Borissova2009}. It can be seen that  
both the $PL(K_\mathrm{s})$ and $PW(K_\mathrm{s})$ relations (red 
lines) derived for T2CEPs in Sect.~\ref{pl} tightly match the 
location of the RR Lyrae stars. On this basis, we decided to proceed 
using also the RR Lyrae with $HST$ parallax to anchor the $ZP$ of the 
$PL(K_\mathrm{s})$ and $PW(K_\mathrm{s},V)$ relations for T2CEPs. 
To this aim, we simply adopted the slopes of the quoted relations from 
Table~\ref{NIR}, corrected for metallicity the $ZP$ for the five RR Lyrae stars 
with $HST$ parallaxes and calculated the weighted average of the 
results in two cases: i) including only stars with $HST$parallax,
namely, $\kappa$ Pav and the five RR Lyrae stars; ii) using the stars at point i)
plus  the two T2CEPs with BW analysis, namely SW Tau and V533 
Cen\footnote{The uncertainties on the $DM$ of these two objects
  were obtained by summing the uncertainties reported in table 4 of
  \citet{Feast2008}.}. 
The results of these procedures are outlined in 
Table~\ref{ZP} (columns 3 and 4) and in Fig.~\ref{calibPL}. For 
comparison, column (2) of Table~\ref{ZP} shows the   
$ZPs$ obtained assuming $DM_{0,LMC} = 18.46\pm0.03$ mag, as derived 
by \citet{Ripepi12b} from LMC CC stars. We choose the work by
\citet{Ripepi12b}  as reference for CCs because: i) these Authors adopted a 
procedure similar to the one adopted in this work; ii) their results
are in excellent agreement with the most recent and accurate
literature findings \citep[see e.g.][and references therein]{Storm2011b,Joner2012,Laney2012,w12,Piet2013,degrijs2014}
An analysis of Table~\ref{ZP} reveals that: i) the inclusion of the two 
stars with BW-based distances does not change significantly the $ZPs$
and ii) there is a difference of at least $\sim$0.1 mag between the $ZPs$
calibrated on the basis of CCs and of Galactic T2CEPs (see Sect.~\ref{discussion}).

\subsection{Comparison with the literature}

The relationships presented in Tables~\ref{NIR} and ~\ref{ZP} can now be compared
to those available in the literature. As mentioned in the introduction,
\citet{Matsunaga2006} and \citet{Matsunaga2009} published the $PL$  
relations in the $JHK_\mathrm{s}$ bands for BL Her and W Vir variables 
hosted by GGCs and the LMC, respectively.  These results can be compared
with ours, provided that we first transform all the $J$ and $K_\mathrm{s}$
magnitudes into the VISTA system. With this aim, we transformed the 
\citet{Matsunaga2006} photometry from 2MASS to VISTA using the equations 
reported in Sect.~\ref{t2ceps}. The results of \citet{Matsunaga2009}
are in the IRSF system, whose $J$ and $K_\mathrm{s}$ can in principle be
transformed to the 
2MASS system \citep[][]{Kato2007}, and in turn, into the VISTA
system. However, this is not possible for the $J$ band, because we
lack $H$-band photometry \citep[see Table 10 in][]{Kato2007}. We can
safely overcome this problem by noting that the $(J-H)$ colour for BL
Her and W Vir stars spans a very narrow range \citep[0.25$<(J-H)<$0.4
mag, see e.g.][]{Matsunaga2011} so that, according to \citet{Kato2007} we can assume
$J$(IRSF)=$J$(2MASS)+(0.005$\pm$0.005). Finally, since our targets
span the range 0.25$<(J-K_\mathrm{s})<$0.6 mag, we obtained: 
$J$(IRSF)=$J$(VISTA)+(0.035$\pm$0.015). As for the $K_\mathrm{s}$, 
the transformation is straightforward:
$K_\mathrm{s}$(IRSF)=$K_\mathrm{s}$(VISTA)+(0.014$\pm$0.001). 

The $PL$ relations by \citet{Matsunaga2006} and \citet{Matsunaga2009}, corrected as discussed
above, are presented in the first four rows of
Table~\ref{comparison}. We can compare directly the $PL(J)$ and
$PL(K_\mathrm{s})$ relations for the LMC 
(lines 2 and 4 in Table~\ref{comparison}) with our
results (lines 1 and 2 in Table~\ref{NIR}). There is a very good
agreement within 1 $\sigma$ errors for the $PL(J)$,
whereas for the $PL(K_\mathrm{s})$ the comparison is slightly worse, especially
concerning the slope of the relation which is discrepant at the 
1.5$\sigma$ level. It is also worth mentioning that the dispersion of our
relations is significantly smaller, as a result of the much better light curve 
sampling of the VMC data.

\begin{table*}
\small 
\caption{Values for the coefficients of the $PL$,
  $PW$ and $PLC$ relations for BL Her and W Vir  
Cepheids taken from the literature.  The $PW$ functions are defined as in Table~\ref{NIR}. 
The errors of $ZP$ take into account the uncertainties in the
transformation of the $J$ and $K_\mathrm{s}$ photometry to the VISTA system (see text for details).}
\label{comparison}
\begin{center}
\begin{tabular}{ccc}
\hline 
\hline 
\noalign{\smallskip}
 method &  Relation & $\sigma$ (mag)  \\
\noalign{\smallskip}
\hline 
\noalign{\smallskip}
\multicolumn{3}{c}{Results by \citet{Matsunaga2006} and \citet{Matsunaga2009}
  transformed to the VISTA system} \\
\noalign{\smallskip}
\hline 
\noalign{\smallskip} 
$PL(J)$  GCs & $M_{J,0}=(-2.23\pm0.05)\log P-(0.84\pm0.03)$ & 0.16 \\ 
\noalign{\smallskip}  
$PL(J)$  LMC & $J_0=(-2.16\pm0.04)\log P+(17.76\pm0.03)$ & 0.21 \\
\noalign{\smallskip} 
$PL(K_\mathrm{s})$  GCs & $M_{K_\mathrm{s,0}}=(-2.41\pm0.05)\log P-(1.11\pm0.03)$&0.14 \\ 
\noalign{\smallskip} 
$PL(K_\mathrm{s})$  LMC & $K_\mathrm{s,0}=(-2.28\pm0.05)\log P+(17.40\pm0.03)$&0.21 \\
\noalign{\smallskip}
\hline 
\noalign{\smallskip}
\multicolumn{3}{c}{Results by \citet{Dicriscienzo2007} transformed to
  the VISTA system} \\
\noalign{\smallskip}
\hline 
\noalign{\smallskip}
$PL(J)$ & $M_{J,0}=(-2.29\pm0.04)\log P-(0.73\pm0.13)$ &  \\
\noalign{\smallskip}  
$PL(K_\mathrm{s})$ & $M_{K_\mathrm{s,0}}=(-2.38\pm0.02)\log P-(1.10\pm0.07)$ & \\
\noalign{\smallskip} 
$PW(J,V)$ & $M_{J}-0.41(V-J)=(-2.37\pm0.02)\log P-(1.15\pm0.08)$& \\
\noalign{\smallskip} 
$PW(K_\mathrm{s},V)$ & $M_{K_\mathrm{s}}-0.13(V-K_\mathrm{s})=(-2.52\pm0.02)\log P-(1.25\pm0.08)$&\\
\noalign{\smallskip} 
$PW(K_\mathrm{s},J)$ &$K_\mathrm{s}-0.69(J-K_\mathrm{s})= (-2.60\pm0.02)\log P-(1.27\pm0.08)$&\\
\noalign{\smallskip} 
\hline 
\end{tabular}
\end{center}
\end{table*}

As for the $PL(J)$ and $PL(K_\mathrm{s})$ derived for GGCs by
\citet{Matsunaga2006}, their slopes are in very good
agreement with ours, which suggest  a ``universal slope''
in the NIR filters, independent of the galactic environment. 
As for the $ZPs$, we can
only compare them for the $PL(K_\mathrm{s})$ relations (see
Table~\ref{ZP}).
We found an excellent agreement when the $ZP$ is
calibrated through the Galactic calibrators (irrespectively of whether
stars with BW measures are included or not), whereas there is a 0.12 mag
discrepancy if the $ZP$ is calibrated by means of the LMC $DM$ coming
from CCs. This occurrence is not surprising, since \citet{Matsunaga2006}
used the $M_V$ vs $[Fe/H]$ relation for RR Lyrae variables by
\citet{Gratton2003} to estimate the distances of the GGCs hosting
T2CEPs and derive their $PL(K_\mathrm{s})$. Hence, the two population II
calibrators, RR Lyrae and T2CEPs, give distance scales in agreement with
each other.  

A similar comparison can be performed with the theoretical predictions
by \citet{Dicriscienzo2007}, who in addition calculated the $PWs$
for all the combinations of magnitudes and colours of interest in this
work. Again, we converted the \citet{Dicriscienzo2007}
results from the \citet{bessellbrett1988} (BB) to the VISTA system.  
To do this, we used the transformations 
 BB-2MASS from \citet{carpenter} and 2MASS-VISTA (see Section 2.1) and
 the same procedure as above to derive:
$J$(BB)=$J$(VISTA)+(0.04$\pm$0.010); 
$K_\mathrm{s}$(BB)=$K_\mathrm{s}$(VISTA)+(0.030$\pm$0.015). Secondly,
since the predicted $PL$ and $PW$ relations mildly depend on metallicity and adopted
a mixing length parameter ($\alpha$\footnote{$\alpha = l/H_p$ is the ratio between the mean free
path of a convective element ($l$) and the pressure scale height
($H_p$). Varying this parameter strongly affects the properties of a star's
outer envelope such as its radius and effective temperature.}), we have to make a choice for these
parameters. We decided to evaluate the relations for
$\alpha$=1.5$\pm$0.5 (to encompass reasonable values for
$\alpha$) and [Fe/H]=$-1.5\pm$0.3 dex as an average value for the LMC
old population \citep[see, e.g.][]{Borissova2004,Borissova2006,Gratton2004,Haschke2012}. 
The uncertainties on these parameters were taken into account in
re-deriving the $ZP$ of the predicted $PL$ and $PW$ relations in
the VISTA system. The result of this  procedure is shown in the second part of
Table~\ref{comparison}. A comparison with Table~\ref{ZP} shows that
both for the $PL(K_\mathrm{s})$ and $PW(K_\mathrm{s},V)$ relations
there is an excellent agreement between ours and theoretical results if the quoted
relationships are calibrated with the Galactic T2CEPs and RR lyrae,
whereas there is a $\sim$0.1 mag discrepancy if we adopt the CC-based $DM$ by \citet{Ripepi12b} 
for the LMC to define the $ZP$. However, if we take into account the
uncertainties this discrepancy results formally not significant within 1$\sigma$.

\section{Discussion}
\label{discussion}

The results reported in Sect.~\ref{section4} allow us to discuss the
distance of the LMC as estimated from NIR observation of the T2CEPs
hosted in this galaxy. Table~\ref{dm} (columns 3 and 4) lists the $DM_{0,LMC}$ calculated
using the different $ZP$ estimates for the $PL(K_\mathrm{s})$ and $PW(K_\mathrm{s},V)$
relations listed in Table~\ref{ZP}.
An inspection of the table reveals that the $DM_{0,LMC}$ calculated 
by means of CCs (column 2 in Table~\ref{dm}) and by means of the
T2CEPs differ by more than $\sim$0.1
mag, even if, formally there is agreement within 1 $\sigma$. Since
both the \citet{Ripepi12b} calibration for CCs and that presented here
for T2CEPs are based on a weighted mix of $HST$ parallaxes and BW
analysis, this discrepancy, albeit only partially significant, seems
to suggest that the distance scales calibrated on pulsating stars
belonging to population I and population II give different results 
\citep[for a recent comprehensive review of the literature and a 
discussion about this argument, see][]{degrijs2014}. 

\begin{figure}
\includegraphics[width=8.5cm]{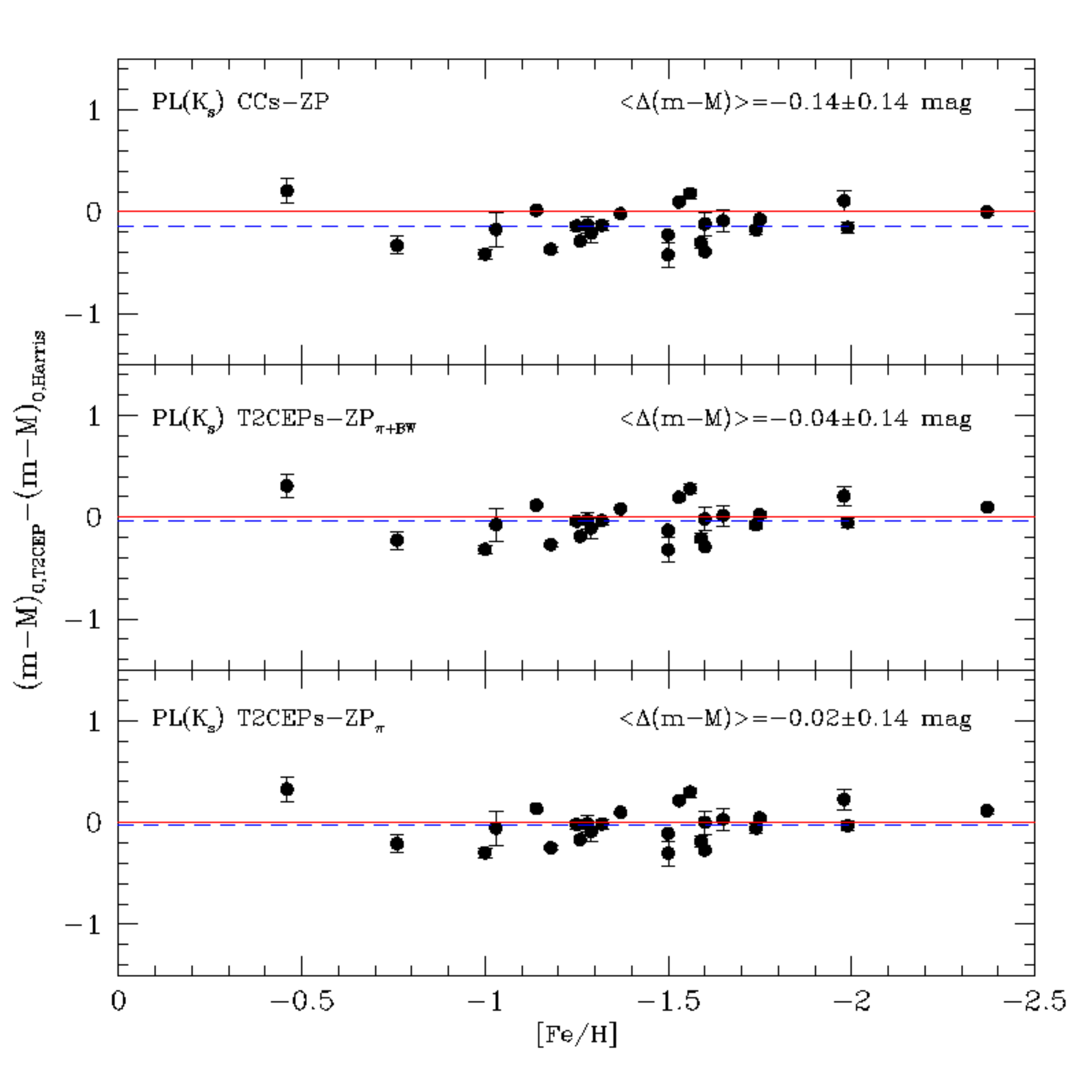}
\caption{Distance modulus differences (this work-\citealt{Harris1996}) for
  a sample of GGCs hosting T2CEPs as a function of [Fe/H]. The dashed
blue line shows the average difference. The solid
  red line show the line with zero difference. The $DMs$
  for the GGCs were estimated adopting the $PL(K_\mathrm{s})$ for T2CEPs
  and $ZP$ determined as follows: (top panel) on the basis of the
  $DM_{0,LMC}$ measured by \citet{Ripepi12b} using LMC CC with VMC NIR
  data; (middle panel) by means of a sample of  Galactic T2CEPs whose distances were
  measured both through $HST$ parallaxes \citep{Benedict2011} and BW
  technique \citep{Feast2008}; (bottom panel) as in the previous case,
  but using objects with $HST$ parallaxes only.}
\label{ggc}
\end{figure}

An additional application of the absolute $PL(K_\mathrm{s})$ relation
for T2CEPs  concerns the distance estimate of GGCs hosting such kind
of pulsators. 
Homogeneous $K_\mathrm{s}$ photometry, as well as period of pulsation 
for most of the known T2CEPs in GGCs  were published by
\citet{Matsunaga2006} (see their Table 2). 
We simply inserted the period of these variables in the  $PL(K_\mathrm{s})$ of
Table~\ref{ZP}, and by difference with the observed magnitudes,
we derived the $DM$ for each GGC. When more than one T2CEP was present
in a cluster, we averaged the resulting $DMs$ (we
excluded from the calculations the variables with periods longer than about 35
d because they are likely neither BL Her nor W Vir variables). 
The result of such a procedure is shown in Fig.~\ref{ggc} where for
each GGC analysed here, we show (as a function of the metal content of
the clusters) the difference between the $DMs$ estimated on the basis of the
three different calibration of the $PL(K_\mathrm{s})$ listed in
Table~\ref{ZP} and the $DMs$ reported by \citet{Harris1996} in
his catalogue of GGCs parameters. In Fig.~\ref{ggc} the average 
discrepancy  in $DMs$ decreases from top to bottom, suggesting that,
even if the statistical significance is low (due to the large
dispersion in $\Delta DM$ values $\sim$ 0.14 mag), 
the distance scale of GGCs, if estimated on the basis of the T2CEPs
hosted in this systems, is more consistent with population II rather
than population I standard candles. This is not particularly surprising since most
of the distances of GGCs in the Harris catalogue are based on RR Lyrae stars.

\begin{table}
\small 
\caption{$DM$ of the LMC estimated on the basis of the different $PL(K_\mathrm{s})$ and
  $PW(K_\mathrm{s})$ relations described in Table~\ref{ZP} (see text).}
\label{dm}
\begin{center}
\begin{tabular}{cccc}
\hline 
\noalign{\smallskip}
Relation  & $DM^{LMC}_{CC}$ & $DM^{LMC}_{\pi+BW}$ & $DM^{LMC}_{\pi}$ \\
\noalign{\smallskip}
(1) & (2) & (3) & (4) \\ 
\hline 
\noalign{\smallskip}
$PL(K_\mathrm{s})$ & $18.46\pm0.04$ & $18.56\pm0.10$& $18.58\pm0.10$\\
\noalign{\smallskip}
$PW(K_\mathrm{s},V)$ & $18.46\pm0.04$ & $18.57\pm0.10$&  $18.59\pm0.10$\\
\noalign{\smallskip}
\hline 
\noalign{\smallskip}
\end{tabular}
\end{center}
\end{table}

\section{Summary}

In the context of the VMC survey, this paper shows the first results
concerning type II Cepheids in the LMC. We presented $J$ and $K_\mathrm{s}$
light curves for 130 pulsators, including 41 BL Her, 62 W Vir (12 pW Vir) and 27 RV Tau variables,
corresponding to 63\%, 63\% (75\%) and 61\% of the known LMC populations
of the three variable classes, respectively. 
The $K_\mathrm{s}$ band light curves are almost always well sampled,
allowing us to obtain accurate spline fits to the data and, in turn,
precise intensity-averaged $\langle K_\mathrm{s} \rangle$ magnitudes for 120 
variables in our sample. As for the $J$ band, only about 1/3 of the $J$ light curves
were sufficiently sampled to allow a satisfactory spline fit to the
data, for the remaining 2/3 of pulsators, the intensity-averaged
$\langle J \rangle$
magnitudes were derived using the $K_\mathrm{s}$ band spline fits
as templates. 
On the basis of this data set for BL Her and W Vir, complemented by the
$\langle V \rangle$ magnitudes
from the OGLE survey, we have built for the first time (apart from
$PL(J)$ and $PL(K_\mathrm{s})$)  a variety of empirical $PL$, $PLC$ and $PW$
relationships, for any combination of the $V, J, K_\mathrm{s}$
filters. 
Several outliers were removed from the calculation of these
relations, and we provided  an explanation for the presence of
these divergent objects. All the quoted $PL$, $PLC$ and $PW$ relationships
were calibrated  in terms of the LMC distance. However, the
availability of absolute $M_V$ and $M_{K_\mathrm{s}}$ for a small sample 
of RR Lyrae and T2CEPs variables based on $HST$ parallaxes allowed us
to obtain an independent absolute calibration of the
$PL(K_\mathrm{s})$ and $PW(K_\mathrm{s},V)$ relationships (the
$PLC(K_\mathrm{s},V)$ is identical to the $PW(K_\mathrm{s},V)$). If
applied to the LMC and to the GGCs hosting T2CEPs, these relations
give distance moduli which are around 0.1 mag longer than those
estimated for Classical Cepheids by means of $HST$ parallaxes 
and BW techniques. However, if we take into account the uncertainties
at their face value, the quoted discrepancy is formally not significant within 1$\sigma$.

\section*{Acknowledgments}

We wish to thank our Referee, Dr. C.D. Laney for his helpful 
and competent review of the manuscript.
V.R. warmly thanks Roberto Molinaro for providing the
program for the spline interpolation of the light curves.

Partial financial support for this work was provided by PRIN-INAF 2011
(P.I. Marcella Marconi) and PRIN MIUR 2011 (P.I. F. Matteucci).
We thank the UK's VISTA Data Flow System comprising the VISTA pipeline
at the Cambridge Astronomy Survey Unit (CASU) and the VISTA Science
Archive at Wide Field Astronomy Unit (Edinburgh) (WFAU) 
for providing calibrated data products supported by the STFC.
This work was partially supported by the Gaia Research for European
Astronomy Training 
(GREAT-ITN) Marie Curie network, funded through the European Union 
Seventh Framework Programme ([FP7/2007-	1312 2013] under 
grant agreement n. 264895).
RdG acknowledges research support from the National
Natural Science Foundation of China (NSFC) through grant 11373010.
This work was partially supported by the Argentinian institutions CONICET and Agencia 
Nacional de Promoci\'on Cient\'{\i}fica y Tecnol\'ogica (ANPCyT).

\newpage

\appendix

\section[]{Light Curves}

\begin{figure*}
\includegraphics[width=16cm]{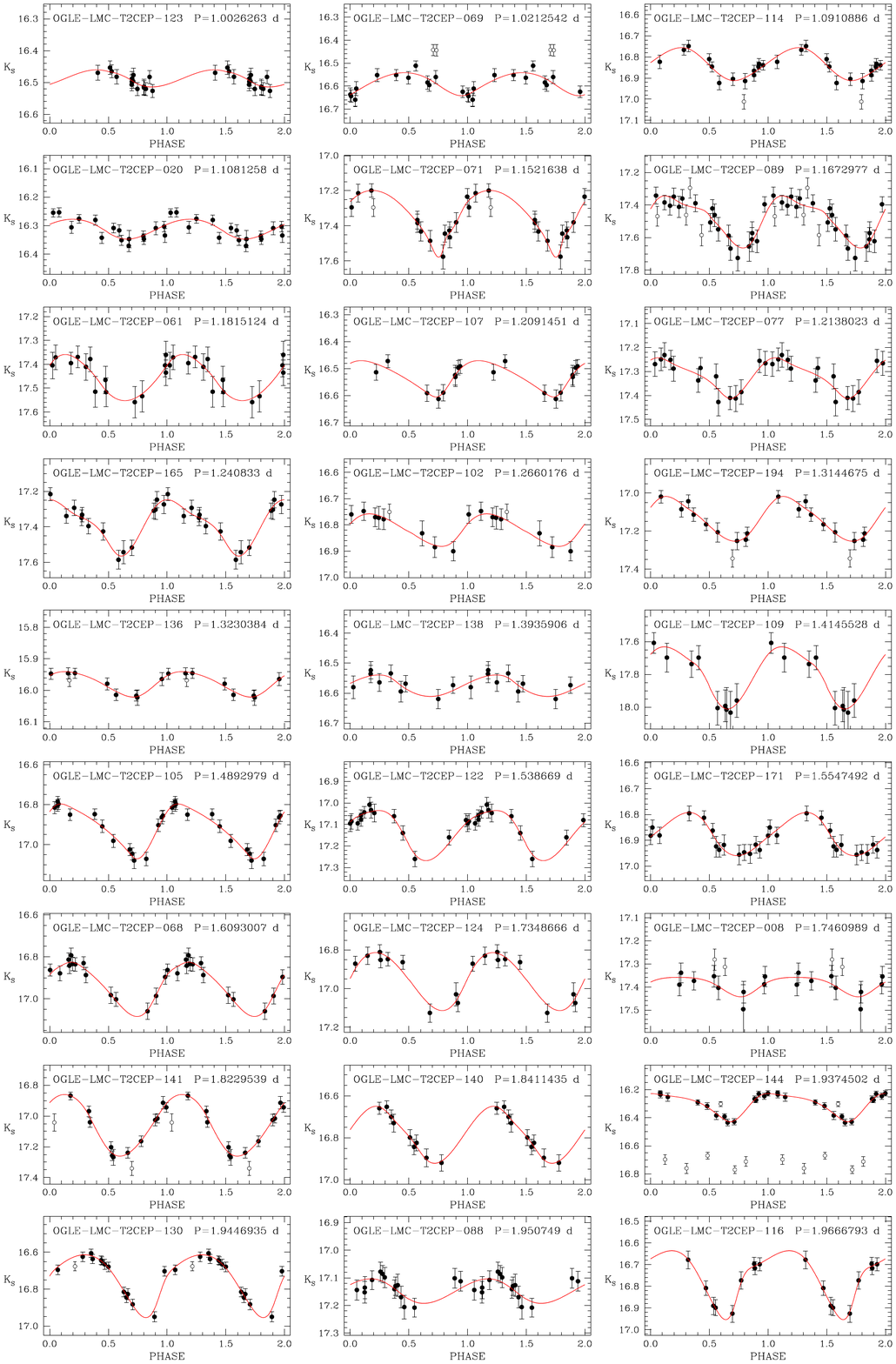}
\caption{$K_\mathrm{s}$-band light curves for T2CEPs with 
  usable data discussed in this paper. Stars are displayed in order of increasing 
period. Filled and open circles represent phase points used or not
used in the fitting procedure, respectively. Solid lines represent best-fitting splines to the data (see 
text). In each panel we report OGLE's identification number  and 
period.} 
\label{figureK}
\end{figure*}

\begin{figure*}
\includegraphics[width=16cm]{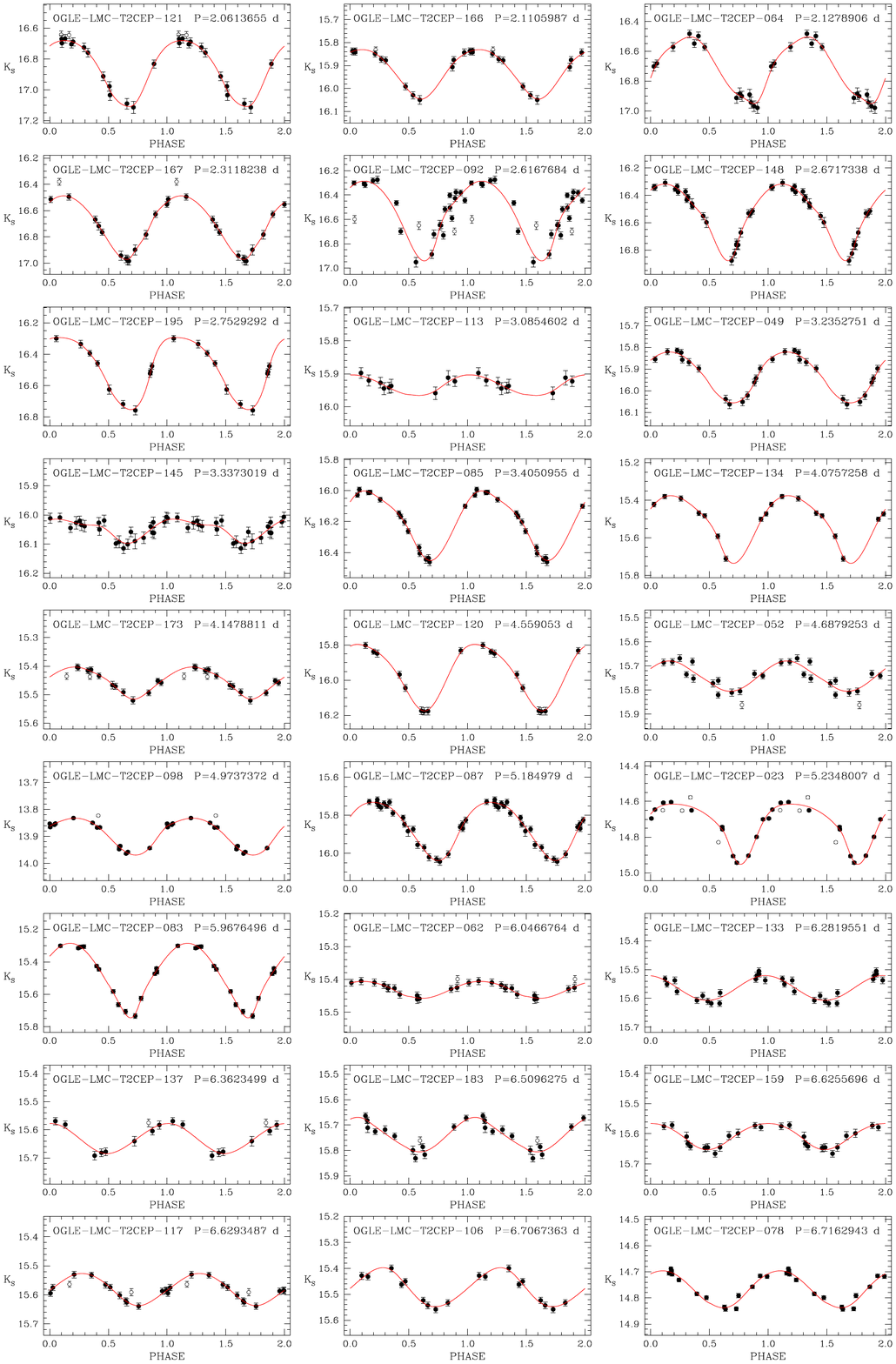}
\contcaption{}
\end{figure*}

\begin{figure*}
\includegraphics[width=16cm]{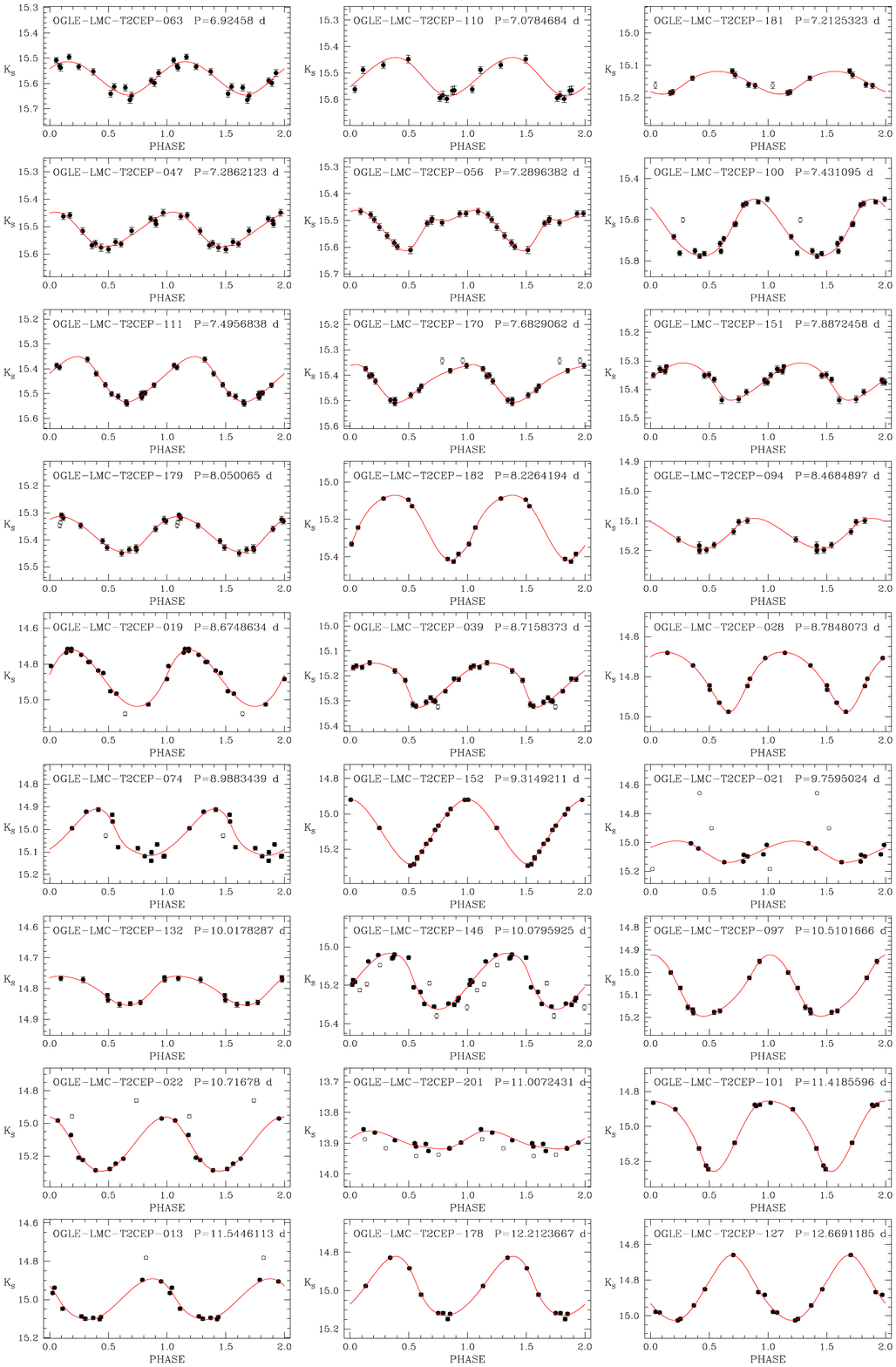}
\contcaption{}
\end{figure*}

\begin{figure*}
\includegraphics[width=16cm]{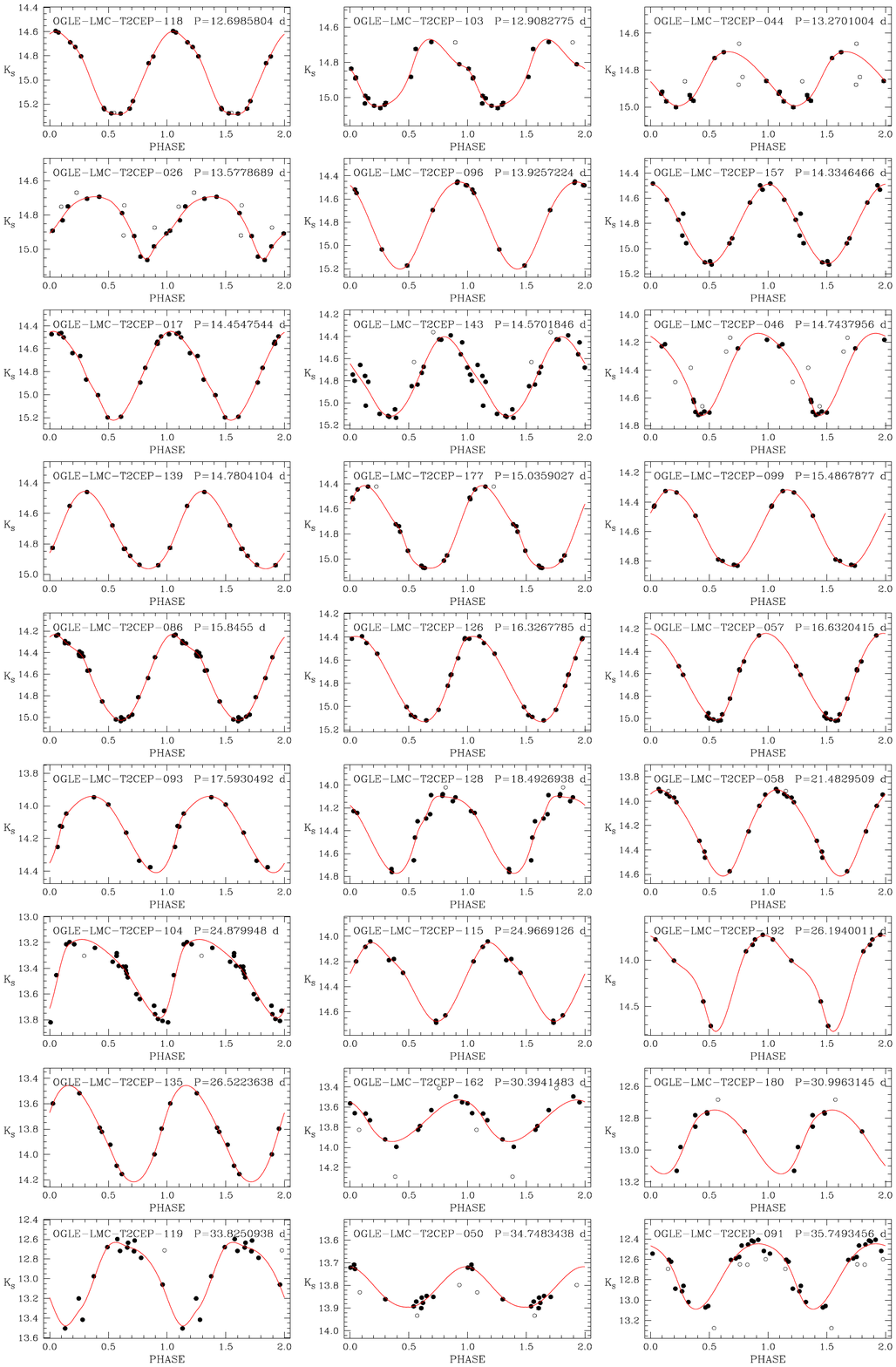}
\contcaption{}
\end{figure*}

\begin{figure*}
\includegraphics[width=16cm]{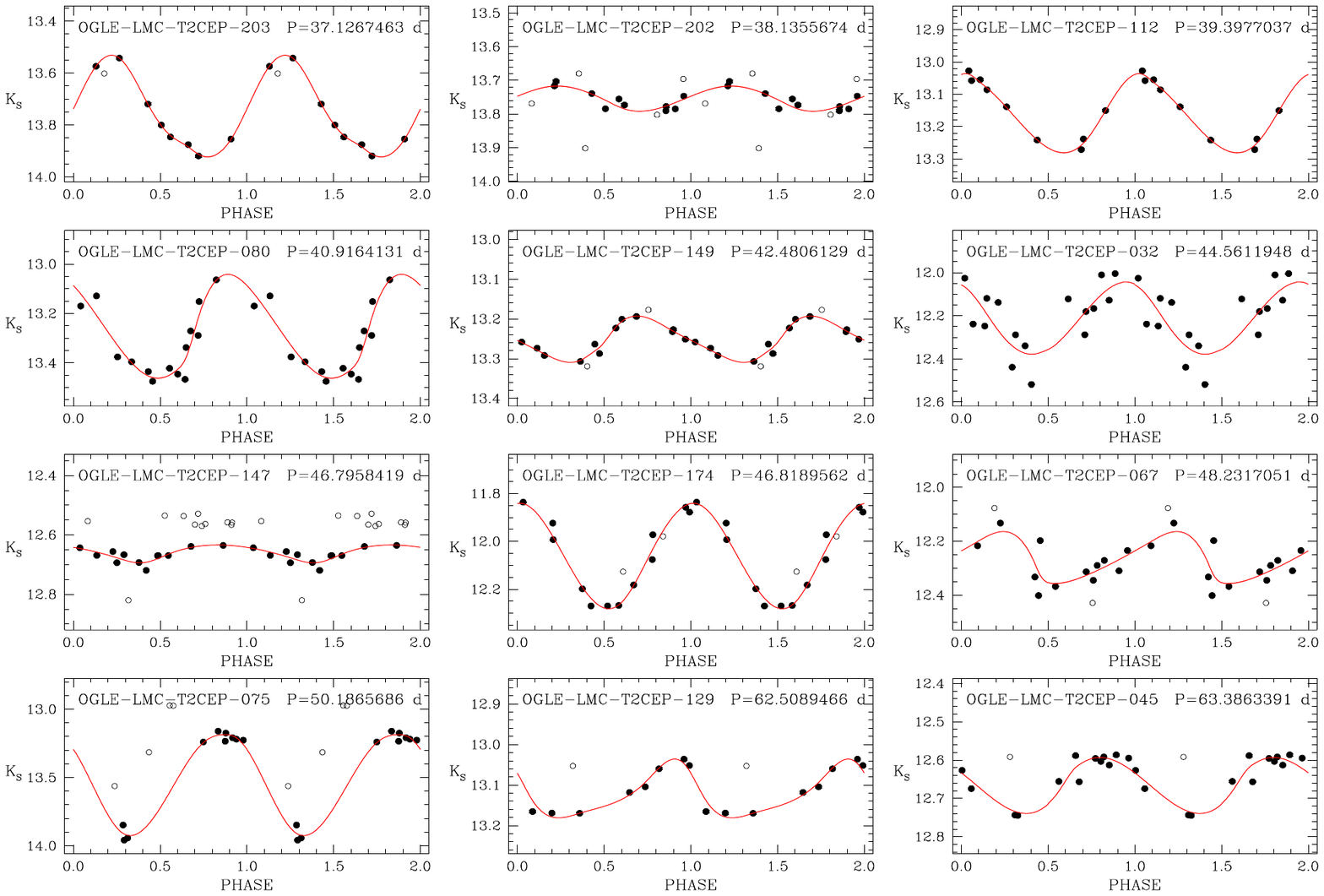}
\contcaption{}
\end{figure*}

\begin{figure*}
\includegraphics[width=16cm]{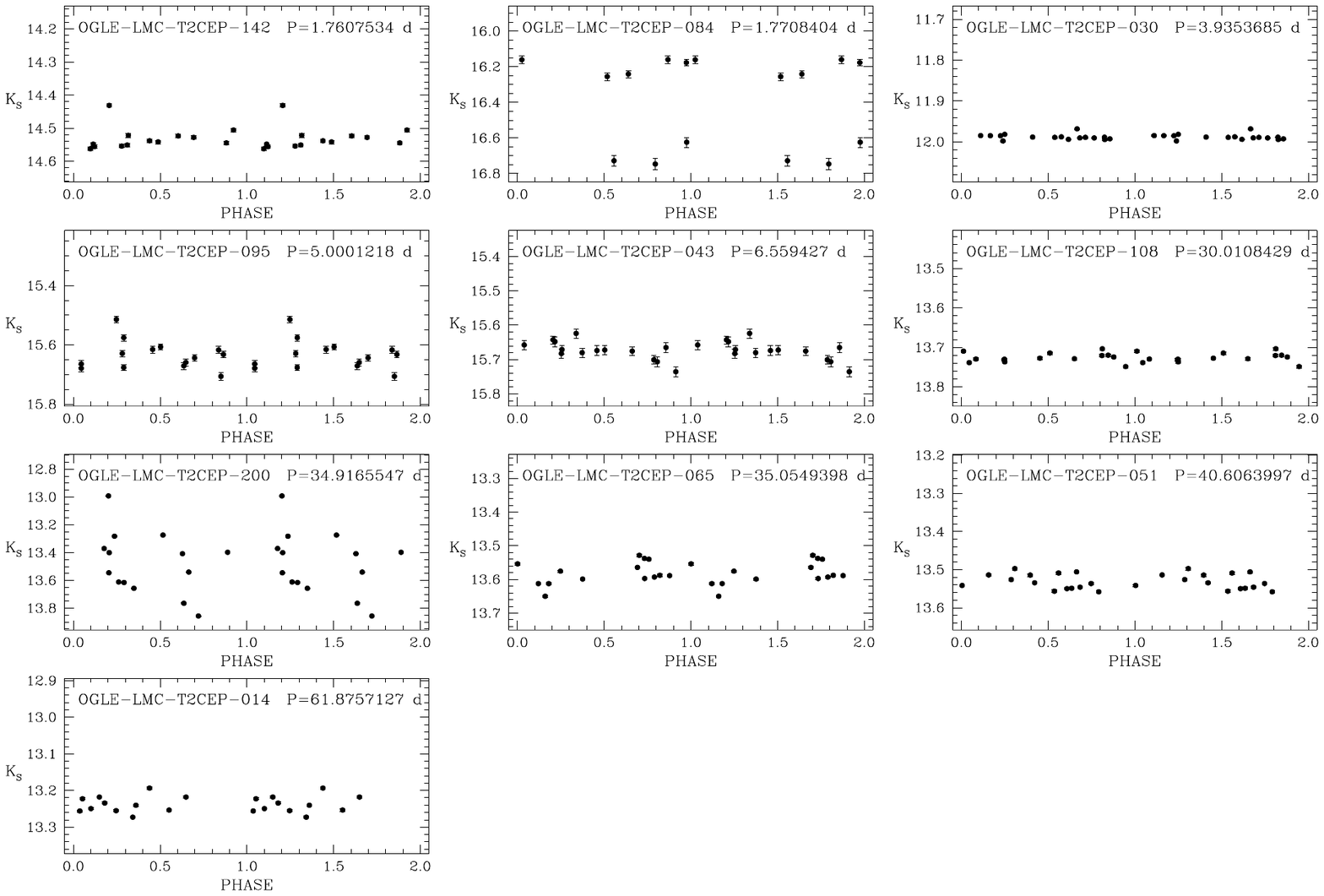}
\caption{$K_\mathrm{s}$--band light curves for problematic stars (see text).} 
\label{figureK_sfigate}
\end{figure*}

\begin{figure*}
\includegraphics[width=16cm]{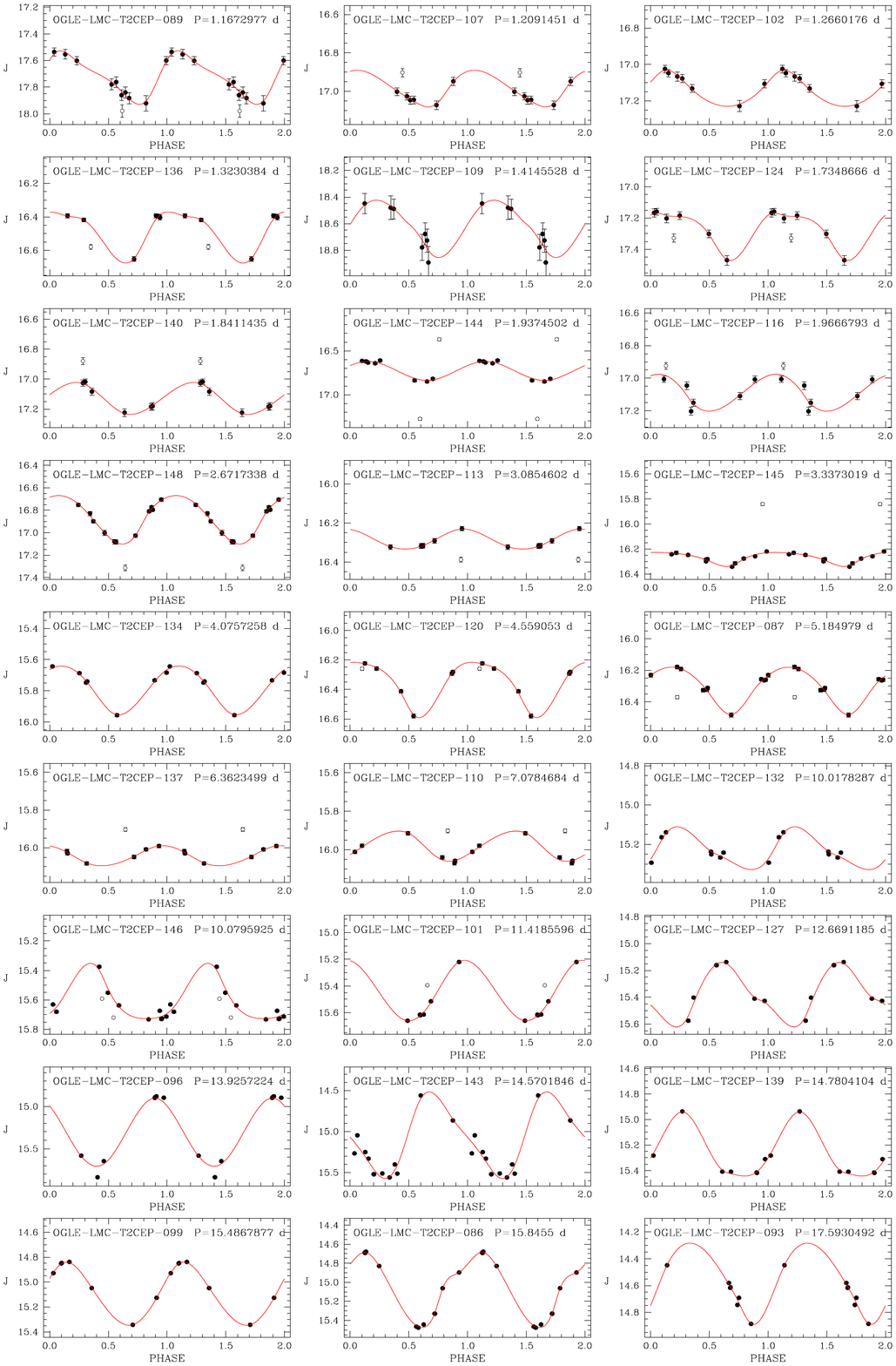}
\caption{$J$-band light curves for T2CEP stars with a sufficient 
  number of epochs to perform the spline fit to the data. Stars are displayed in order of increasing 
period. Solid lines represent spline best-fits to the data (see 
text). In each panel we report OGLE's identification number  and 
period.} 
\label{figureJspline}
\end{figure*}

\begin{figure*}
\includegraphics[width=16cm]{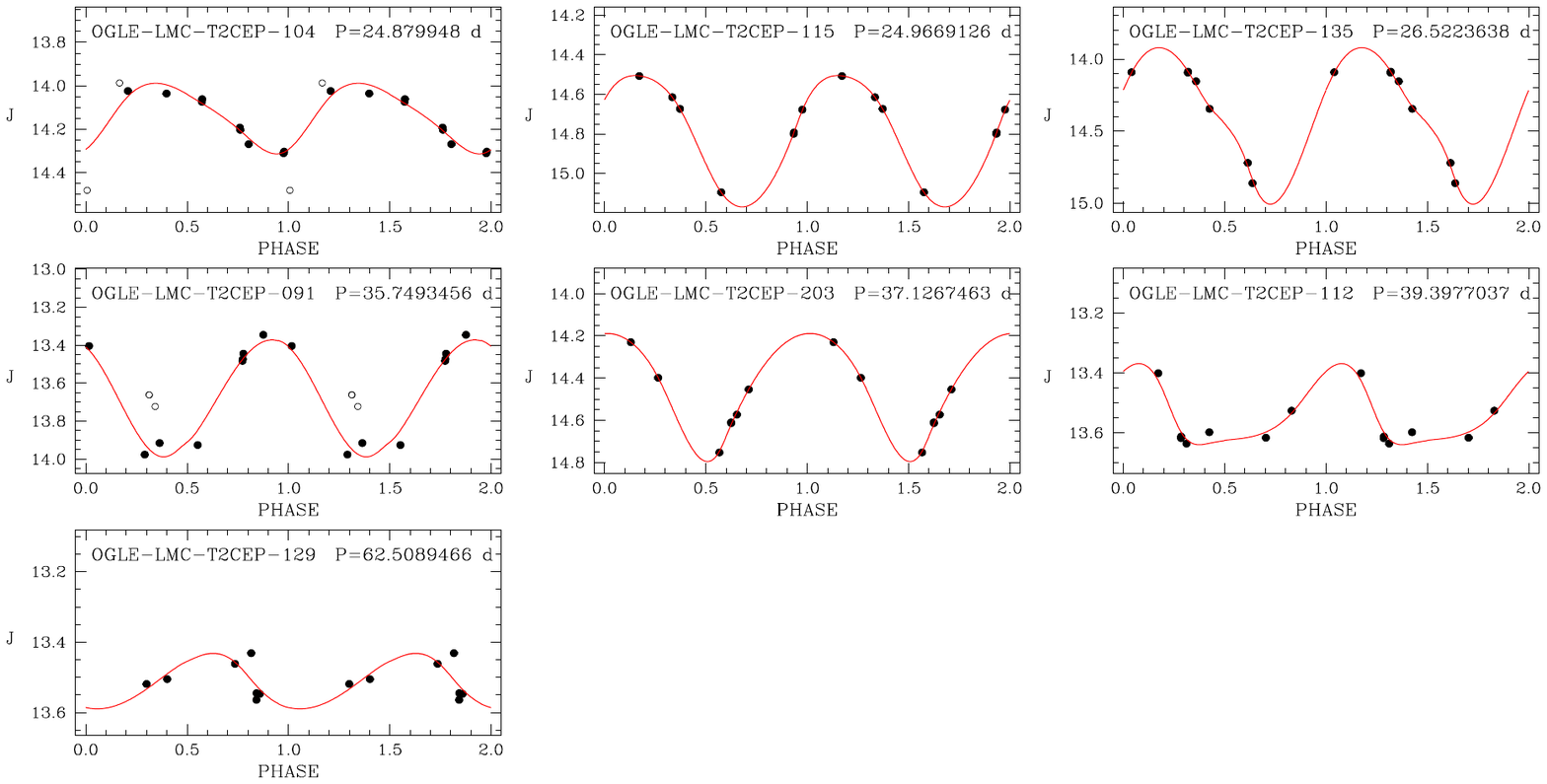}
\contcaption{}
\end{figure*}

\begin{figure*}
\includegraphics[width=16cm]{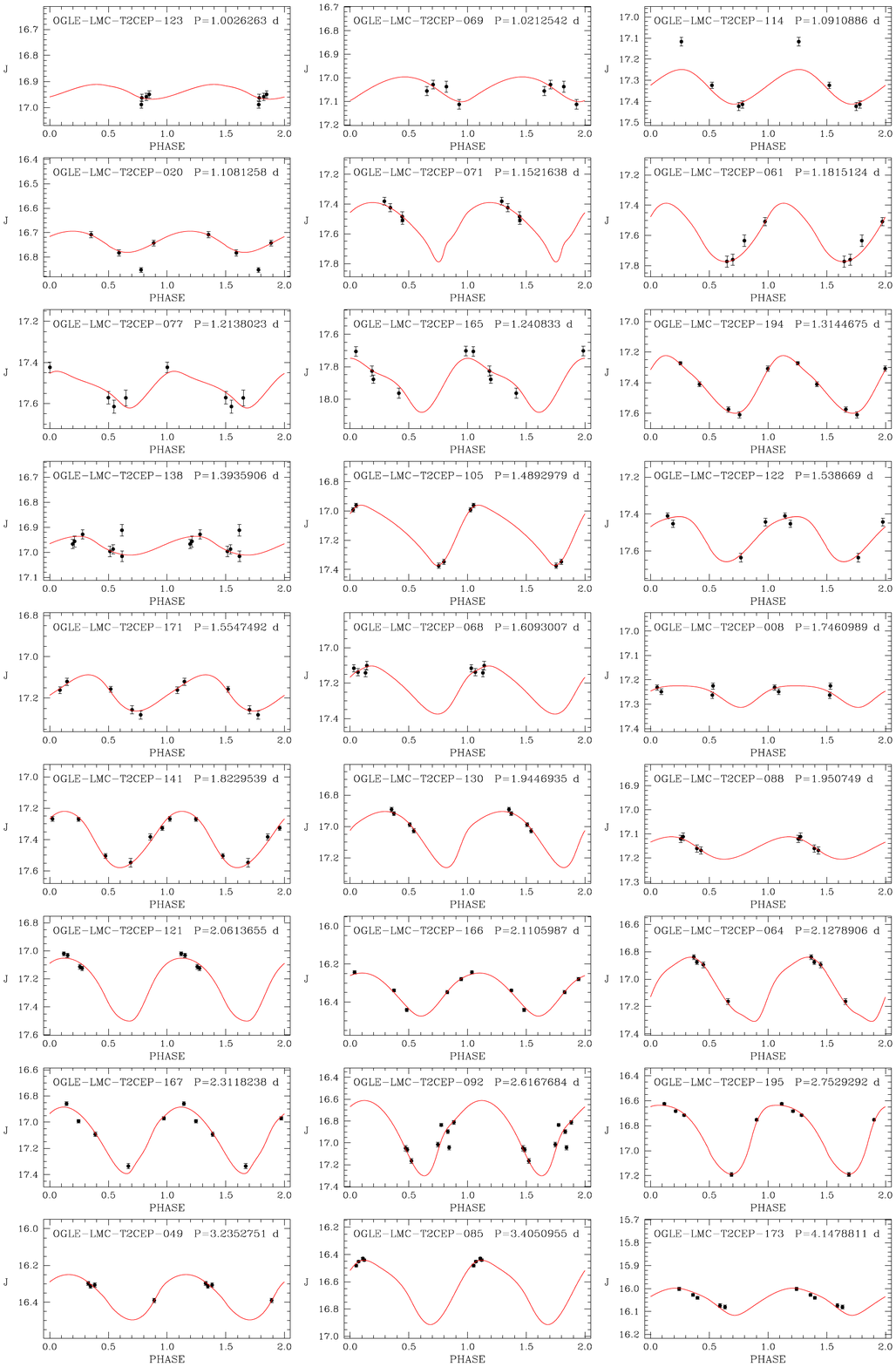}
\caption{J--band light curves for T2CEP stars not possessing a sufficient 
  number of epochs to perform the spline fit to the data and for which
  template fitting was used (see text). Stars are displayed in order of increasing 
period. Solid lines represent spline best-fits to the data (see 
text). In each panel we report OGLE's identification number  and 
period.} 
\label{figureJtempl}
\end{figure*}

\begin{figure*}
\includegraphics[width=16cm]{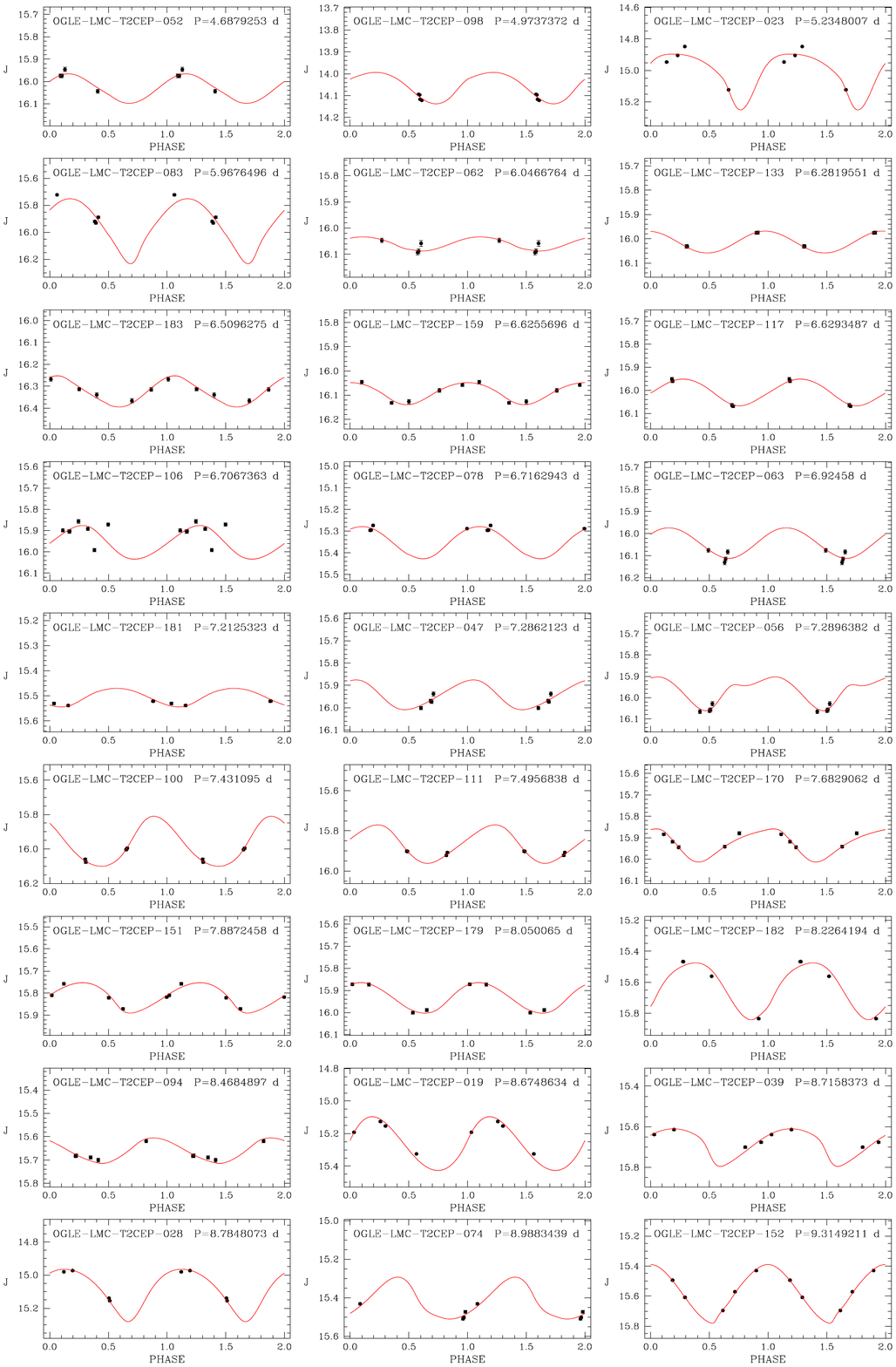}
\contcaption{}
\end{figure*}

\begin{figure*}
\includegraphics[width=16cm]{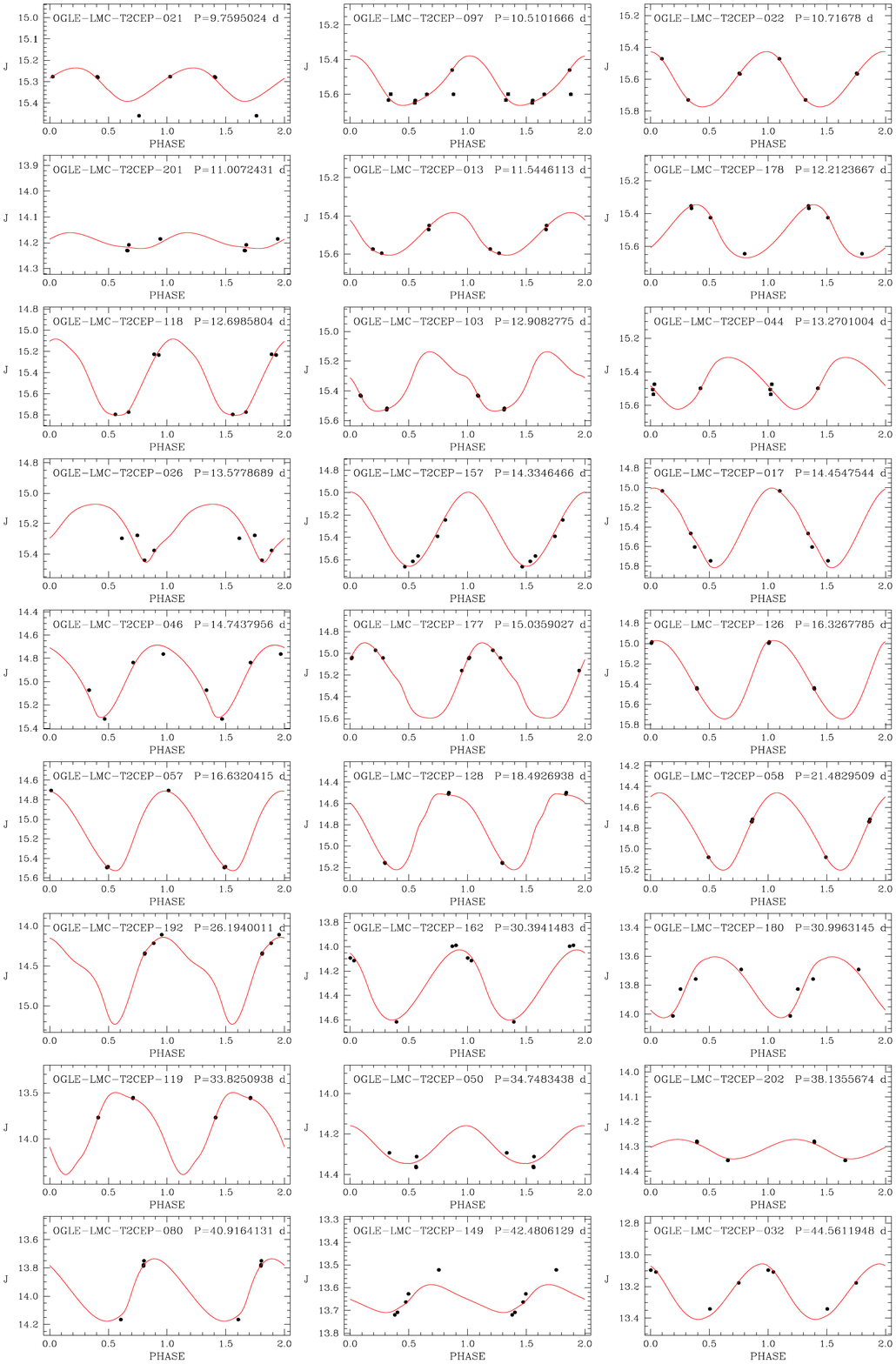}
\contcaption{}
\end{figure*}

\begin{figure*}
\includegraphics[width=16cm]{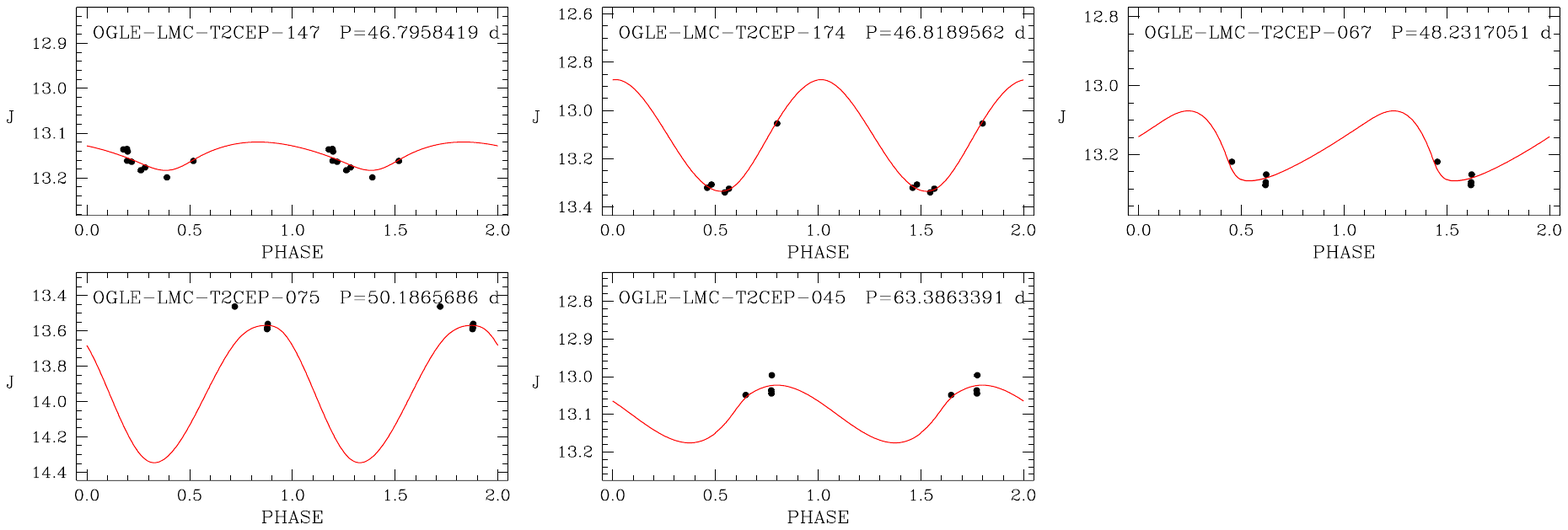}
\contcaption{}
\end{figure*}

\begin{figure*}
\includegraphics[width=16cm]{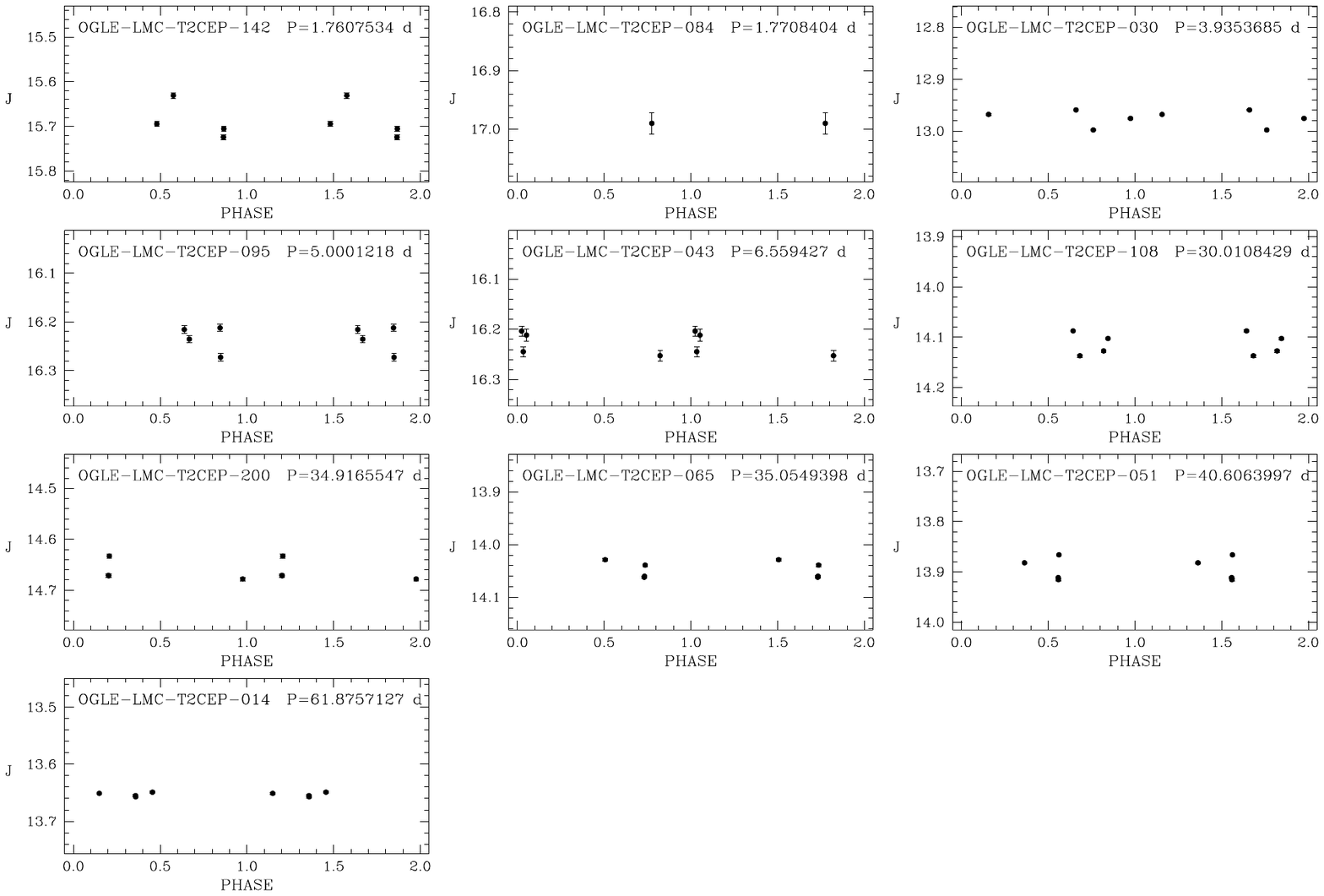}
\caption{Light curves for stars showing problems in the $J$- and $K_\mathrm{s}$-band (see text).} 
\label{figureJ_sfigate}
\end{figure*}


\end{document}